%% file: main.tex
\documentclass{article}

\usepackage{iclr2027_conference,times}
\input{math_commands.tex}

\usepackage{url}
\usepackage{graphicx}
\usepackage{booktabs}
\usepackage{array}
\usepackage{amssymb}
\usepackage{colortbl}
\usepackage{tabularx}
\usepackage{float}
\usepackage{multirow}
\usepackage{placeins}

\usepackage{fontawesome5}
\usepackage[hypertexnames=false]{hyperref}
\hypersetup{hidelinks}

\definecolor{seisDatasetOrange}{HTML}{F59E0B}
\definecolor{seisProjectBlue}{HTML}{2563EB}
\definecolor{seisMetricHeat}{HTML}{3B82F6}
\definecolor{seisRandomHeat}{HTML}{3B82F6}
\definecolor{seisInterpolationHeat}{HTML}{16A34A}
\definecolor{seisSurfaceHeat}{HTML}{14B8A6}
\definecolor{seisMultipleHeat}{HTML}{F97316}
\definecolor{seisDeblendingHeat}{HTML}{A855F7}
\definecolor{seisFirstArrivalHeat}{HTML}{EF4444}
\newcolumntype{L}[1]{>{\raggedright\arraybackslash}p{#1}}
\newcolumntype{C}[1]{>{\centering\arraybackslash}p{#1}}
\newcolumntype{M}[1]{>{\raggedright\arraybackslash}m{#1}}
\newcolumntype{N}[1]{>{\centering\arraybackslash}m{#1}}
\newcolumntype{Y}{>{\raggedright\arraybackslash}X}

\newcommand{\yes}{\(\checkmark\)}
\newcommand{\no}{\(\times\)}
\newcommand{\currentHeat}{seisMetricHeat}
\newcommand{\taskheat}[2]{\gdef\currentHeat{#1}#2}
\newcommand{\heatbest}[1]{\cellcolor{\currentHeat!50}\textbf{#1}}
\newcommand{\heatgood}[1]{\cellcolor{\currentHeat!34}#1}
\newcommand{\heatmid}[1]{\cellcolor{\currentHeat!22}#1}
\newcommand{\heatlow}[1]{\cellcolor{\currentHeat!12}#1}
\newcommand{\heatbase}[1]{\cellcolor{\currentHeat!5}#1}

\renewenvironment{abstract}{\vskip.02in\centerline{\large\sc Abstract}\vspace{-2ex}\begin{quote}}{\par\end{quote}\vskip 1ex}

\title{SPBench: A Multi-Task Evaluation Benchmark for Exploration Seismic Processing}

\author{Qi Liu$^{1}$, Tianxiang Gao$^{2}$, Zhitong Cheng$^{2}$, Chen Zhang$^{2}$, Peng Hu$^{2}$, Wei Gao$^{2}$, Jianwei Ma$^{1,2}$\\[1mm]
$^{1}$School of Earth and Space Sciences, Institute for Artificial Intelligence, Peking University, Beijing, China\\
$^{2}$School of Mathematics and Center of Geophysics, Harbin Institute of Technology, Harbin, Heilongjiang, China}

\iclrfinalcopy

\begin{document}

\maketitle
\vspace*{-1.4em}

\begin{figure}[H]
\centering
\setlength{\abovecaptionskip}{3pt}
\includegraphics[width=0.94\linewidth]{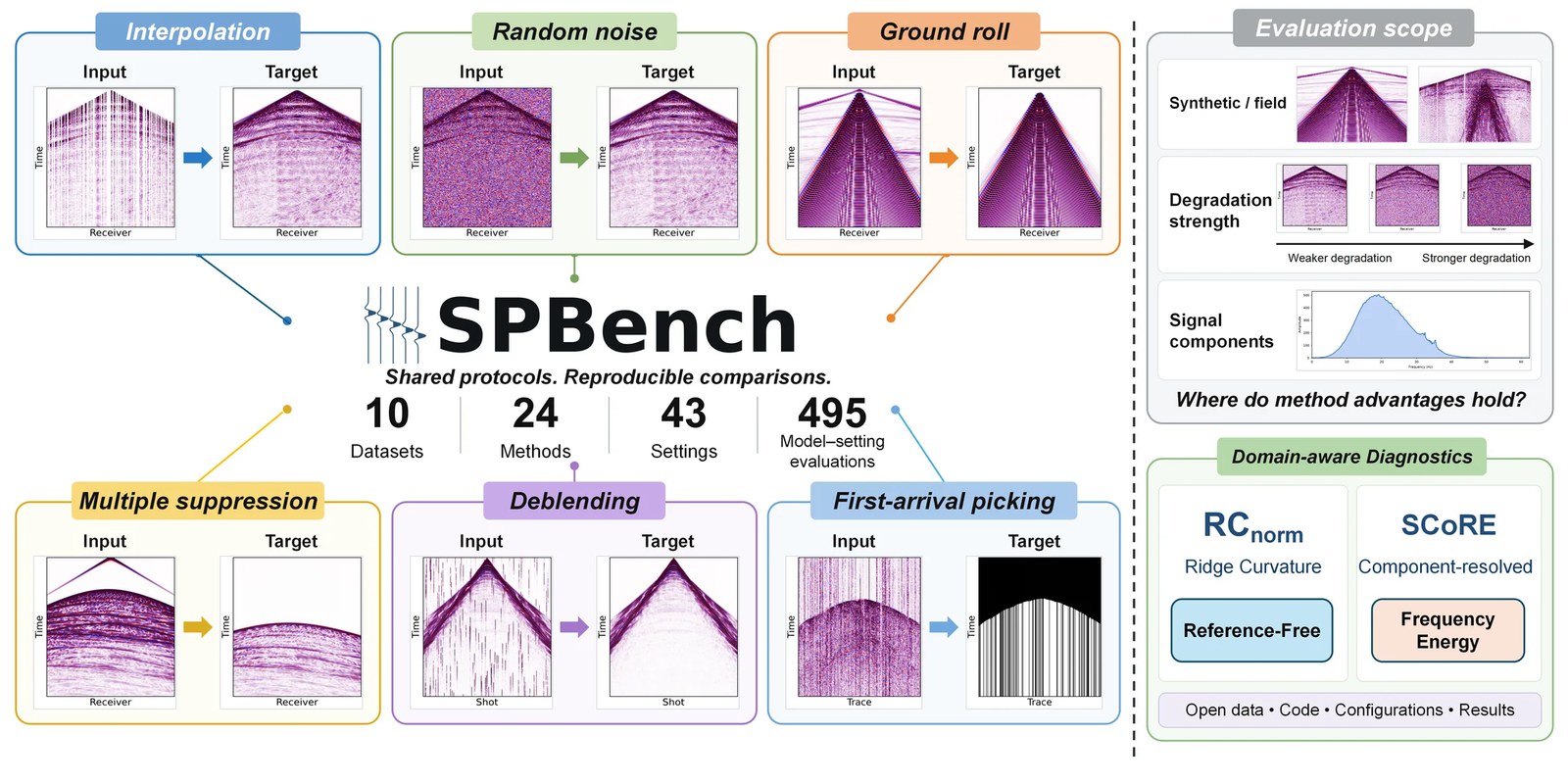}
\caption{\textsc{SPBench} overview. \textsc{SPBench} covers six exploration seismic processing tasks under shared protocols, with open datasets, implementations, configurations, and results. Project page: \url{https://seismicprocessingbenchmark.github.io/SPBench/}.}
\label{fig:overview}
\vspace{-0.5em}
\end{figure}

\begin{abstract}
Exploration seismic processing underpins subsurface imaging and resource exploration, where data quality shapes downstream interpretation and decision making. Learning-based methods have advanced rapidly, but their results remain difficult to compare across studies. Our literature survey of 368 papers reveals widespread reliance on private or difficult-to-reproduce datasets, and only 25 provide public code. This makes it difficult to attribute reported gains to model design rather than differences in experimental settings. To fill this gap, we introduce the Seismic Processing Benchmark (\textsc{SPBench}). \textsc{SPBench} covers six representative processing tasks, including random noise attenuation, trace interpolation, ground-roll noise suppression, multiple suppression, deblending, and first-arrival picking. We reproduce 24 supervised methods on 10 datasets under 43 standardized settings and release datasets, implementations, configurations, evaluation scripts, and results. Global scores hide frequency- and energy-dependent behavior, and per-trace pick errors ignore spatial continuity. We therefore introduce signal-component-resolved evaluation (SCoRE) for reconstruction and the reference-free ridge-curvature score ($\mathrm{RC}_{\mathrm{norm}}$) for first-arrival picking. Our analyses reveal three patterns. Synthetic rankings do not reliably predict field rankings, and the agreement varies by task when models train within each setting. As degradation strengthens, rankings reorder more under coherent ground roll than under random-like interference. The reference-free ridge score tracks MAE-based model rankings in the evaluated settings (mean Kendall correlation 0.881 across three field surveys), while SCoRE reveals component-dependent differences hidden by global scores. Together, these resources provide a fair and reproducible basis for comparing learning-based seismic-processing methods. The findings further characterize how method advantages vary across the evaluated conditions, and inform model selection.
\end{abstract}

\section{Introduction}

Oil and gas exploration relies on indirect geophysical measurements to infer subsurface structures and reservoir properties. Among these observations, exploration seismic data provide one of the most important sources of information for subsurface imaging, inversion, and geological interpretation \citep{yilmaz2001seismic}. These data are wavefields recorded at the surface (Appendix~\ref{app:primer}). The quality of seismic data processing therefore directly affects later interpretation and exploration decisions. Machine learning has recently become a central tool in this pipeline, with neural networks increasingly used for seismic processing, interpretation, inversion, and foundation-model workflows \citep{bergen2019machine,yu2021deeplearning,wu2019faultseg3d,harsuko2022storseismic,sheng2025seismicfoundation,liu2025foundation}. This growth has created a comparability problem. The literature contains many sophisticated architectures, but the field lacks shared evaluation infrastructure. It is difficult to tell which methods generalize, and whether reported gains reflect model design or differences in data, implementation, and evaluation.

We quantify this fragmentation with a curated corpus of 368 supervised deep-learning papers published between 2018 and 2026 in journals that regularly publish machine-learning research for exploration seismic processing (scope and criteria in Appendix~\ref{app:survey}). Figure~\ref{fig:reproducibility-landscape} summarizes the corpus. Two barriers stand out. First, most studies rely on private or difficult-to-reconstruct datasets with often incompletely specified generation procedures (Figure~\ref{fig:reproducibility-landscape}(a)). Second, released code is rare. Only 25 of the 368 papers include a public code link (Figure~\ref{fig:reproducibility-landscape}(b)). These patterns make published setups hard to reproduce and genuine progress hard to isolate.

To address these barriers, we introduce the Seismic Processing Benchmark (\textsc{SPBench}), an open multi-task benchmark for learning-based exploration seismic processing (Figure~\ref{fig:overview}). \textsc{SPBench} improves comparability at three levels. Data construction, method implementations, and evaluation protocols are standardized across six representative tasks, including random noise attenuation, trace interpolation, ground-roll noise suppression, multiple suppression, deblending, and first-arrival picking \citep{liu2023generativeinterpolation,li2024robustdenoising,yang2023softattentiongroundroll,wang2023surfacerelatedmultiple,wang2023iterativedeblending,stcharles2023hardpicks}. Four tasks pair synthetic settings with field settings, and three tasks provide several controlled degradation levels, so methods can be compared across data conditions and degradation strengths. For multiple suppression, deblending, and ground-roll noise suppression, we release new paired synthetic datasets. The benchmark provides 10 datasets under 43 standardized settings. We implement and evaluate 24 supervised methods and report 495 model-setting evaluations over three random seeds. All datasets, implementations, splits, configurations, evaluation scripts, and per-seed results are released.

How the results are scored is part of the comparison itself. Global metrics such as the signal-to-noise ratio (SNR) and the mean absolute error (MAE) compress performance into a single number and ignore the physical structure of the data. Within one survey area, traces with similar source-receiver offsets have similar first-arrival times, so predicted picks should form a spatially continuous travel-time ridge. Building on this prior, the reference-free ridge-curvature score ($\mathrm{RC}_{\mathrm{norm}}$) measures the geometric continuity of predicted ridges, and on three field surveys it agrees with pick-error rankings at a mean Kendall rank correlation of 0.881. A second diagnostic, signal-component-resolved evaluation (SCoRE), resolves reconstruction fidelity across frequency bands with frequency-binned fidelity and recovery evaluation (FB-FRE) and across local reference-energy levels with energy-binned fidelity evaluation (EB-FE).

\textsc{SPBench} evaluates not only which methods perform well, but also how their relative advantages change across data settings, degradation strengths, and evaluation criteria. We decompose this question into three measurements. How consistent are method rankings between corresponding synthetic and field settings? How stable are rankings as degradation severity increases? Do the proposed diagnostics track the standard scores? The answers show that advantages are real but conditional. Between corresponding synthetic and field settings, where every method is trained and tested within the setting, ranking agreement varies by task, from a shared leading method to a complete turnover of the top three. Rankings remain moderately stable as random-like interference strengthens, but reorder under coherent ground roll. The ridge score agrees with pick-error rankings at 0.881, and the SCoRE partitions agree with SNR only to a degree. These advantages should therefore be read together with the conditions under which they were measured.

\begin{figure}[t]
\centering
\begin{tabular}{@{}c@{\hspace{0.03\linewidth}}c@{}}
\includegraphics[width=0.49\linewidth]{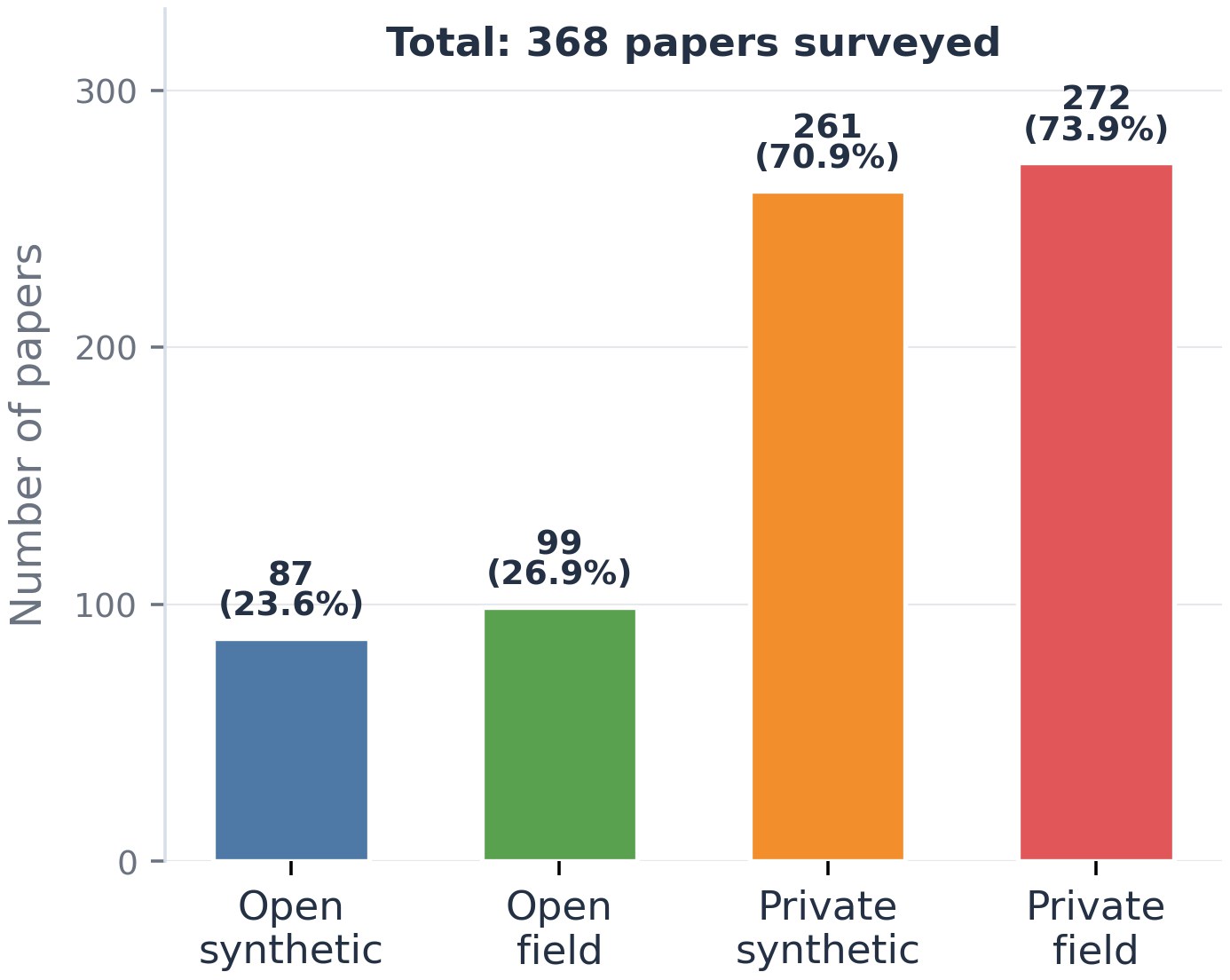} &
\includegraphics[width=0.38\linewidth]{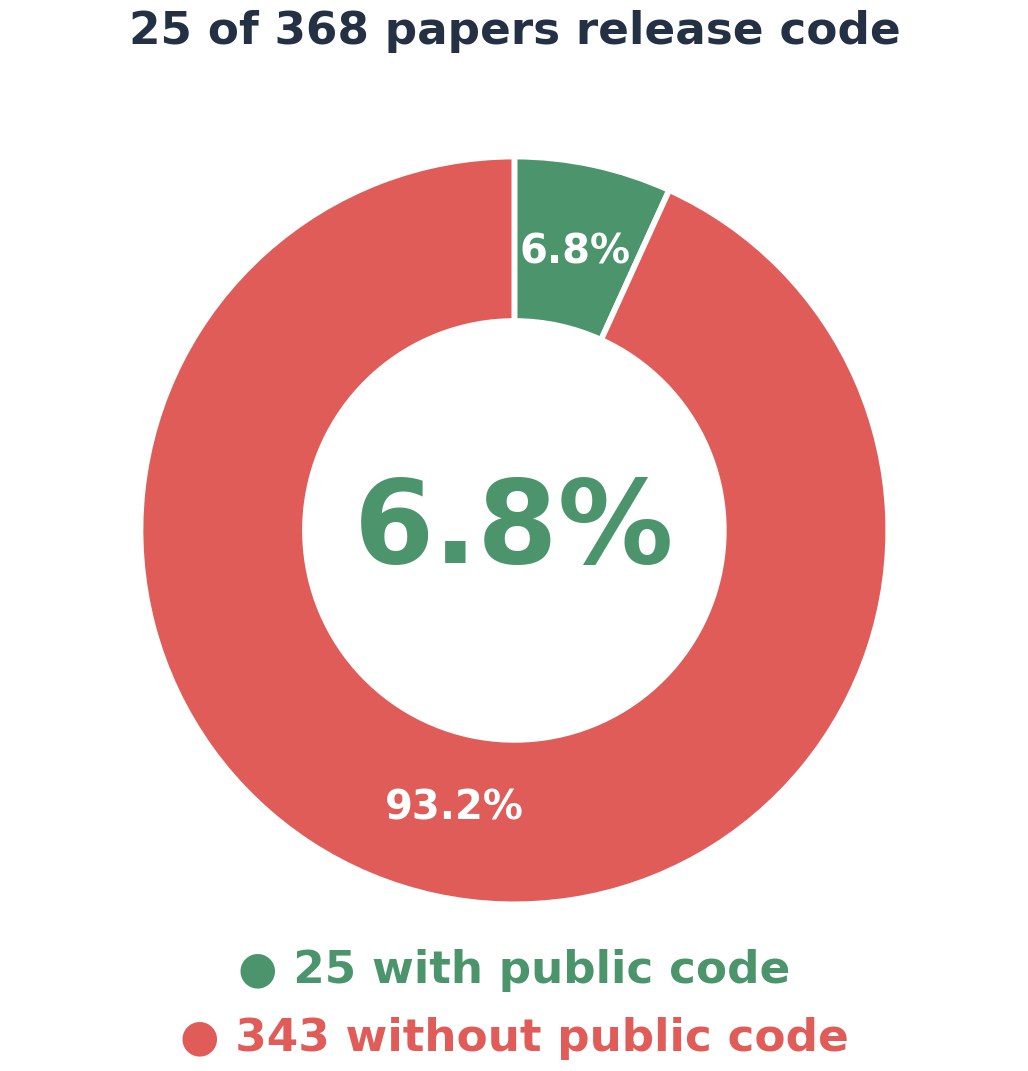} \\
{\small\textbf{(a) Dataset-source usage}} & {\small\textbf{(b) Code availability}}
\end{tabular}
\caption{Fragmented reproducibility in seismic-processing deep learning, measured on the 368-paper survey corpus. (a) Multi-label dataset-source usage. A paper may use several source types. (b) Only 25 of 368 papers include a public code URL.}
\label{fig:reproducibility-landscape}
\end{figure}

\section{Related Work}

\begin{table}[!t]
\caption{Comparison of representative exploration-geophysics benchmarks and resources.}
\label{tab:coverage}
\centering
\scriptsize
\setlength{\tabcolsep}{1.2pt}
\renewcommand{\arraystretch}{1.0}
\setlength{\extrarowheight}{2pt}
\begin{tabular*}{\linewidth}{@{\extracolsep{\fill}}M{0.27\linewidth}M{0.15\linewidth}N{0.06\linewidth}N{0.085\linewidth}N{0.042\linewidth}N{0.042\linewidth}N{0.075\linewidth}N{0.10\linewidth}@{}}
\toprule
Benchmark / reference & Primary scope & Type & Baselines & Code & Data & Params & \shortstack{Domain\\metrics} \\
\midrule
FaultSeg3D~\citep{wu2019faultseg3d} & Fault detection & Single & \no & \yes & \yes & \no & \no \\
OpenFWI~\citep{deng2022openfwi} & Acoustic FWI & Single & 4 methods & \yes & \yes & \yes & \no \\
E-FWI~\citep{feng2023efwi} & Elastic FWI & Single & 4 methods & \yes & \yes & \no & \no \\
HardPicks~\citep{stcharles2023hardpicks} & First-arrival picking & Single & 3 methods & \yes & \yes & \no & \no \\
Swell-noise~\citep{barros2024swell} & Denoising & Single & 2 methods & \no & \yes & \no & \no \\
OpenSeisML~\citep{bhar2026openseisml} & Inversion & Single & \no & \no & \yes & \no & \no \\
CIG-Bench~\citep{dou2026cigbench} & Multi-task interpretation & Multi & 8 methods & \yes & \yes & \yes & \no \\
\midrule
\rowcolor{black!8}\textbf{\textsc{SPBench}} & \textbf{Multi-task seismic processing} & \textbf{Multi} & \textbf{24 methods} & \yes & \yes & \textbf{$>$1000} & \textbf{\shortstack{SCoRE\\$\mathrm{RC}_{\mathrm{norm}}$}} \\
\bottomrule
\end{tabular*}
\end{table}

Scientific machine-learning benchmarks increasingly promote comparable research by pairing shared datasets with reproduced baselines, evaluation protocols, and task-relevant metrics~\citep{takamoto2022pdebench,ohana2024well,rasp2023weatherbench2,hu2026realpdebench}. Exploration-geophysics benchmark resources have begun to appear, but they target tasks that are mostly adjacent to seismic processing. OpenFWI~\citep{deng2022openfwi} and E-FWI~\citep{feng2023efwi} focus on acoustic and elastic full-waveform inversion, with public datasets and reproduced baselines. OpenSeisML~\citep{bhar2026openseisml} contributes real seismic and well-log data for inversion, but is primarily a data resource rather than a full evaluation suite with released code and parameters. FaultSeg3D~\citep{wu2019faultseg3d} and CIG-Bench~\citep{dou2026cigbench} address seismic interpretation, ranging from fault detection to multi-task subsurface interpretation. These resources are complementary to \textsc{SPBench} rather than overlapping. The inversion resources estimate subsurface models directly from recorded wavefields, and the interpretation resources analyze migrated images and volumes. \textsc{SPBench} instead standardizes the processing tasks that prepare data for imaging, inversion, and interpretation. HardPicks~\citep{stcharles2023hardpicks} and the swell-noise benchmark~\citep{barros2024swell} provide valuable single-task resources for first-arrival picking and denoising. Existing open resources still lack task-specific synthetic datasets and standardized evaluation for multiple suppression, seismic deblending, and ground-roll noise suppression. As Table~\ref{tab:coverage} summarizes, \textsc{SPBench} unifies six processing tasks under one shared protocol, pairs its datasets with 24 reproduced supervised baselines and 495 released configurations, and adds the seismic-specific diagnostics SCoRE and $\mathrm{RC}_{\mathrm{norm}}$, which existing resources do not provide.

\section{\textsc{SPBench} Overview}

\subsection{Tasks and Data}

\textsc{SPBench} defines 43 settings across six representative seismic-processing tasks and 10 datasets. All six tasks share one mathematical structure and differ only in the degradation operator and the prediction target. Let $x$ denote a clean shot record and $\mathcal{C}_t$ a task-specific degradation operator. The observed input is
\begin{equation}
\tilde{x} = \mathcal{C}_t(x), \qquad \mathcal{C}_t(x) \in \{\, x + n,\;\; M \odot x,\;\; x + \alpha g,\;\; x + m,\;\; \bm{B}x \,\},
\label{eq:task-operator}
\end{equation}
covering additive random noise $n$ (random-noise attenuation), a missing-trace mask $M$ with $\odot$ denoting element-wise multiplication (interpolation), scaled ground roll $\alpha g$ (ground-roll suppression), multiple energy $m$ (multiple suppression), and the blending operator $\bm{B}$ (deblending). First-arrival picking uses the recorded data directly and is annotated rather than degraded. Each method learns a mapping $f_\theta : \tilde{x} \mapsto \hat{y}$ with parameters $\theta$, and the prediction target is chosen per task or method design. Most methods predict the clean signal directly. For ground-roll and multiple suppression, methods predict the degradation component, and the clean signal is recovered by subtraction as $\hat{x} = \tilde{x} - \hat{y}$. All methods are trained and evaluated under shared train/validation/test splits. Appendix~\ref{app:tasks-data} describes the datasets and forward modeling, Appendix~\ref{app:experiment-design} details the experiment design, and Appendix~\ref{app:primer} explains how the benchmark data are recorded and simulated.

\paragraph{Seismic interpolation.}
Seismic interpolation restores absent traces so that a sparsely sampled acquisition approximates a spatially continuous wavefield. The benchmark uses the SEG C3 synthetic dataset and the Mobil AVO Viking Graben Line 12 field dataset, both openly distributed through the SEG open-data catalog\footnote{SEG open-data catalog: \url{https://wiki.seg.org/wiki/Open_data}}. Here SEG denotes the Society of Exploration Geophysicists and AVO denotes amplitude versus offset. We design three families of masks, random missing, uniform missing, and consecutive missing, each at several missing levels. This task contributes 16 benchmark settings.

\paragraph{Random noise attenuation.}
Random noise attenuation suppresses incoherent fluctuations while retaining coherent reflections, diffractions, and weak geological events. The additive noise $n$ is seeded Gaussian or Poisson noise injected at a prescribed input SNR. The benchmark builds on the same two SEG open datasets as the interpolation task, with three SNR levels per noise family. This task contributes 12 benchmark settings.

\paragraph{Ground-roll noise suppression.}
Ground-roll noise suppression removes high-amplitude, low-velocity ground roll that masks reflection energy in land seismic records. The degradation $g$ is elastically modeled ground roll scaled by an amplitude multiplier $\alpha$. Methods predict $\hat{g}$, and the clean signal is recovered as $\hat{x} = \tilde{x} - \hat{g}$. We release the \textsc{SPBench} ground-roll datasets, comprising a synthetic set built on SEG C3 data and a companion field set. The synthetic set covers five multiplier levels. This task contributes 6 settings.

\paragraph{Multiple suppression.}
Multiple suppression separates primary reflections that directly describe subsurface interfaces from repeated-bounce energy that can produce misleading structural events. The degradation component $m$ is the non-primary energy whose exact composition is specified in Appendix~\ref{app:tasks-data}. Methods predict $\hat{m}$, and the primaries are recovered as $\hat{x} = \tilde{x} - \hat{m}$. We release the \textsc{SPBench} multiples dataset, whose forward-modeling construction is detailed in Appendix~\ref{app:tasks-data}. This task contributes a single benchmark setting.

\paragraph{Deblending.}
Deblending disentangles overlapping wavefields recorded during simultaneous-source acquisition into the responses of individual shots. The blending operator $\bm{B}$ sums source responses shifted by random firing times. We release the \textsc{SPBench} deblending datasets, comprising a synthetic set of randomly blended common-receiver gathers with retained unblended references and a companion semi-synthetic field set built on Mobil AVO records. The synthetic settings cover three blending levels. This task contributes 4 settings.

\paragraph{First-arrival picking.}
First-arrival picking identifies the earliest physically meaningful seismic onset on each trace, providing travel-time observations for statics correction and subsurface velocity estimation. The task maps a record $x$ to per-trace arrival times $t^*$, realized by predicting a binary step mask per trace and reading $t^*$ at its zero-to-one transition. The benchmark uses the Brunswick, Halfmile, and Lalor surveys from HardPicks~\citep{stcharles2023hardpicks}, evaluated both per survey and on their combined evaluation set. This task contributes 4 benchmark settings.

\subsection{Domain-Aware Evaluation Metrics}
\label{sec:domain-aware-metrics}

The benchmark reports standard reconstruction and picking metrics, including SNR, peak signal-to-noise ratio (PSNR), MAE, root mean squared error (RMSE), and tolerance-based hit rates (H@$k$). These metrics summarize how much error remains globally, but not how reconstruction fidelity varies across signal components or whether predictions preserve the spatial continuity of seismic events across traces. \textsc{SPBench} therefore adds two kinds of domain-aware diagnostics alongside the standard scores. They complement rather than replace the aggregate metrics.

\paragraph{Signal-component-resolved evaluation (SCoRE).}
SCoRE measures fidelity within partitions of the reference signal instead of only globally. Its two instances partition by different signal properties and report the same quantity per partition $\mathcal{S}$, the partition SNR
\begin{equation}
\mathrm{SNR}_{\mathcal{S}} = 10\log_{10}\frac{\|r_{\mathcal{S}}\|_2^2}{\|p_{\mathcal{S}}-r_{\mathcal{S}}\|_2^2+\varepsilon},
\end{equation}
where $r_{\mathcal{S}}$ and $p_{\mathcal{S}}$ are the reference and prediction restricted to the partition and $\varepsilon$ is a small positive constant for numerical stability. FB-FRE partitions by frequency. It estimates the effective frequency band from the reference power spectrum, divides it into four adaptive subbands (low, mid, high, very high), and restricts the signals through frequency masks, with $M_b$ the mask of subband $b$ and RFFT and IRFFT the real discrete Fourier transform along time and its inverse,
\begin{equation}
r_{\mathcal{S}}=r_b=\mathrm{IRFFT}\!\left(M_b \cdot \mathrm{RFFT}(r)\right),\quad
p_{\mathcal{S}}=p_b=\mathrm{IRFFT}\!\left(M_b \cdot \mathrm{RFFT}(p)\right).
\end{equation}
A large gap between low- and high-frequency scores indicates that high-frequency detail is being removed together with noise. EB-FE partitions by local reference energy. Global scores can be dominated by strong reflections, so per-regime behavior is invisible in them. EB-FE builds a smoothed reference-energy map and groups samples into energy-percentile bins (very weak, weak, medium, strong), and each bin directly forms $\mathcal{S}$. Low-energy bins serve as a proxy for weak-signal regimes rather than explicitly identifying weak geological events. Both instances require a clean reference, and per-partition rankings agree with the global SNR ranking only to a degree (Appendix~\ref{app:metrics}). Full construction details are given in Appendix~\ref{app:metrics}.

\paragraph{First-arrival ridge continuity.}
Standard picking metrics score each trace independently. They do not use a basic physical prior. Within one survey area, traces with similar source-receiver offsets have similar first-arrival times, so the predicted picks should form a spatially continuous travel-time ridge. The ridge curvature score ($\mathrm{RC}_{\mathrm{norm}}$) measures how much the predicted ridge wiggles along the offset direction. It divides the offset range into $B$ equal-width bins, represents each bin by its median predicted arrival, and averages the absolute slope change along the resulting ridge,
\begin{equation}
\mathrm{RC}_{\mathrm{norm}} = \frac{X_{\mathrm{range}}^{2}}{T_{\mathrm{range}}}\cdot\frac{1}{B-2}\sum_{j=3}^{B}\lvert c_j\rvert,\qquad c_j=\frac{d_j-d_{j-1}}{x_j-x_{j-1}},\qquad d_j=\frac{t_j-t_{j-1}}{x_j-x_{j-1}},
\end{equation}
where $\{(x_j,t_j)\}_{j=1}^{B}$ are the ridge points in offset-time coordinates, $X_{\mathrm{range}}$ is the offset range of the valid picks and $T_{\mathrm{range}}$ the time range of the ridge, and the leading factor removes first-order differences in spatial and temporal scale. An affine ridge scores zero, and locally changing slopes increase the value. A low value therefore indicates a smooth aggregated ridge, not accurate picks on every trace. The median binning suppresses outlier traces by design, which adds robustness and also marks the diagnostic boundary. The metric is reference-free. It uses the geometry of the prediction itself, so it complements label-based accuracy metrics. Rearranging a fixed set of per-trace errors along the offset axis leaves MAE, RMSE, the mean bias error (MBE), and every H@$k$ unchanged by construction, while $\mathrm{RC}_{\mathrm{norm}}$ changes, which isolates the geometric information it captures. Full construction details are given in Appendix~\ref{app:metrics}.

\subsection{Reproduced Baselines}

\textsc{SPBench} reproduces 24 supervised deep-learning methods under shared protocols within each task. The baseline set combines four common architectures, U-Net~\citep{chai2020unet}, ResUNet~\citep{zhang2018resunet}, DnCNN~\citep{yu2021deeplearning}, and attention-gated U-Net~\citep{oktay2018attentionunet}, with 20 task-specific methods. The common architectures provide reference points for comparison across tasks, while the task-specific methods represent designs developed for individual processing problems. Together, they span representative model families, including convolutional and residual networks~\citep{liao2023fbresnet,wang2022cbdrdn}, U-Net variants~\citep{zhang2024unetpp,wang2025qunet}, attention-based networks and Transformers~\citep{yu2022anet,jiang2023stunet}, generative adversarial networks~\citep{tao2022sagan,yuan2020gan}, diffusion models~\citep{li2024cddpm}, and model-driven or physics-constrained networks~\citep{wu2024spnet,pham2022physics}. This combination supports comparisons of how general-purpose and specialized designs perform across the evaluated settings. Appendix~\ref{app:methods} provides the complete method list, references, descriptions, and per-task assignments.

\section{Experimental Setup and Results}

\subsection{Evaluation Protocol and Main Results}
\label{sec:protocol-results}

All methods are evaluated under a shared protocol. One benchmark setting pairs a dataset with a task and, where applicable, a controlled degradation configuration. Every method in a setting uses the same train/validation/test split, the same input-target construction, and the same evaluation scripts, and validation data are used only for model selection and early stopping. We call one method evaluated on one setting an evaluation and one complete train-and-test execution with a single seed a run. Each method--setting configuration is trained independently with three random seeds. Reported results are means over the three seeds. Hyperparameters follow the source papers, and for widely used architectures such as U-Net, multiple parameter groups appear as width variants in the result tables. Reported numbers are therefore not directly comparable to values in the original papers. Setting construction, corruption protocols, and split details are given in Appendix~\ref{app:experiment-design}. The 495 configurations therefore release more than 1,000 trained parameter sets. Each reported result traces to one released configuration and to the per-seed results of its three runs, all queryable on the project page.

Tables~\ref{tab:main-results} and~\ref{tab:main-picking-results} list the top eight methods per representative setting from the synchronized public leaderboard, with standard deviations over the three seeds in small type. First-arrival picking is reported separately because its MAE and tolerance-based hit rates are not commensurate with reconstruction metrics. Complete results appear in Appendix~\ref{app:complete-results}. The remainder of this section varies one evaluation condition at a time and asks whether the conclusions hold. Section~\ref{sec:transfer} varies the data condition, Section~\ref{sec:robustness} the degradation strength and interference type, and Section~\ref{sec:metrics-validation} the metric itself.

\input{tables/main_results_tables.tex}

\subsection{Do Synthetic Rankings Predict Field Rankings?}
\label{sec:transfer}

\begin{figure}[t]
\centering
\includegraphics[width=0.95\linewidth]{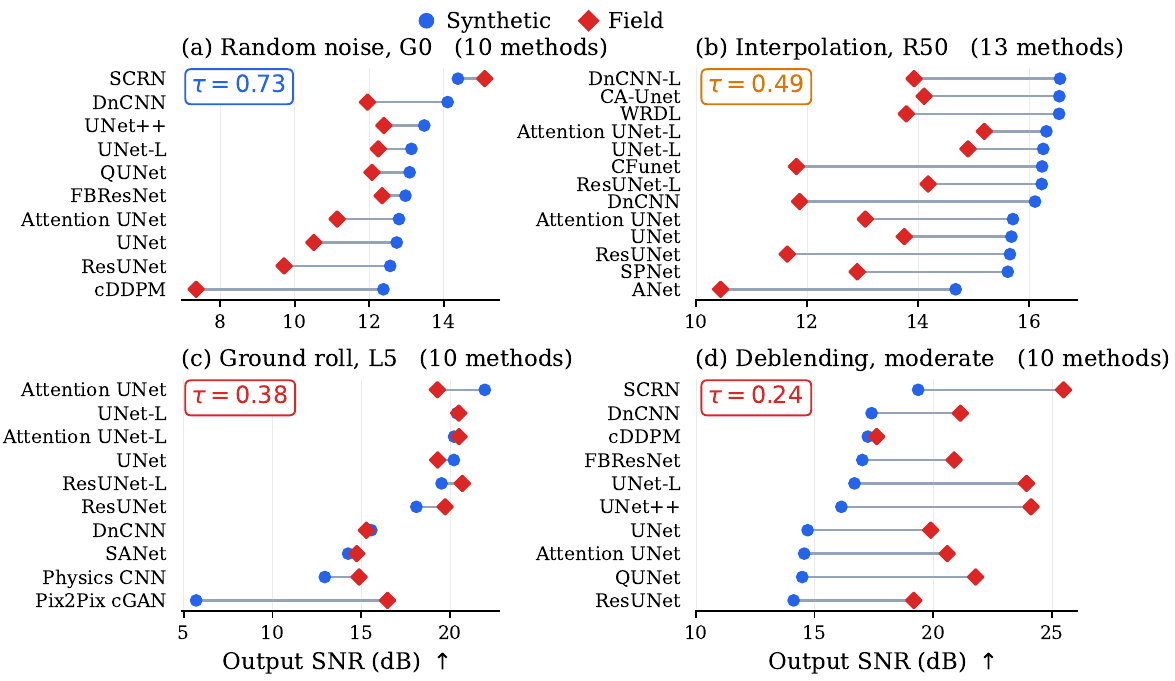}
\caption{Synthetic-to-field performance under matched protocols. Each point reports the mean output SNR over three random seeds, and horizontal segments connect the same method across paired synthetic and field settings. Methods are ordered by their synthetic-setting performance, with the best-performing method placed at the top. Kendall's $\tau_b$ measures the agreement between the two method rankings. Models are trained and evaluated separately within each setting; the comparison therefore measures ranking consistency rather than cross-domain generalization. Only methods evaluated in both settings are included.}
\label{fig:transfer-pairs}
\end{figure}

Does performance on the synthetic setting predict performance on the corresponding field setting? Figure~\ref{fig:transfer-pairs} shows every method evaluated on both settings of each pair, sorted by synthetic performance. All methods are trained and tested within each setting (Appendix~\ref{app:experiment-design}), so this comparison concerns ranking consistency rather than cross-domain deployment. We quantify ranking agreement by the Kendall rank correlation $\tau$, which is the concordant minus the discordant fraction of all method pairs (Appendix~\ref{app:metrics}) and varies widely across tasks. On random noise the rankings partially survive ($\tau = 0.73$), yet 70\% of methods still lose more than 1 dB in absolute SNR. On interpolation the correlation drops to $\tau = 0.49$ and every method loses performance, with a median drop of 2.7 dB. On ground-roll suppression the correlation drops to $\tau = 0.38$, so the synthetic ranking is only a weak guide to the field ranking. On deblending the direction inverts and every method gains on the field set (median +5.6 dB), yet the ranking agreement is the weakest of the four pairs ($\tau = 0.24$). The leading method is shared for random noise and deblending (SCRN), while the top three interpolation methods turn over completely. Absolute gaps partly reflect reference construction, so we read them as setting-difficulty differences rather than deployment degradation. Synthetic performance therefore does not, by itself, reliably predict field performance. \textbf{Takeaway.} Synthetic rankings carry only partial information about field rankings, and the degree varies by task, from a shared leading method to a complete turnover of the top three. Model claims should therefore be validated on matched synthetic and field settings. Classic architectures remain strong baselines: DnCNN ranks in the top three of all three synthetic deblending settings and U-Net variants lead interpolation in 9 of its 16 settings.

\subsection{How Robust Are Models to Noise Strength?}
\label{sec:robustness}

\begin{figure}[t]
\centering
\includegraphics[width=\linewidth]{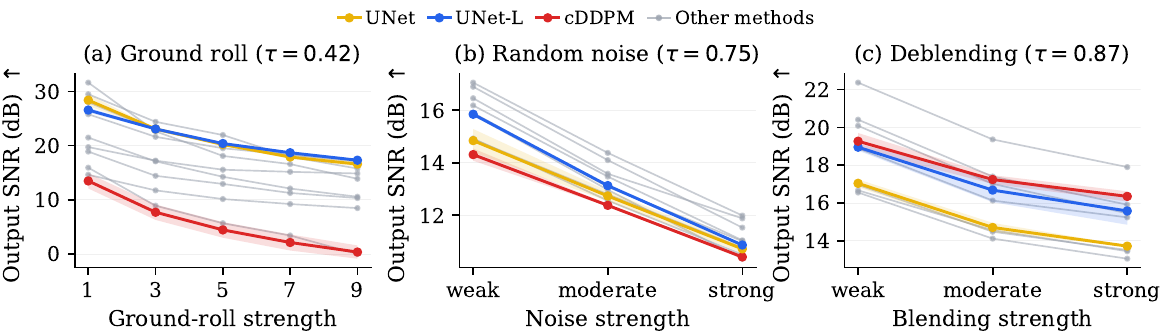}
\caption{Output SNR across degradation strengths for the three multi-level tasks. Curves show means over three random seeds; shaded bands give one standard deviation for the highlighted UNet (yellow), its width-increased variant UNet-L (dark blue), and cDDPM (red); all other methods are in light gray. Panel titles report the Kendall $\tau$ between endpoint levels. (a) Rankings partially reorder as ground-roll strength grows: cDDPM collapses to 0.4 dB, and UNet-L matches UNet at L3 and leads from L5 onward. (b,c) UNet-L leads UNet at every noise and blending level.}
\label{fig:robustness}
\end{figure}

Figure~\ref{fig:robustness} tracks output SNR as degradation strengthens in the three tasks that provide multiple levels. On random noise (Figure~\ref{fig:robustness}(b)) and on deblending (Figure~\ref{fig:robustness}(c)), absolute scores degrade almost in parallel and rankings remain moderately stable. The Kendall correlation between the weakest and strongest levels is 0.75 for Gaussian noise and 0.87 for deblending, and SCRN leads at every noise level. Ground roll behaves differently: rankings partially reorder as strength grows, with $\tau = 0.42$ between levels 1 and 9. ResUNet falls from first at 31.6 dB to seventh at 13.9 dB, and cDDPM collapses from 13.5 dB to 0.4 dB. The width-increased variant UNet-L (Appendix~\ref{app:methods}) behaves oppositely, starting 1.8 dB below UNet at L1, matching it at L3, and ending 0.7 dB ahead at L9. Under random noise and blending, UNet-L leads UNet at every level. The random-noise and ground-roll settings share the same clean reference shots (SEG C3 shots 1--9), so this contrast is observed between interference types on the same background. One plausible account is morphological. Ground roll forms dispersive, low-frequency, high-amplitude cones that overlap reflection events, unlike the random-like interference of the other two tasks. Random noise is incoherent by construction, and blended energy appears as incoherent vertical bursts in the common-receiver gathers used here, a form that classical random-noise suppression flows are designed to attenuate. This is an association, not a tested mechanism. \textbf{Takeaway.} Ranking stability under growing interference is a separate selection dimension. It is moderate under random-like corruption but weak under coherent ground roll, so it must be assessed per interference type.

\subsection{Can We Rank Picking Models Without Reference Labels?}
\label{sec:metrics-validation}

\begin{table}[t]
\centering
\footnotesize
\setlength{\tabcolsep}{4pt}
\caption{Kendall rank correlation of the reference-free ridge score with MAE-based and hit-rate-based model rankings (rows 1--2) and between its two aggregation rules (row 3), per field picking survey over the ten evaluated methods.}
\label{tab:rcnorm-validation}
\begin{tabular}{@{}lcccc@{}}
\toprule
Ranking pair & Brunswick & Halfmile & Lalor & Mean \\
\midrule
$\mathrm{RC}_{\mathrm{norm}}$ vs pick error (MAE) & 0.778 & 0.911 & 0.956 & 0.881 \\
$\mathrm{RC}_{\mathrm{norm}}$ vs $1-$H@5 & 0.689 & 0.899 & 0.822 & 0.803 \\
Shot macro vs trace weighted & 0.956 & 0.956 & 1.000 & 0.970 \\
\bottomrule
\end{tabular}
\end{table}

MAE and hit rates both require reference picks, and on field data many traces lack usable labels entirely (Figure~\ref{fig:picking-examples}, Appendix~\ref{app:tasks-data}). $\mathrm{RC}_{\mathrm{norm}}$ avoids this dependency because it reads the geometry of the prediction itself. It is useful for model selection only if its model rankings agree with labeled evaluation, which we test directly. On each of the three field surveys, we rank the ten evaluated methods by mean $\mathrm{RC}_{\mathrm{norm}}$ and by MAE, and compute the Kendall rank correlation between the two rankings (Table~\ref{tab:rcnorm-validation}). The agreement is strong on every survey, with a mean of 0.881, and holds under a hit-rate-based view as well ($1-$H@5, mean 0.803). It varies by survey, and Brunswick is weakest at 0.778. We therefore read the score per survey, not as a universal ranking guarantee. The two aggregation choices, shot-level macro averaging and trace-count weighting, rank the models nearly identically (mean 0.970), so the conclusions depend on neither the accuracy metric nor the aggregation rule.

$\mathrm{RC}_{\mathrm{norm}}$ has a clear boundary. It is blind to a global time shift, so a smooth but systematically offset prediction can still score well. This blindness is visible in the tolerance sweep (Table~\ref{tab:rcnorm-sweep}), where the agreement falls monotonically from 0.862 at H@9 to 0.244 at H@1. Where reference picks exist, MAE and hit rates remain the primary accuracy measures, and $\mathrm{RC}_{\mathrm{norm}}$ complements them. \textbf{Takeaway.} Across the ten evaluated methods on three field surveys, $\mathrm{RC}_{\mathrm{norm}}$ rankings show strong agreement with MAE rankings, with a mean Kendall correlation of 0.881. These results support its use as a reference-free auxiliary criterion for model comparison in the evaluated settings. $\mathrm{RC}_{\mathrm{norm}}$ measures ridge geometry and does not establish absolute picking accuracy.

\section{Conclusion}

In this work, to address the limited comparability of learning-based seismic processing studies, we introduce \textsc{SPBench}, an open multi-task benchmark. It incorporates six tasks, 10 datasets, 43 settings, 24 methods, 495 model-setting evaluations, and two diagnostic protocols. Our analyses show that conclusions are conditional on the data, the degradation, and the metric. Synthetic rankings carry only partial information about field rankings. Ranking stability under growing interference is a separate selection dimension, moderate under random-like corruption but weaker under coherent ground roll. For first-arrival picking, the reference-free ridge score ranks models consistently with pick-error evaluation without any labels. \textsc{SPBench} currently centers on supervised deep learning over two-dimensional gathers, and classical flows, foundation models, and three-dimensional acquisition remain to be covered. Future versions will add field datasets, baselines, and hidden-test evaluation. We hope \textsc{SPBench} will provide a more comprehensive and fair basis for comparing these methods and spur robust learning-based seismic processing on field data.

\subsection*{AI use statement}

In this work, generative AI tools assisted with developing the benchmark code framework, implementing and reproducing baseline methods, writing data-processing code, and preparing tables from computed experimental results. We also used these tools for language editing, LaTeX formatting, visualization prototyping, feedback on evaluation methodology and experimental design, and assistance with interpreting and presenting results. The authors assessed this feedback and made the final research decisions. Benchmark measurements were obtained by executing the experimental pipeline. The authors reviewed AI-assisted text and figures, executed and checked AI-assisted code, checked reproduced implementations against the source methods, and verified reported numerical results against experimental records. The authors take responsibility for the accuracy, originality, and integrity of the manuscript and accompanying artifacts.

\subsection*{Ethics statement}

This work studies benchmark infrastructure for exploration seismic data processing. Potential risks include benchmark overfitting, misuse of proprietary data, and overstating generalization from synthetic or restricted datasets. We mitigate these risks by documenting data provenance, separating provisional estimates from audited statistics, and emphasizing transparent evaluation protocols.

\subsection*{Reproducibility statement}

The benchmark is designed around fixed splits, public evaluation scripts, baseline implementations, and versioned benchmark artifacts. All code, parameter configurations, benchmark data, model artifacts, metric implementations, and leaderboard resources are available at \url{https://seismicprocessingbenchmark.github.io/SPBench/}. Each released configuration records the data split, degradation realization, normalization, input construction, training schedule, optimizer, stopping rule, and random seeds for one run, so every leaderboard number can be audited against its exact parameter setting. All resources are provided anonymously at the project page.

\bibliography{references}
\bibliographystyle{iclr2027_conference}

\clearpage
\appendix
\section*{Appendix}

\section{Exploration Seismic Data as Recorded and Simulated Wavefields}
\label{app:primer}

This appendix is written for readers from the machine-learning community. It explains what exploration seismic data are and how they relate to the wave equation. The key point is that seismic data are samples of a wavefield. Some of the data in this benchmark are measured in the field, and some are computed by solving a partial differential equation. The two are connected by the same physics.

Exploration seismology probes the subsurface with controlled sources. An active source, such as an air-gun array at sea or a vibrator truck on land, injects mechanical energy into the Earth. The energy travels downward as elastic waves, and part of it returns after reflection, refraction, and diffraction at subsurface interfaces. Receivers at the surface record the returning wavefield over time. A shot gather is one of the primary data representations used in this benchmark, collecting the recordings for one source position. Each trace is one receiver's time series, so a shot gather is a spatiotemporal sample of the true wavefield of the Earth at the receiver locations. Other arrangements of the same wavefield, such as common-receiver gathers and three-dimensional receiver-grid volumes, are used where a task requires them. The field datasets in this benchmark, such as the Mobil AVO Viking Graben line and the field first-arrival data, are actual measurements obtained in this way. They are observations of the real Earth, not model outputs.

\begin{figure}[!tb]
\centering
\setlength{\abovecaptionskip}{3pt}
\begin{tabular}{@{}c@{}}
\includegraphics[height=0.21\textheight,keepaspectratio]{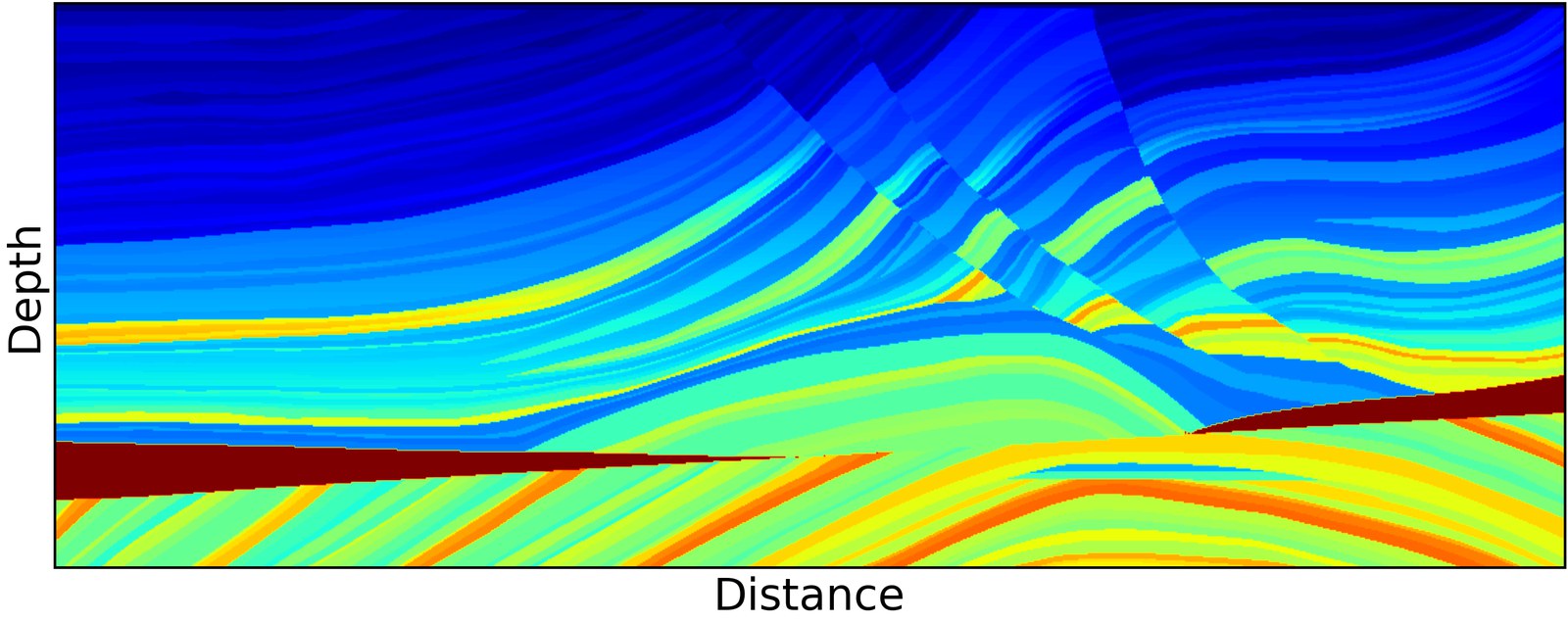}\\
{\small\textbf{(a) Subsurface velocity model}}\\[1mm]
\includegraphics[height=0.21\textheight,keepaspectratio]{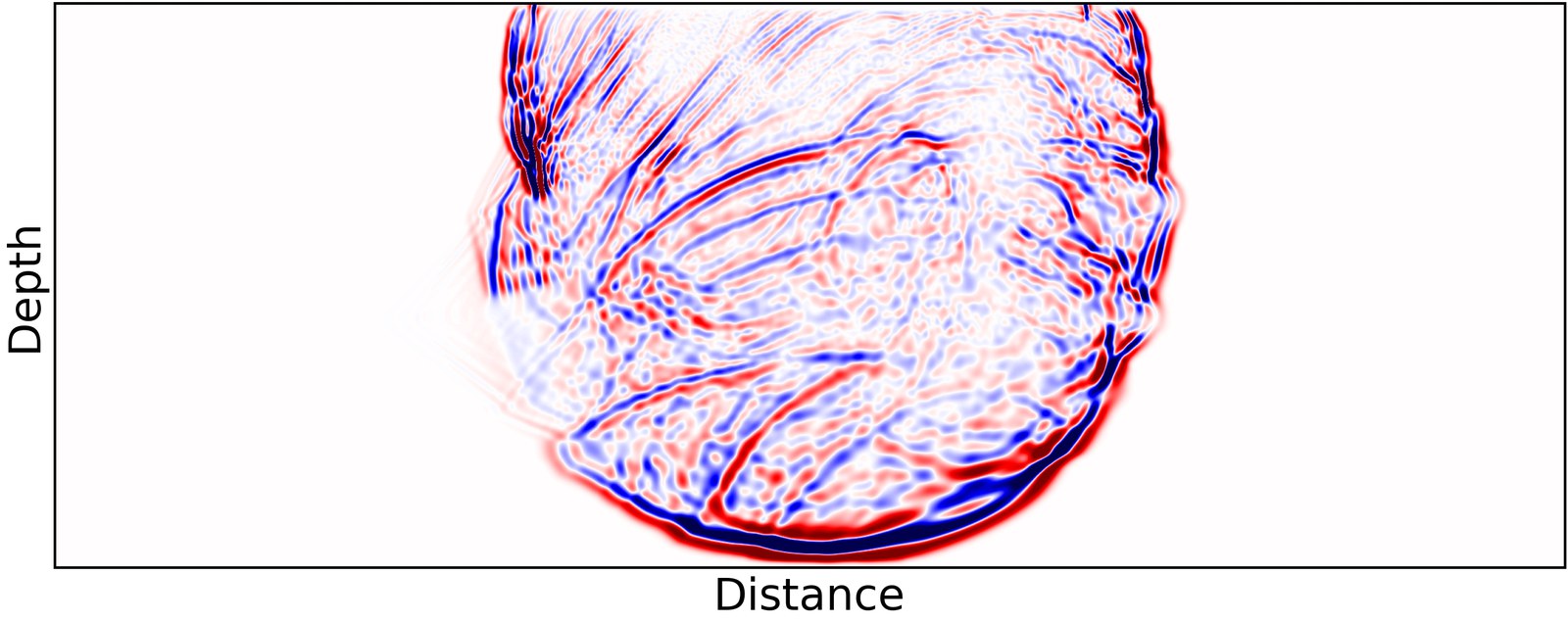}\\
{\small\textbf{(b) Wavefield snapshot}}\\[1mm]
\includegraphics[height=0.21\textheight,keepaspectratio]{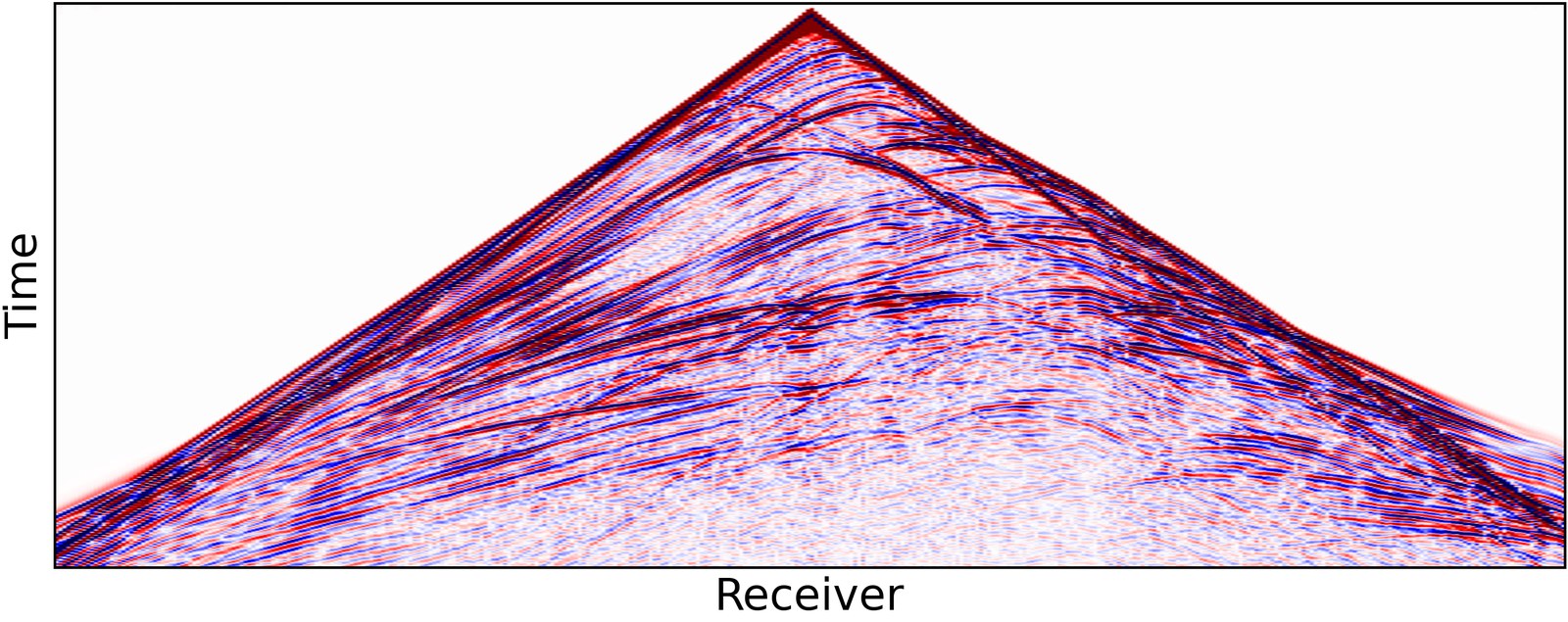}\\
{\small\textbf{(c) Recorded shot gather}}\\
\end{tabular}
\caption{The chain from a subsurface model to recorded data. (a) A velocity model (Marmousi) specifies $v(\bm{x})$, the coefficient of the wave equation. (b) A numerical solution of the wave equation gives the propagating wavefield. (c) Surface receivers sample the wavefield over time, forming the shot gather that processing methods operate on.}
\label{fig:physics-primer}
\end{figure}

The same wavefield can also be computed. Wave propagation in the subsurface is governed by partial differential equations. In the simplest useful approximation, the pressure field $u(\bm{x},t)$ in a medium with wave speed $v(\bm{x})$ satisfies the constant-density acoustic wave equation
\begin{equation}
\frac{1}{v(\bm{x})^2}\,\frac{\partial^2 u}{\partial t^2} \;=\; \nabla^2 u \;+\; s(\bm{x},t),
\label{eq:acoustic}
\end{equation}
Given a velocity model $v(\bm{x})$ and a source term $s(\bm{x},t)$, solving this equation numerically is called forward modeling. It produces a simulated wavefield, and sampling that wavefield at the receiver positions produces a synthetic shot gather. Writing $\bm{x}_s$ for the source position and $\bm{x}_r$ for the receiver position, the recorded data are the wavefield sampled at the receivers, $d(\bm{x}_r, t; \bm{x}_s) = u(\bm{x}_r, t; \bm{x}_s)$. Figure~\ref{fig:physics-primer} illustrates this chain from a given model to a simulated wavefield to a recorded gather. More complete descriptions use the elastic wave equation, which also supports the dispersive surface-wave modes known as ground roll. The synthetic datasets in this benchmark are produced exactly in this way, by acoustic or elastic forward modeling on given velocity models.

This is the same data-generation paradigm as in AI-for-PDE benchmarks, where training and test data are numerical solutions of partial differential equations \citep{takamoto2022pdebench,ohana2024well}. The synthetic component of \textsc{SPBench} is therefore directly comparable to AI-for-PDE data. The velocity model plays the role of the PDE coefficient field, the source term is the input, and the shot gather is a partial observation of the solution. Exploration seismology, however, also provides large amounts of real measured data governed by the same physics. \textsc{SPBench} contains both. Methods are trained separately within each setting and evaluated on PDE-simulated data and on real measured data under matched protocols, so the benchmark can also measure how rankings on simulation data relate to rankings on observed data (Section~\ref{sec:transfer}).

Each benchmark task is an operation on this wavefield. Random noise attenuation separates the deterministic wavefield from incoherent measurement noise. Trace interpolation restores a spatially undersampled wavefield. Ground-roll suppression removes the dispersive surface modes of the elastic equation that mask the weaker reflected body waves. Multiple suppression removes energy that has reflected more than once between interfaces. Deblending separates overlapping wavefields produced by simultaneous sources. First-arrival picking reads the traveltimes of the direct and refracted waves. Processing quality determines how much of the physics recorded in the data remains usable for imaging and inversion.

\FloatBarrier
\section{Literature Survey Scope and Selection Criteria}
\label{app:survey}

The reproducibility analysis in the Introduction and Figure~\ref{fig:reproducibility-landscape} is based on a curated literature corpus. This appendix documents its scope, selection criteria, and annotation procedure.

\paragraph{Scope.} We searched and screened papers published between January 2018 and May 2026 in nine English-language journals that regularly publish machine-learning research for exploration seismic data processing: \emph{Geophysics}, \emph{IEEE Transactions on Geoscience and Remote Sensing} (TGRS), \emph{Petroleum Science}, \emph{Computers \& Geosciences}, \emph{Journal of Geophysics and Engineering} (JGE), \emph{IEEE Geoscience and Remote Sensing Letters} (GRSL), \emph{Geophysical Prospecting}, \emph{Journal of Seismic Exploration} (JSE), and \emph{Journal of Applied Geophysics} (JAG), together with Chinese-language journals in the same field. Table~\ref{tab:survey-journals} summarizes the resulting corpus composition.

\begin{table}[H]
\centering
\small
\setlength{\tabcolsep}{8pt}
\caption{Composition of the 368-paper survey corpus by journal, with the share of each venue in the total.}
\label{tab:survey-journals}
\begin{tabular}{@{}lcc@{}}
\toprule
Journal & Papers & Share \\
\midrule
IEEE Transactions on Geoscience and Remote Sensing (TGRS) & 130 & 35.3\% \\
IEEE Geoscience and Remote Sensing Letters (GRSL) & 63 & 17.1\% \\
Geophysics & 61 & 16.6\% \\
Geophysical Prospecting & 30 & 8.2\% \\
Journal of Geophysics and Engineering (JGE) & 20 & 5.4\% \\
Journal of Seismic Exploration (JSE) & 12 & 3.3\% \\
Petroleum Science & 8 & 2.2\% \\
Computers \& Geosciences & 8 & 2.2\% \\
Journal of Applied Geophysics (JAG) & 4 & 1.1\% \\
Chinese-language journals & 32 & 8.7\% \\
\midrule
Total & 368 & 100\% \\
\bottomrule
\end{tabular}
\end{table}

\paragraph{Selection criteria.} We include papers that propose or evaluate supervised deep-learning methods for exploration seismic processing, covering random noise attenuation, trace interpolation, ground-roll noise suppression, multiple suppression, deblending, first-arrival picking, and closely related processing tasks. Supervised methods are defined as models trained on paired inputs and targets, such as degraded and reference data pairs or annotated arrival times. Papers based purely on unsupervised or self-supervised learning are outside the scope of this survey and are not counted. Each candidate paper was screened by title and abstract and then verified in full text.

\paragraph{Annotation.} For every included paper we record the task labels, the dataset sources as multi-label categories (open synthetic, open field, private synthetic, private field), and whether a public code URL is provided. This procedure yields the 368-paper corpus reported in the Introduction, where Figure~\ref{fig:reproducibility-landscape} summarizes the resulting dataset-source usage and code availability.

\FloatBarrier
\section{Dataset Details}
\label{app:tasks-data}

\begin{figure}[!tb]
\centering
\setlength{\abovecaptionskip}{4pt}
\begin{tabular}{@{}c@{\hspace{0.02\linewidth}}c@{\hspace{0.02\linewidth}}c@{}}
\includegraphics[width=0.31\linewidth]{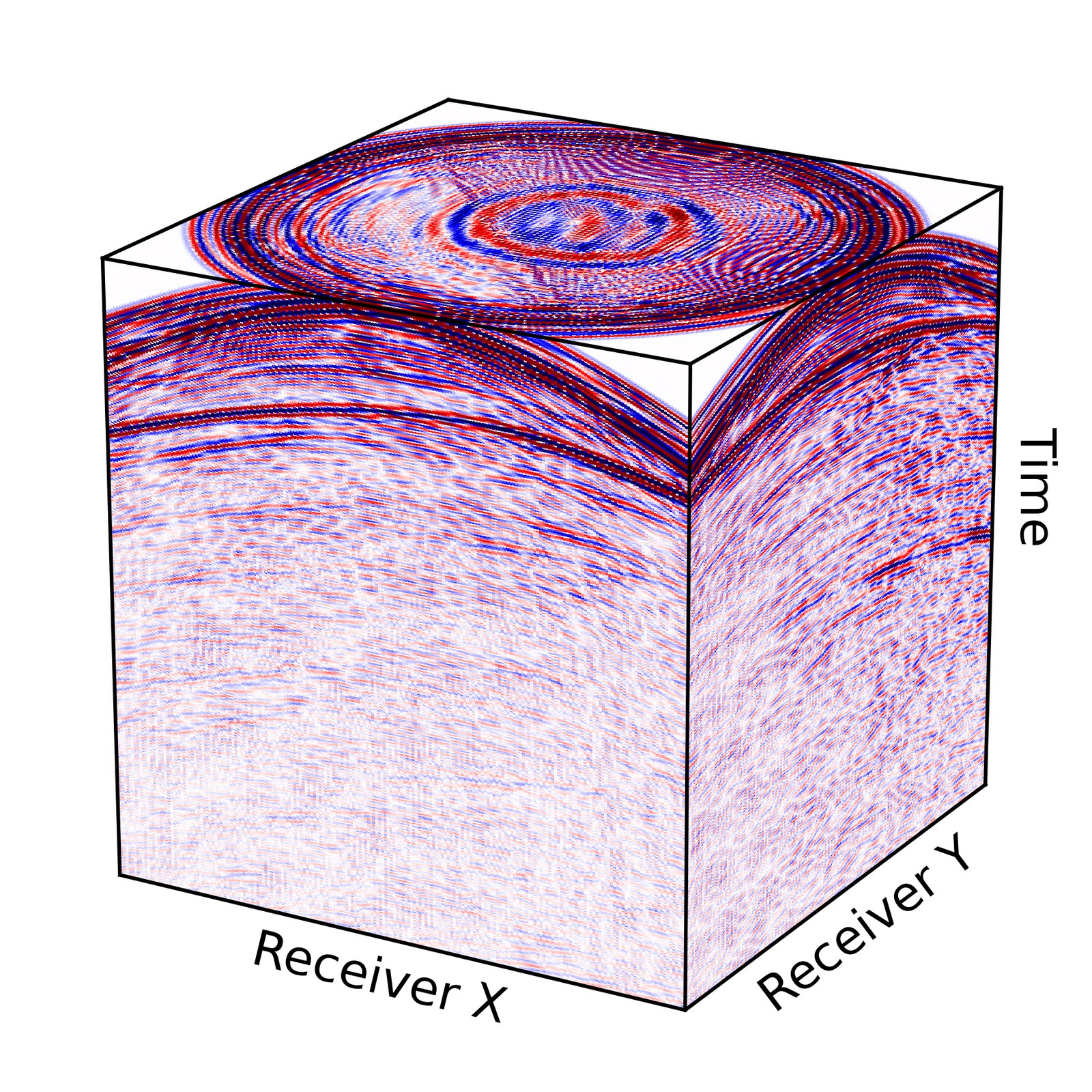} &
\includegraphics[width=0.31\linewidth]{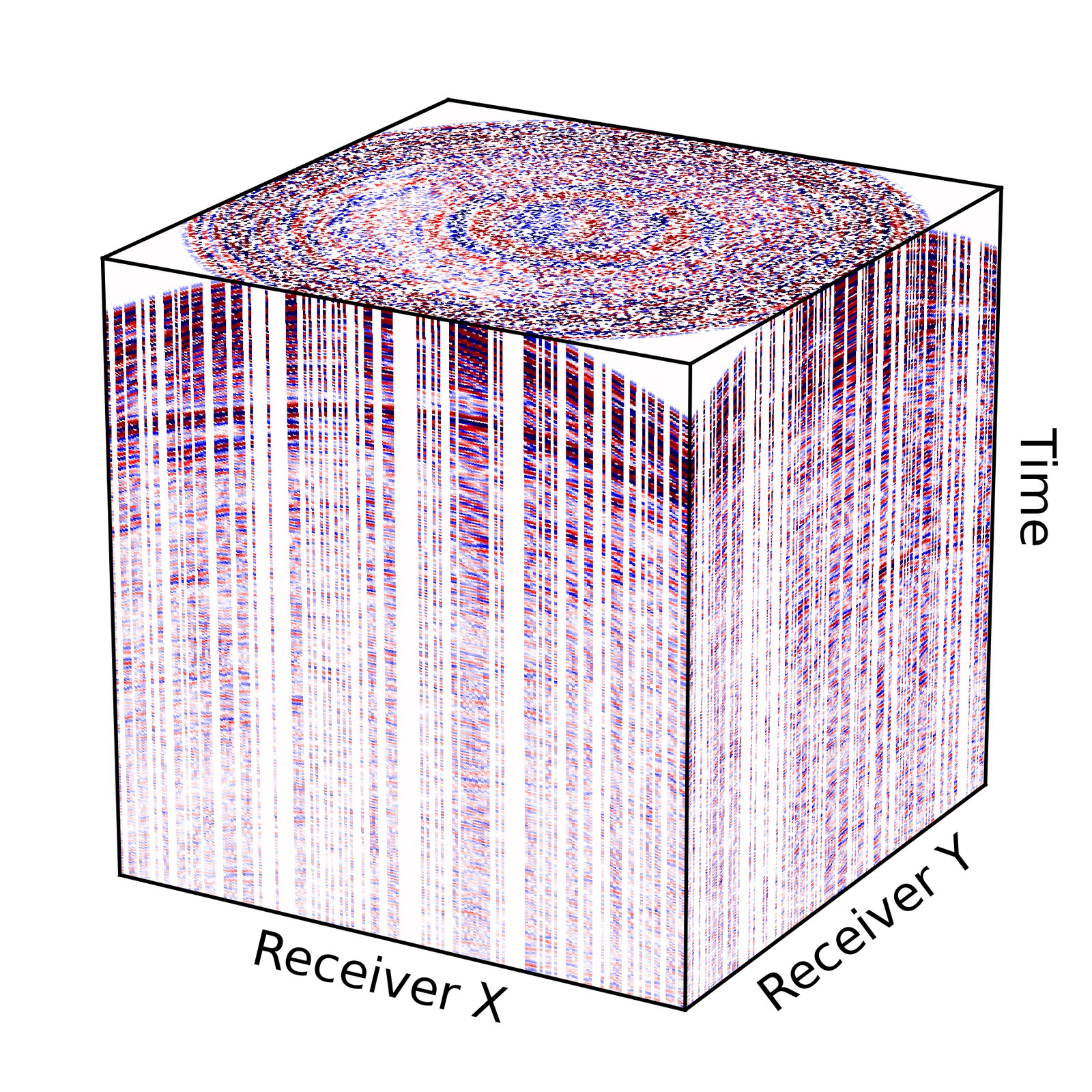} &
\includegraphics[width=0.31\linewidth]{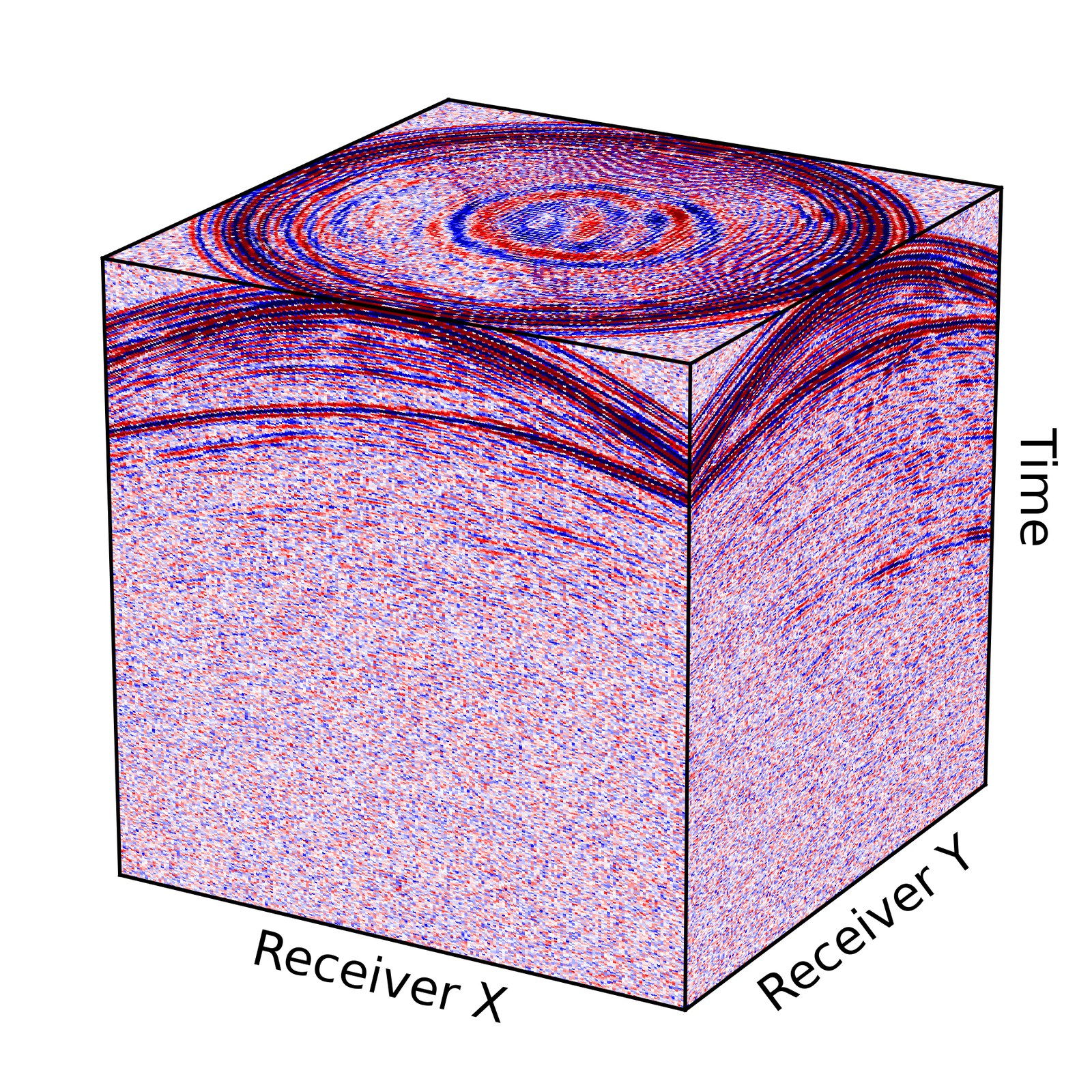} \\
{\small\textbf{(a) Reference}} & {\small\textbf{(b) Interpolation input}} & {\small\textbf{(c) Random-noise input}} \\
\end{tabular}
\caption{Example SEG C3 benchmark data. The clean reference data (a) are corrupted with missing traces to construct interpolation inputs (b) and with additive random noise to construct denoising inputs (c).}
\label{fig:segc3-examples}
\end{figure}

\paragraph{SEG C3 datasets.}
SEG C3 is an openly distributed synthetic dataset from the SEG open-data catalog.\footnote{SEG C3 45-shot page: \url{https://wiki.seg.org/wiki/SEG_C3_45_shot}} It is a 45-shot subset of the SEG/EAGE salt model, originally extracted by Sandia National Laboratory for tests of the SALVO finite-difference imaging program and later reformatted by ARCO into uniform pre-stack SEG-Y files ordered by sail line. It is one of the most commonly used public datasets in seismic data processing research, particularly in interpolation and noise attenuation studies. Each shot is recorded on a $201 \times 201$ receiver grid with 625 time samples per trace at an 8~ms sampling interval. In \textsc{SPBench} it provides the base data for the interpolation and random-noise attenuation tasks. Example benchmark inputs built on SEG C3 are shown in Figure~\ref{fig:segc3-examples}.

\begin{figure}[!tb]
\centering
\setlength{\abovecaptionskip}{4pt}
\begin{tabular}{@{}c@{\hspace{0.03\linewidth}}c@{}}
\includegraphics[width=0.42\linewidth]{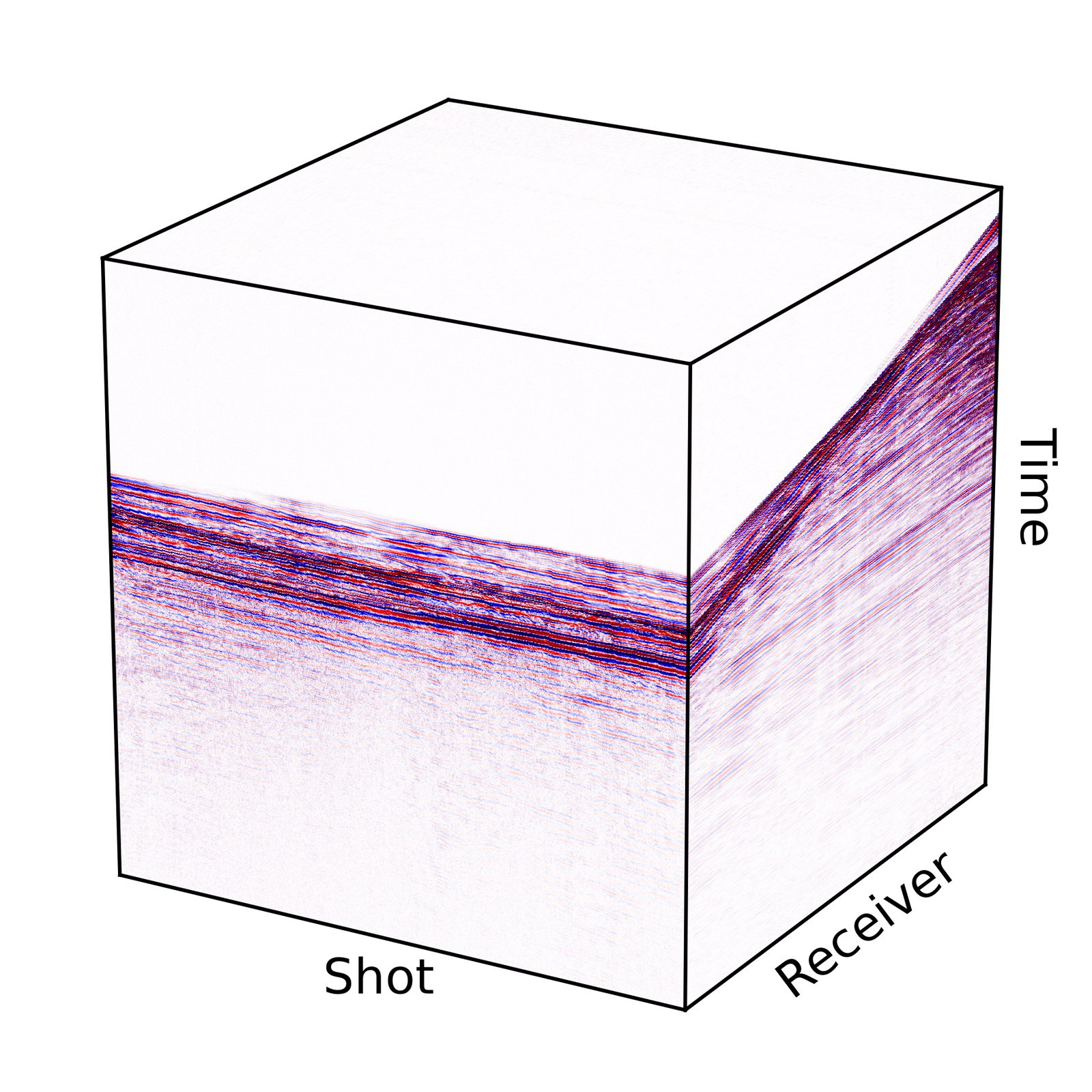} &
\includegraphics[width=0.42\linewidth]{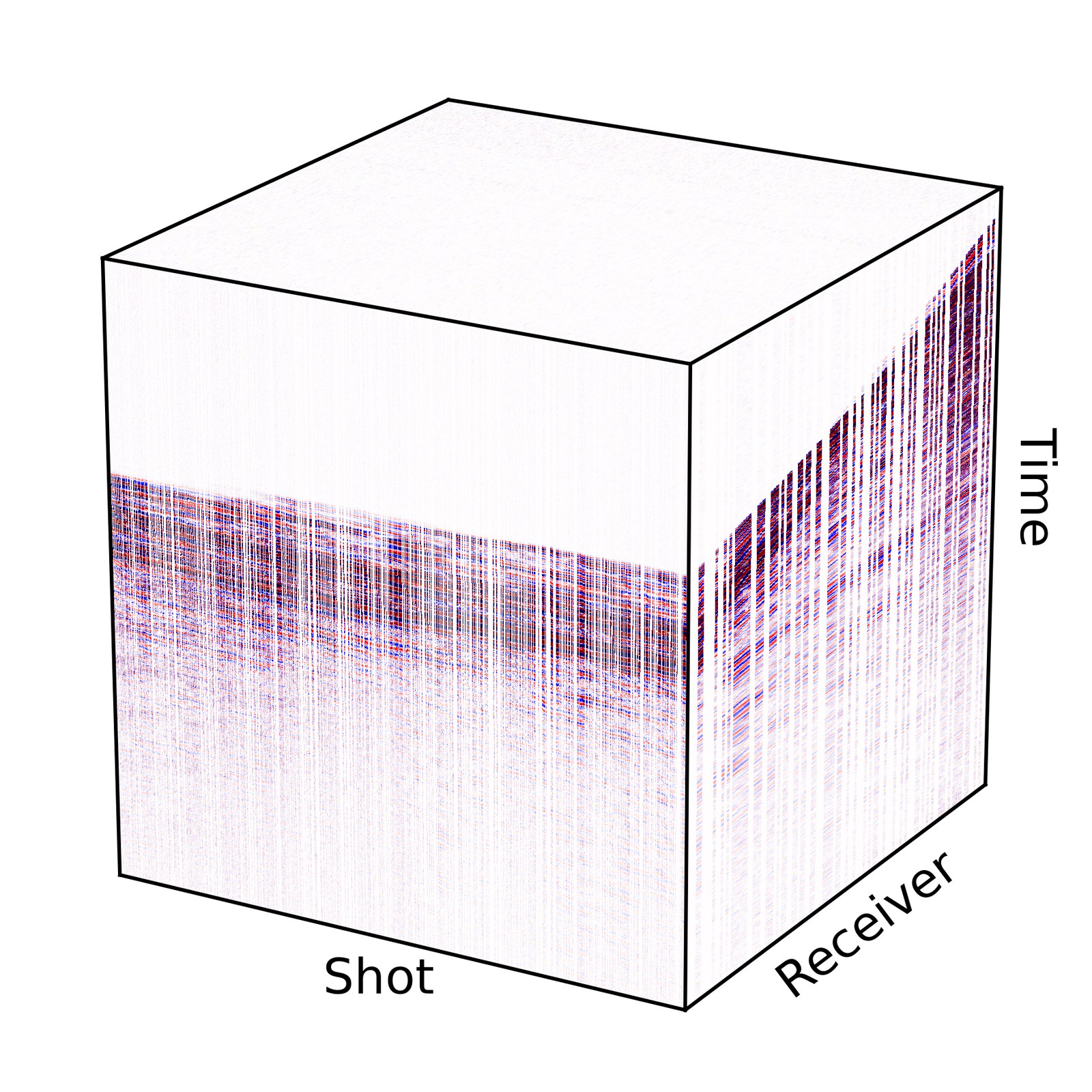} \\
{\small\textbf{(a) Field data}} & {\small\textbf{(b) Interpolation input}} \\[1mm]
\includegraphics[width=0.42\linewidth]{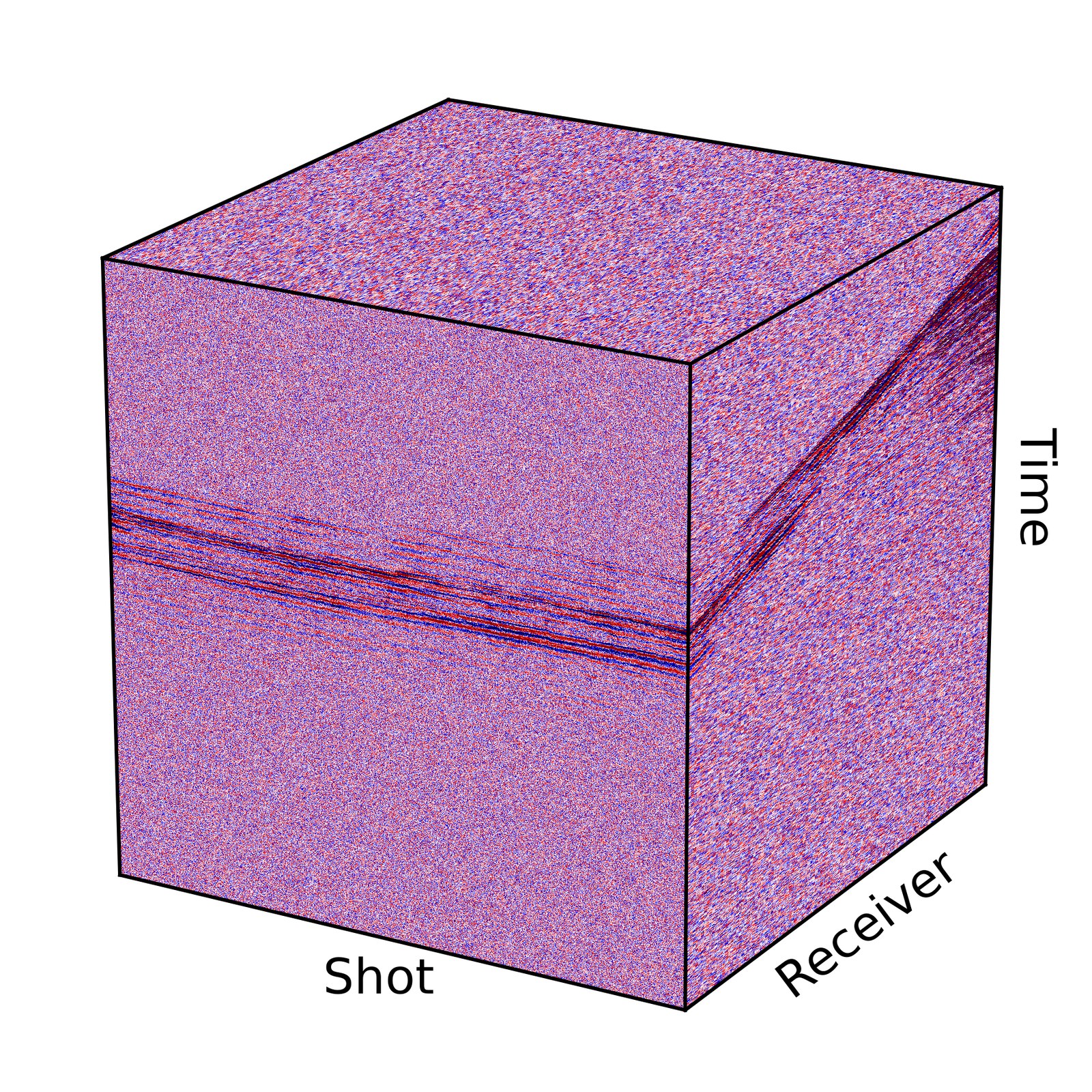} &
\includegraphics[width=0.42\linewidth]{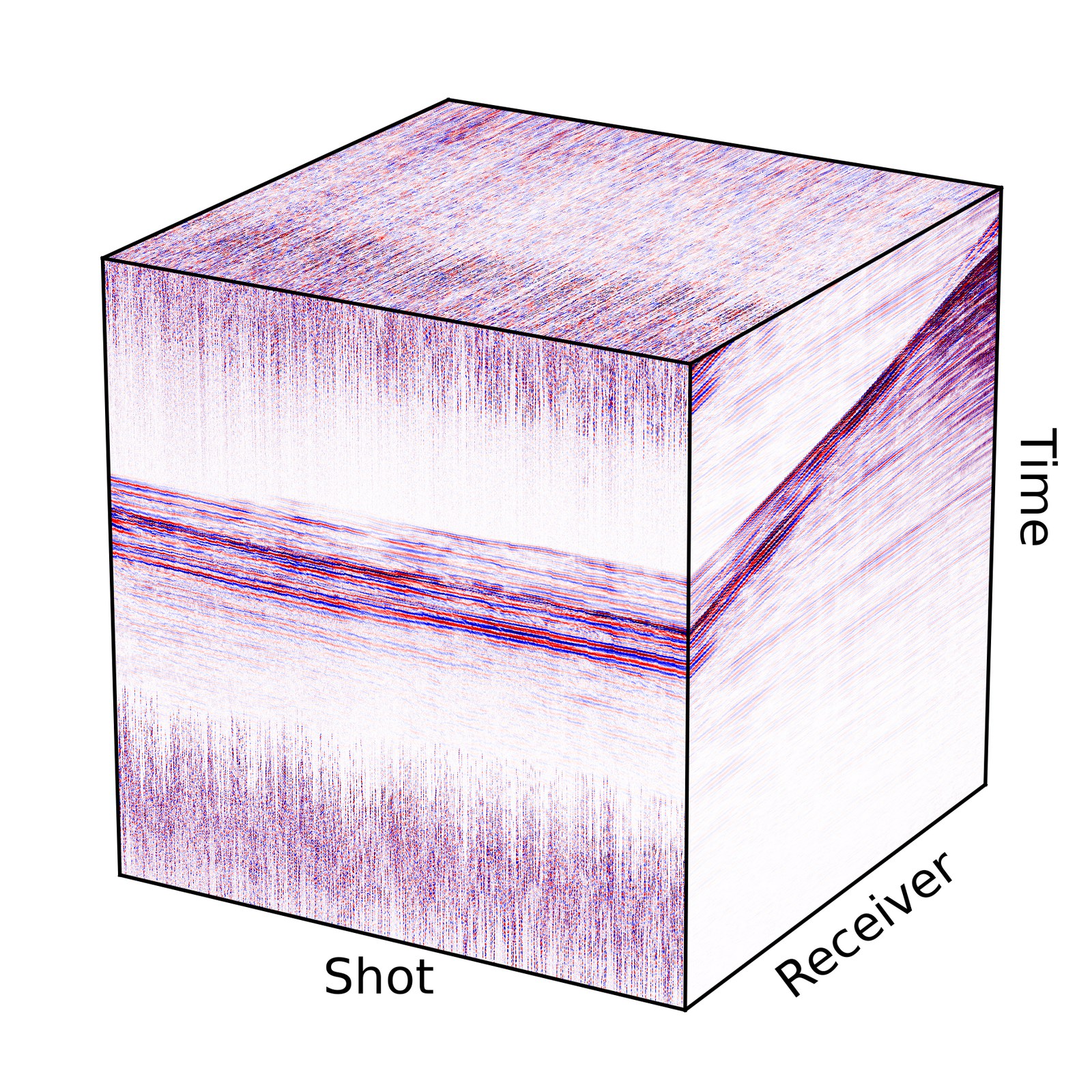} \\
{\small\textbf{(c) Random-noise input}} & {\small\textbf{(d) Deblending input}} \\
\end{tabular}
\caption{Example Mobil AVO benchmark data. The field records (a) serve as the base data for interpolation inputs with missing traces (b), random-noise inputs (c), and the semi-synthetic field deblending setting (d).}
\label{fig:mobilavo-examples}
\end{figure}

\paragraph{Mobil AVO datasets.}
The Mobil AVO Viking Graben Line 12 dataset is an openly distributed field dataset from the SEG open-data catalog~\citep{madiba2003viking}. It is an east-west marine 2D line from the North Viking Graben in the North Sea, containing 1001 shot records, and was originally released for the 1994 SEG workshop on the comparison of seismic inversion methods. It has since become one of the most commonly used public field datasets in seismic processing, inversion, and AVO studies. In \textsc{SPBench} it provides the field-data base for the interpolation and random-noise attenuation tasks, and for a semi-synthetic field deblending setting. Figure~\ref{fig:mobilavo-examples} shows the field data and the derived benchmark inputs.

\begin{figure}[!tb]
\centering
\setlength{\abovecaptionskip}{4pt}
\begin{tabular}{@{}c@{\hspace{0.03\linewidth}}c@{}}
\includegraphics[height=0.25\textheight]{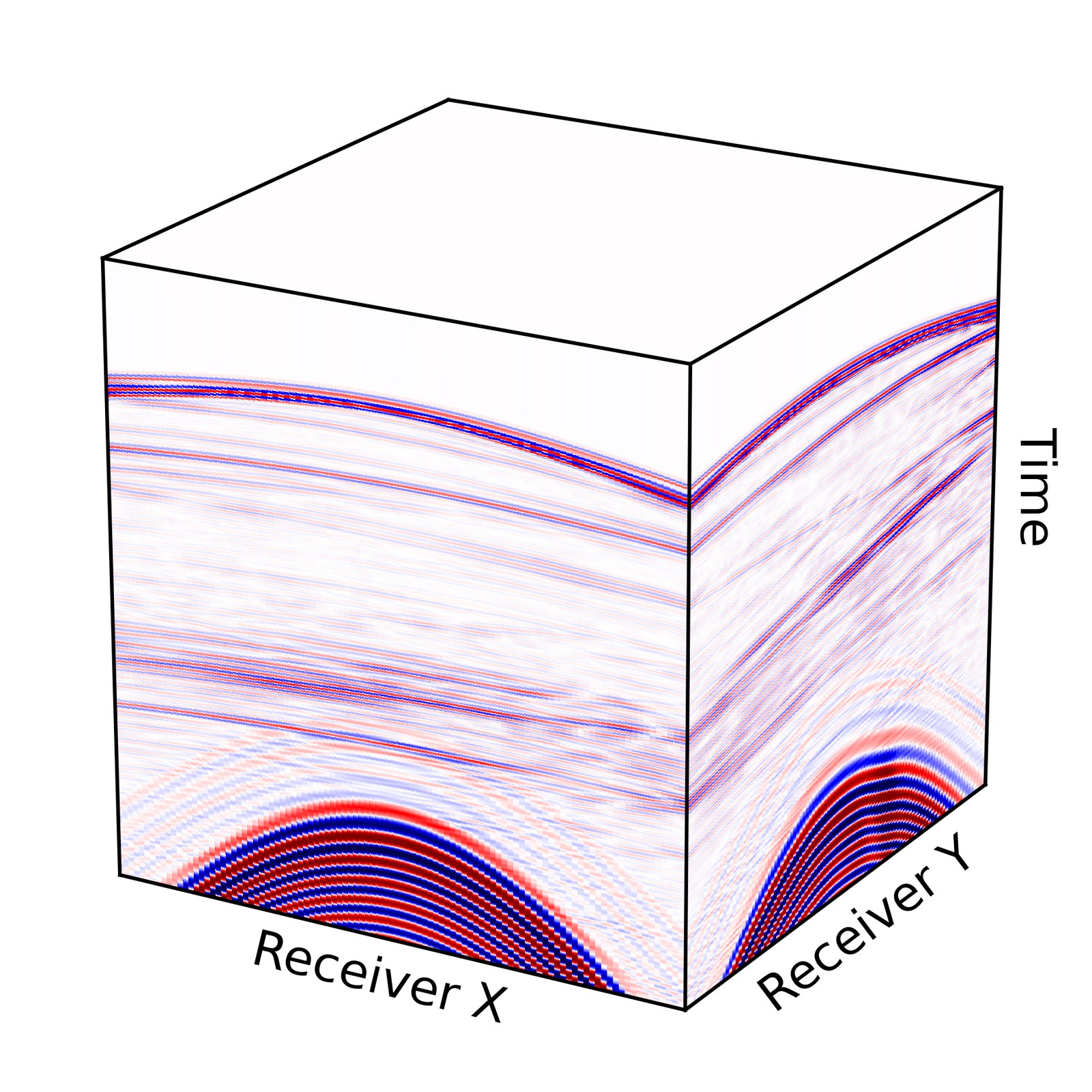} &
\includegraphics[height=0.25\textheight]{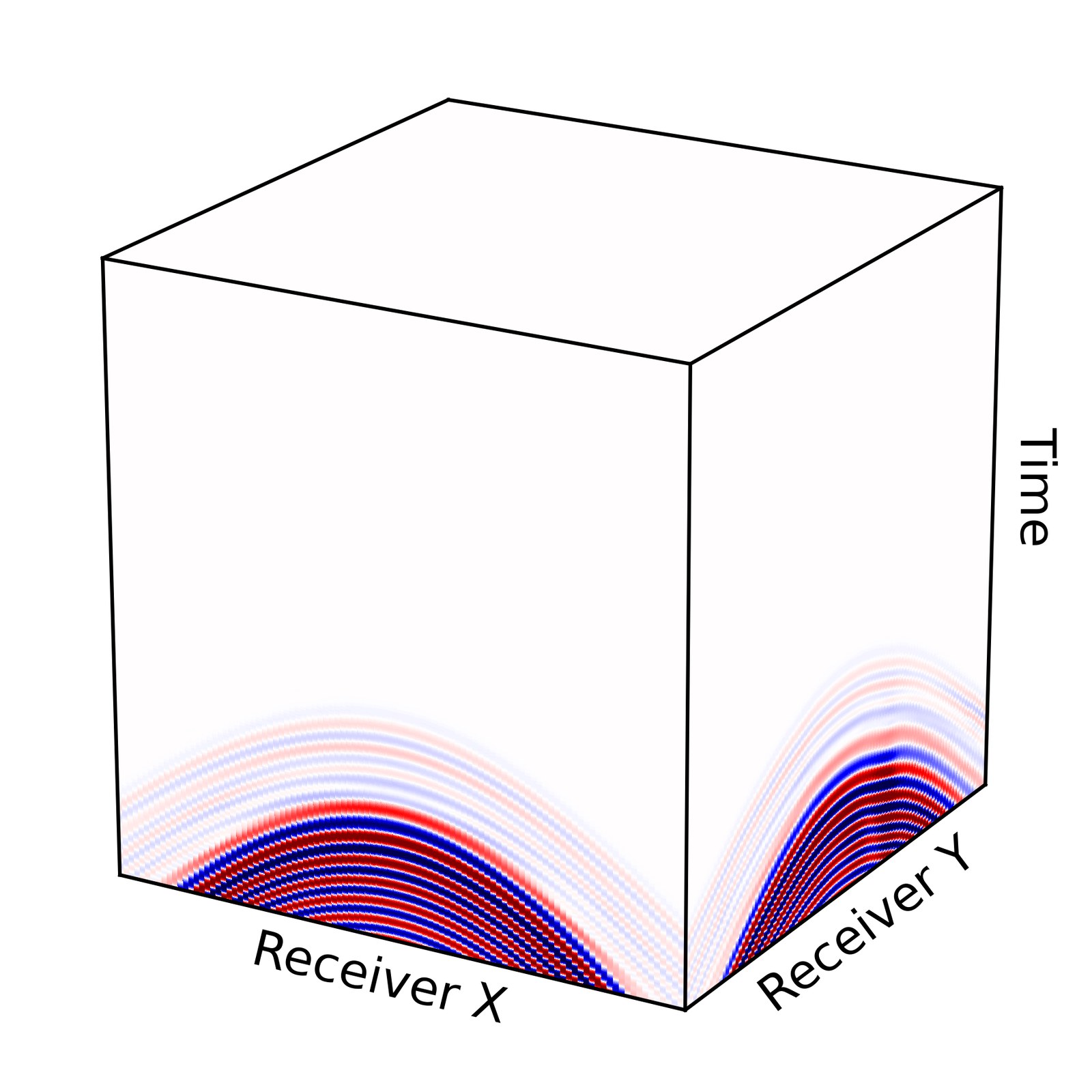} \\
{\small\textbf{(a) Synthetic input}} & {\small\textbf{(b) Ground-roll label}} \\[1mm]
\includegraphics[height=0.25\textheight]{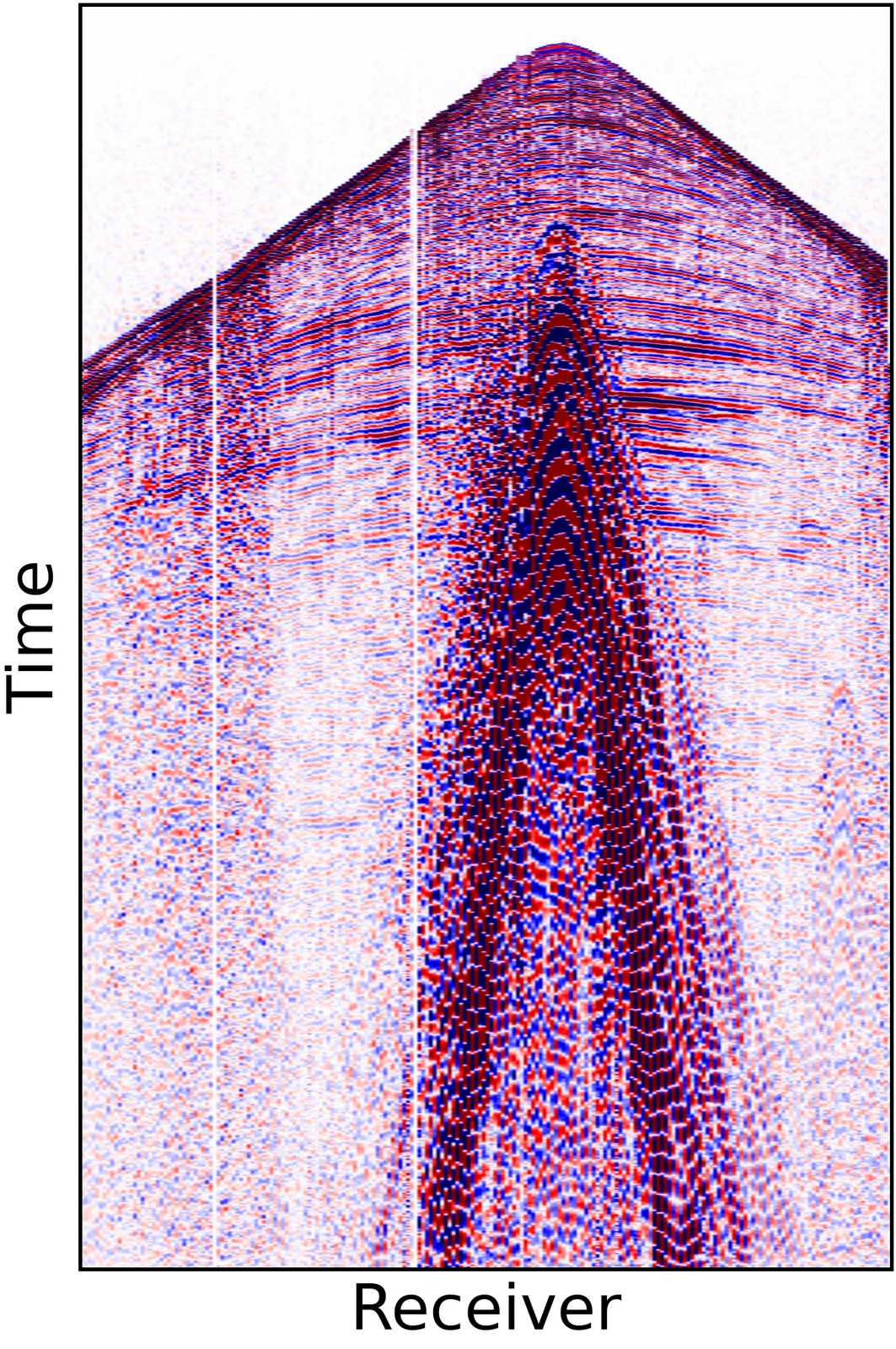} &
\includegraphics[height=0.25\textheight]{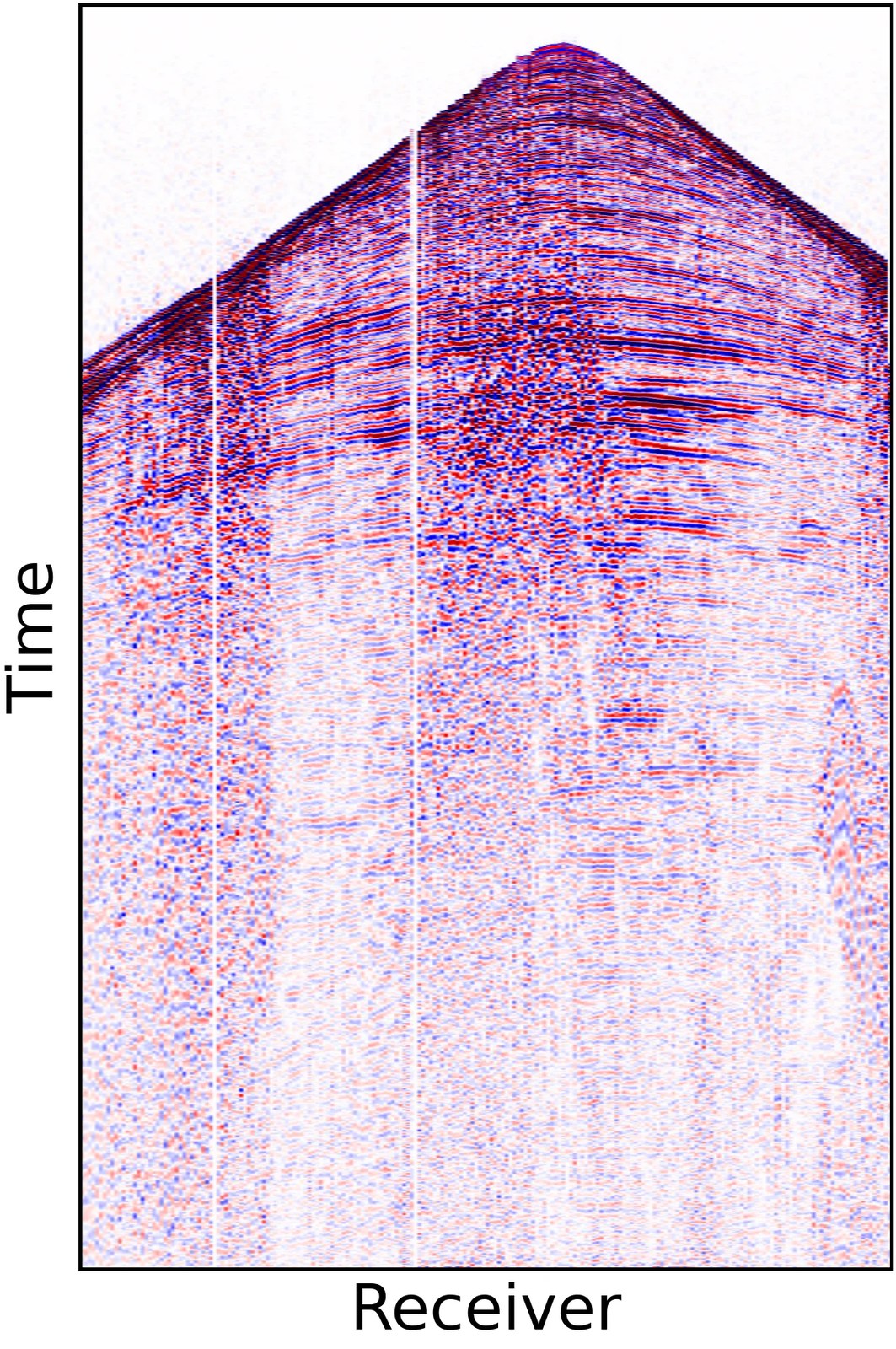} \\
{\small\textbf{(c) Field input}} & {\small\textbf{(d) Expert-processed reference}} \\
\end{tabular}
\caption{Examples from the \textsc{SPBench} ground-roll datasets. Top row, synthetic set: the contaminated input (a) superposes modeled ground roll onto clean reflection data, and the ground-roll label (b) serves as the prediction target. Bottom row, field set: the field shot record (c) contains strong dispersive ground roll, and (d) is the reference produced by expert processing.}
\label{fig:surface-wave-examples}
\end{figure}

\paragraph{\textsc{SPBench} ground-roll datasets.}
The \textsc{SPBench} ground-roll datasets are new datasets for ground-roll noise suppression, comprising a synthetic set and a field set. The synthetic set is built on the SEG C3 dataset and contains nine shots, each recorded on a $201 \times 201$ receiver grid with 625 time samples per trace. Clean reflection data are modeled with the acoustic wave equation
\begin{equation*}
\frac{1}{c^2(\bm{x})}\frac{\partial^2 p}{\partial t^2} = \nabla^2 p + s,
\end{equation*}
for the given survey geometry, where $c(\bm{x})$ is the wave speed, $p$ the pressure field, and $s$ the source term, and ground roll is modeled for the same geometry with the isotropic elastic wave equation
\begin{equation*}
\rho \frac{\partial^2 \bm{u}}{\partial t^2} = (\lambda + \mu)\,\nabla(\nabla \cdot \bm{u}) + \mu \nabla^2 \bm{u},
\end{equation*}
under a free-surface boundary condition, with density $\rho$, Lam\'e parameters $\lambda$ and $\mu$, and displacement $\bm{u}$. The elastic records are sorted by offset, and reflections are muted by exploiting their higher apparent velocity relative to the ground roll, yielding relatively pure ground-roll records. The traces are then restored to their original ordering, scaled to different amplitudes, and added to the clean acoustic data. The companion field set contains real acquisition records and extends the ground-roll task beyond synthetic conditions. Figure~\ref{fig:surface-wave-examples} shows example inputs and labels from the synthetic and field sets.

\begin{figure}[!tb]
\centering
\setlength{\abovecaptionskip}{4pt}
\begin{tabular}{@{}c@{\hspace{0.03\linewidth}}c@{}}
\includegraphics[width=0.42\linewidth]{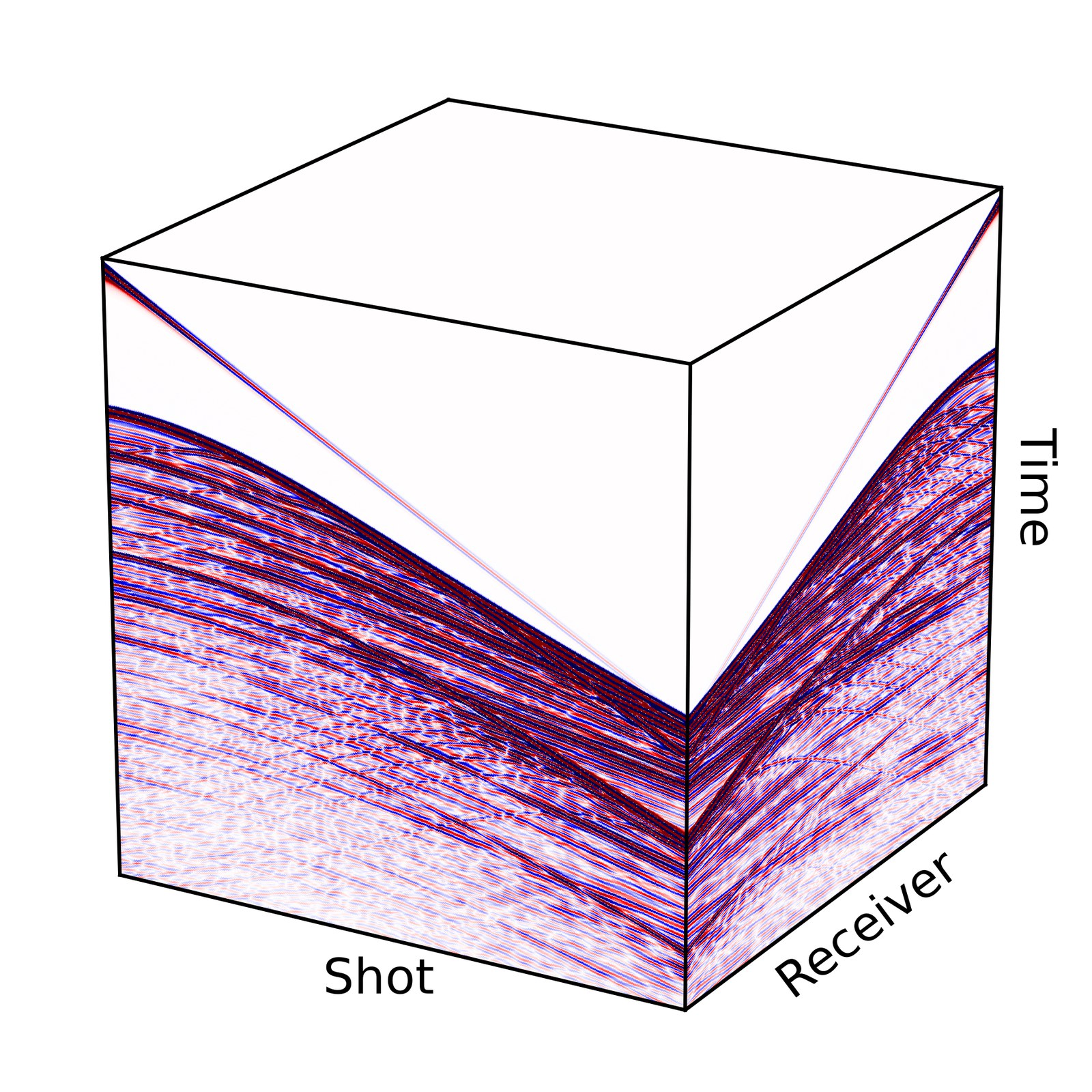} &
\includegraphics[width=0.42\linewidth]{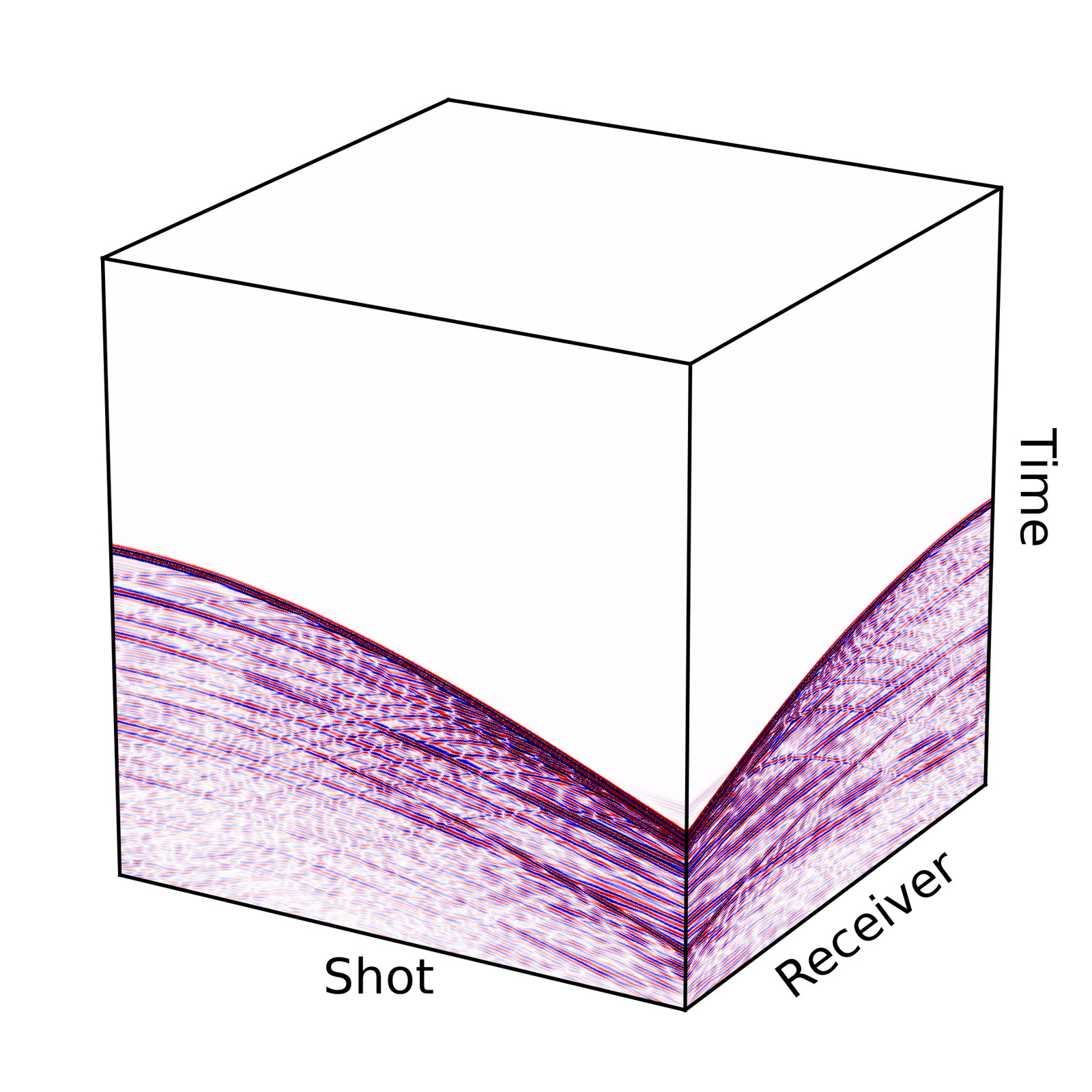} \\
{\small\textbf{(a) Contaminated input}} & {\small\textbf{(b) Multiple label}} \\
\end{tabular}
\caption{Example from the \textsc{SPBench} multiples dataset: a multiple-contaminated input gather (a) and the aligned additive multiple label (b) used as the prediction target.}
\label{fig:multiples-examples}
\end{figure}

\paragraph{\textsc{SPBench} multiples dataset.}
The \textsc{SPBench} multiples dataset is a new synthetic dataset for multiple suppression, forward-modeled on the SEAM velocity model.\footnote{SEG Advanced Modeling (SEAM) program: \url{https://www.seg.org/SEAM}} Two wavefields are simulated for each shot. The total data are modeled with a free surface, so they contain primaries, source and receiver ghosts, surface-related multiples, and the direct wave. The primary references are modeled without a free surface using the mirror source/receiver method: four acoustic simulations with mirrored source and receiver depths are superposed to synthesize the primary reflection plus source and receiver ghosts,
\begin{equation*}
p = p_1 + p_4 - p_2 - p_3,
\end{equation*}
where $p_1$ is the primary reflection, $p_2$ and $p_3$ are the receiver-side and source-side ghosts, and $p_4$ is the source-plus-receiver ghost. The same combinations are used to compute the direct wave for the primary model. The true surface-related multiples are then obtained by subtracting the primaries from the total data after both have had their direct waves removed. The released set contains paired shot gathers with aligned additive multiple labels. Figure~\ref{fig:multiples-examples} shows an example input and its multiple label.

\begin{figure}[!tb]
\centering
\setlength{\abovecaptionskip}{4pt}
\begin{tabular}{@{}c@{\hspace{0.03\linewidth}}c@{}}
\includegraphics[width=0.42\linewidth]{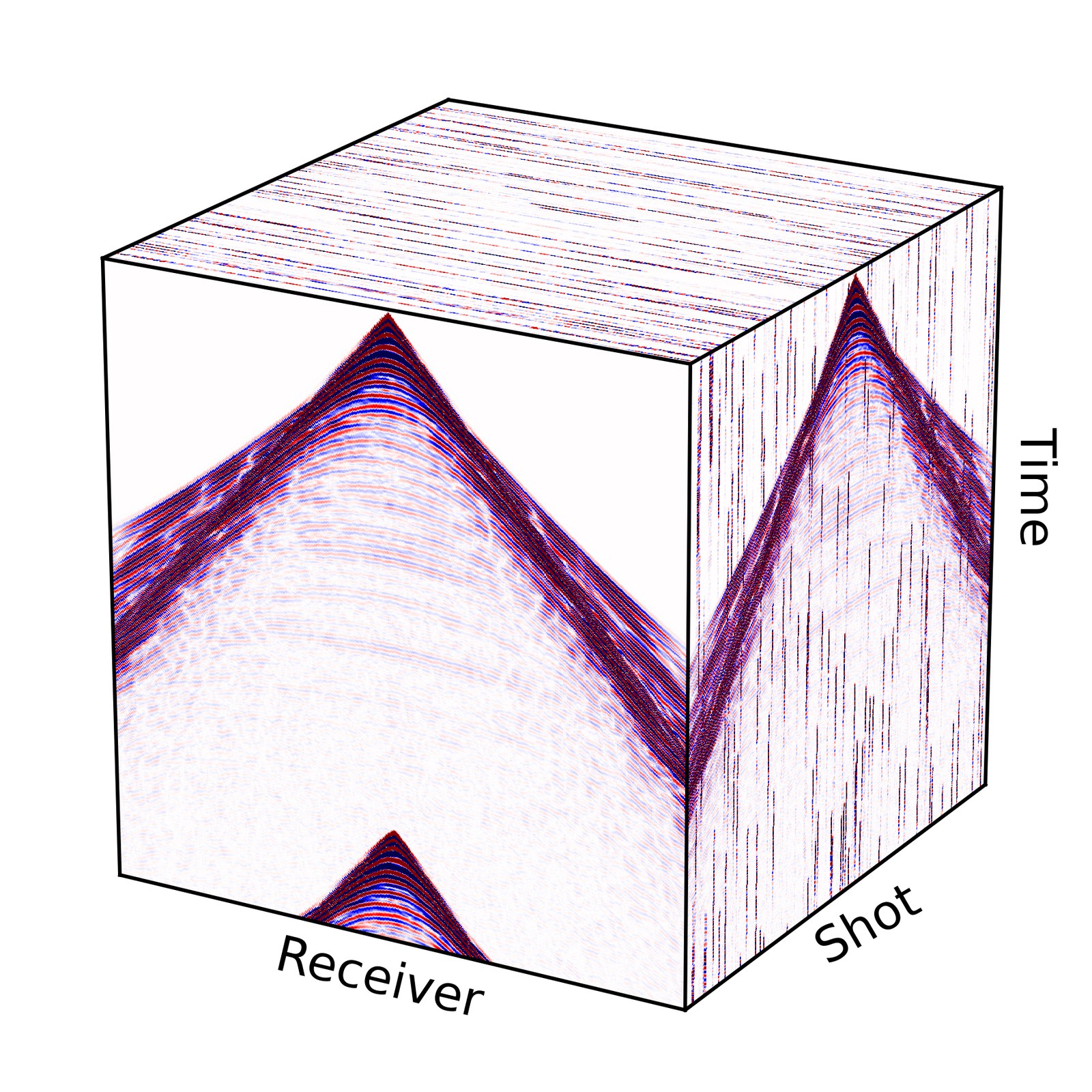} &
\includegraphics[width=0.42\linewidth]{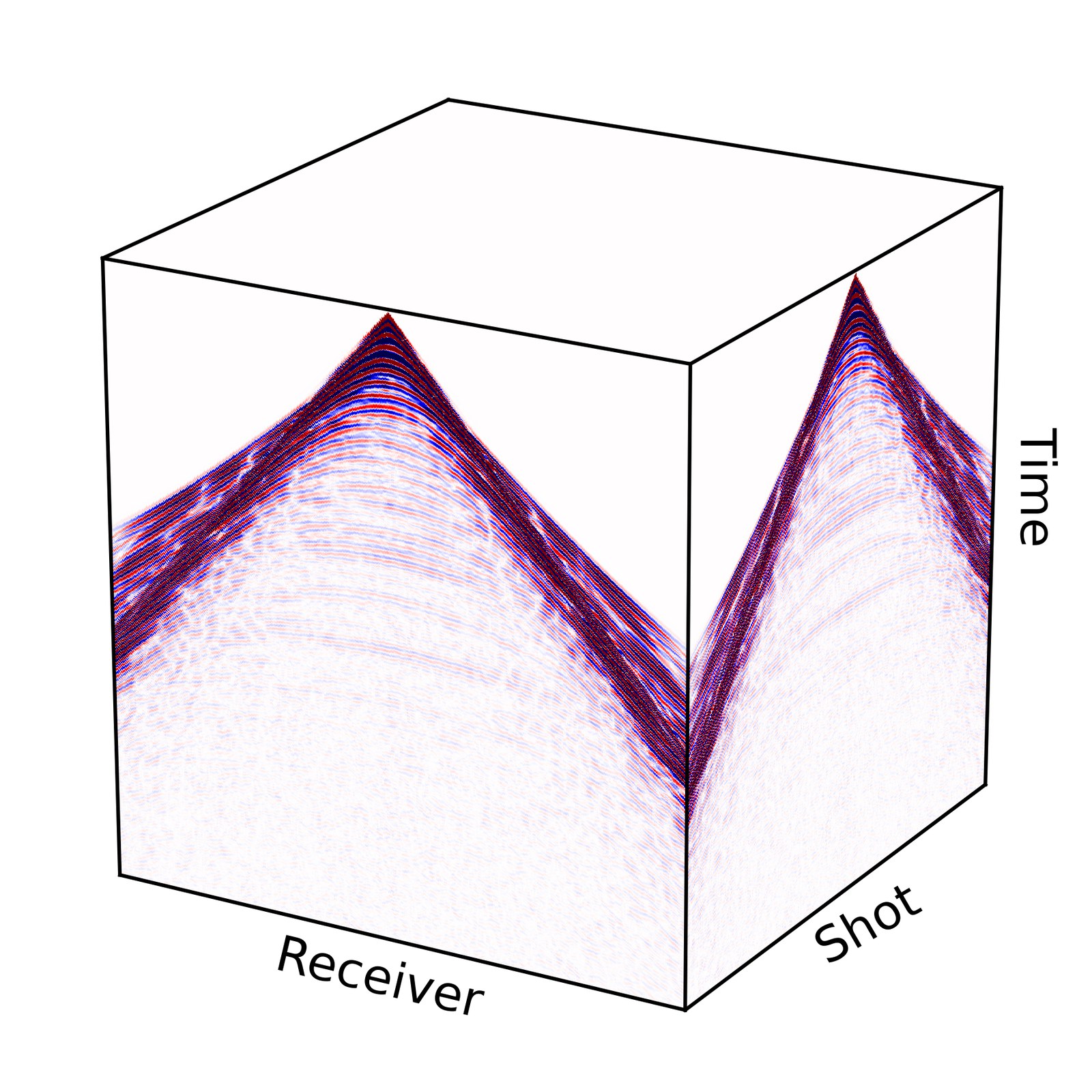} \\
{\small\textbf{(a) Blended input}} & {\small\textbf{(b) Unblended reference}} \\
\end{tabular}
\caption{Example from the \textsc{SPBench} deblending datasets: a blended common-receiver gather (a) and the retained unblended reference (b).}
\label{fig:deblending-examples}
\end{figure}

\paragraph{\textsc{SPBench} deblending datasets.}
The \textsc{SPBench} deblending datasets are new deblending datasets comprising a synthetic set and a field set. The synthetic set is constructed from single-shot seismic records generated by 3D acoustic forward modeling, containing 270 shots and 386 receivers with 2000 time samples per trace at a 2~ms sampling interval. Blended data are synthesized in the common-receiver domain by time-shifting and summing the single-source records at each receiver,
\begin{equation*}
b(t;r) = \sum_{i=1}^{N_s} u(t - \tau_i; s_i; r),
\end{equation*}
where $s_i$ and $\tau_i$ denote the position and firing time of the $i$-th source and $N_s$ is the number of sources in the record. The original unblended records are retained as references, and all data are arranged as common-receiver gathers. The companion field set is a semi-synthetic field-data setting constructed by applying synthetic blending to real Mobil AVO Viking Graben Line 12 records, rather than a true simultaneous-source acquisition. Figure~\ref{fig:deblending-examples} shows an example blended input and its unblended reference.

\begin{figure}[!tb]
\centering
\setlength{\abovecaptionskip}{4pt}
\begin{tabular}{@{}c@{\hspace{0.02\linewidth}}c@{\hspace{0.02\linewidth}}c@{}}
\includegraphics[height=0.20\textheight]{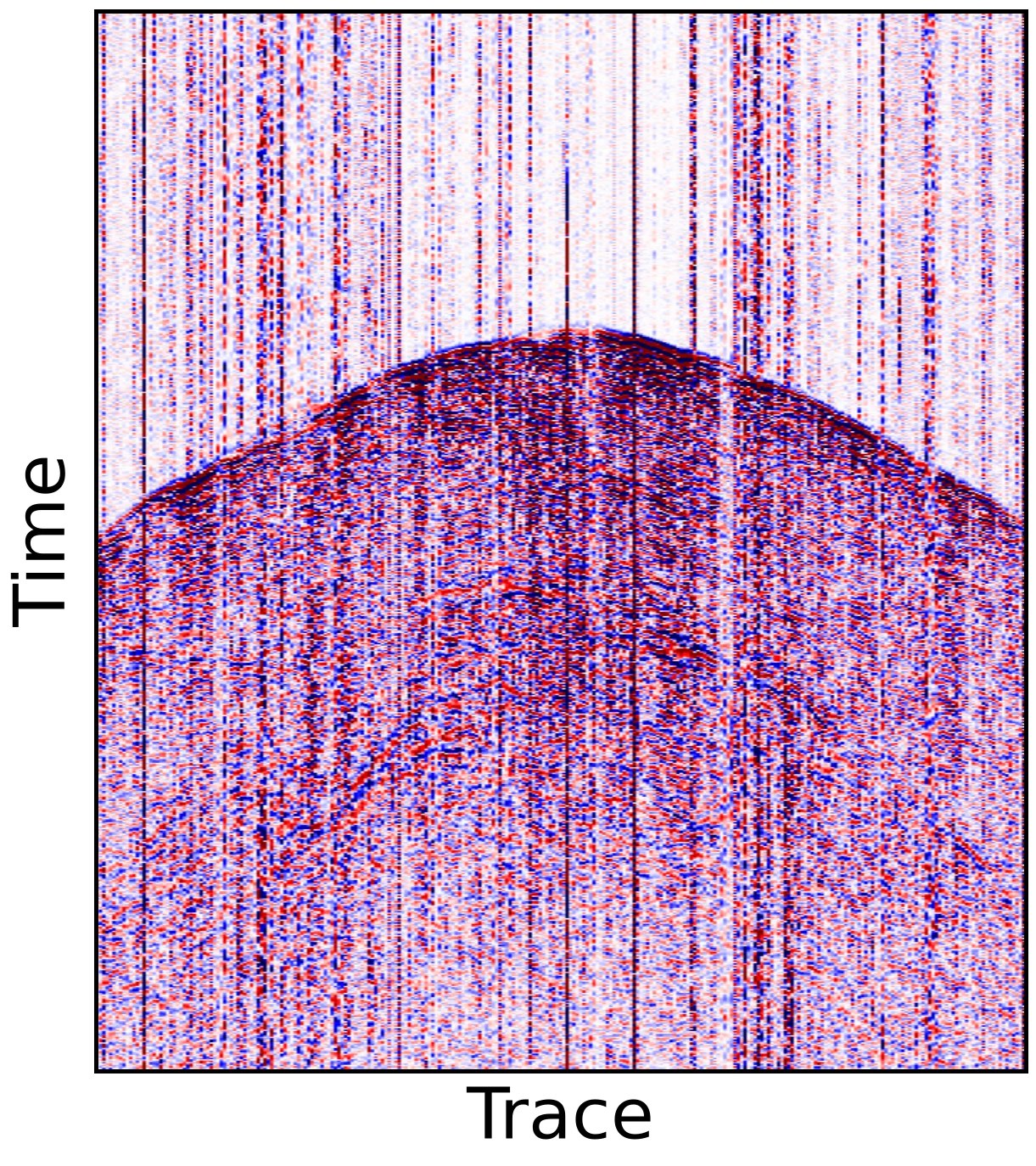} &
\includegraphics[height=0.20\textheight]{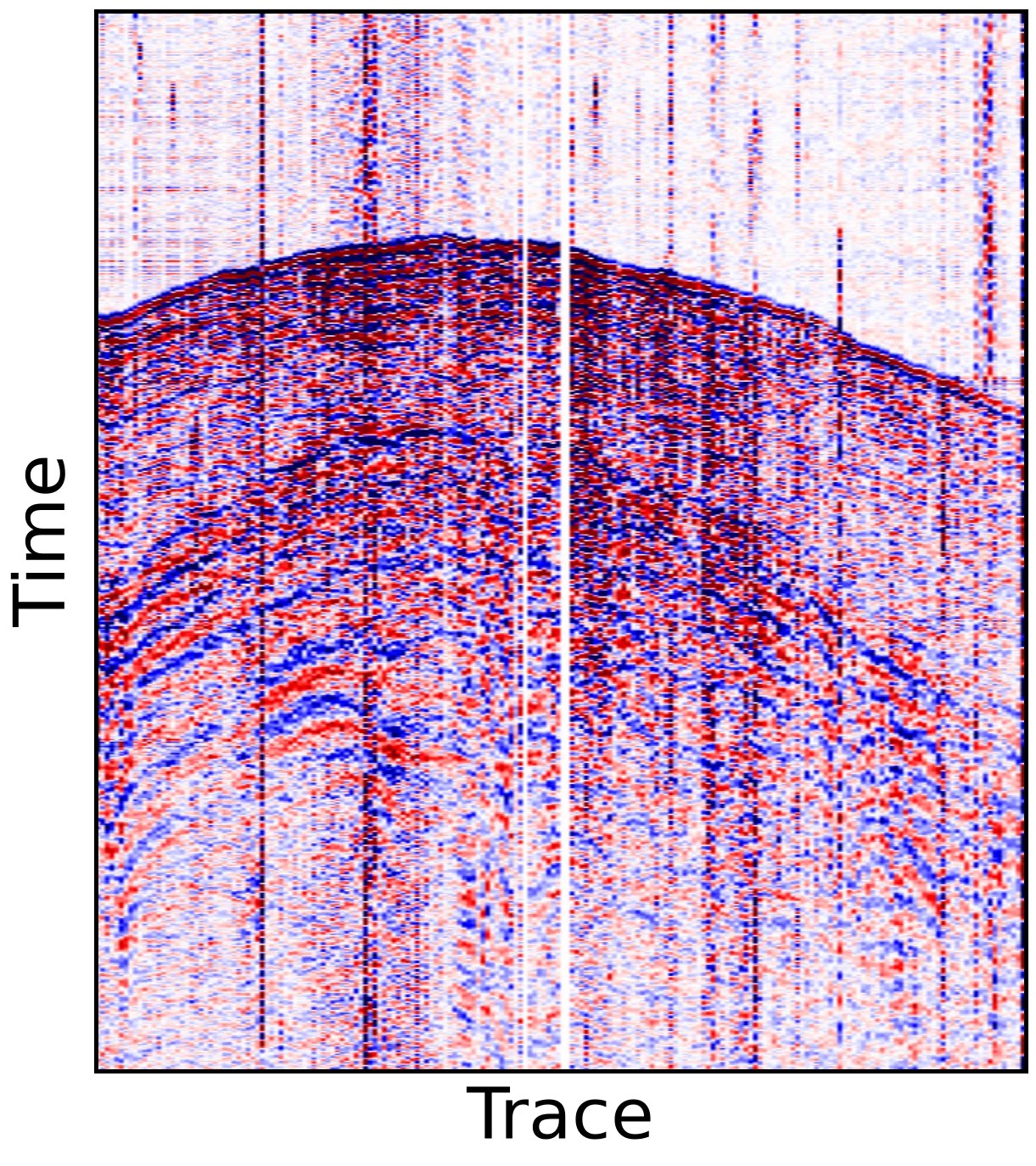} &
\includegraphics[height=0.20\textheight]{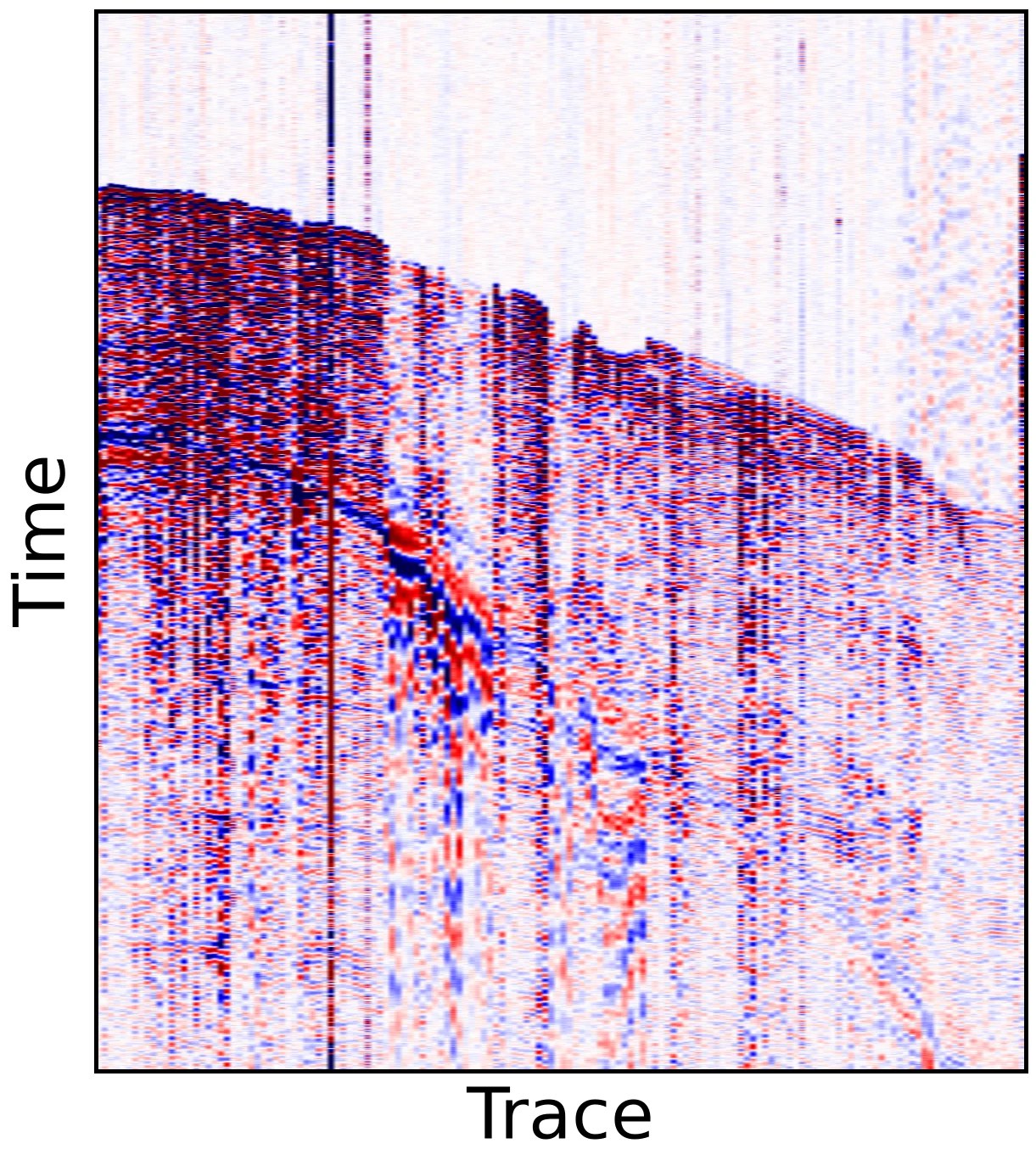} \\
{\small\textbf{(a) Brunswick input}} & {\small\textbf{(b) Halfmile input}} & {\small\textbf{(c) Lalor input}} \\[1mm]
\includegraphics[height=0.20\textheight]{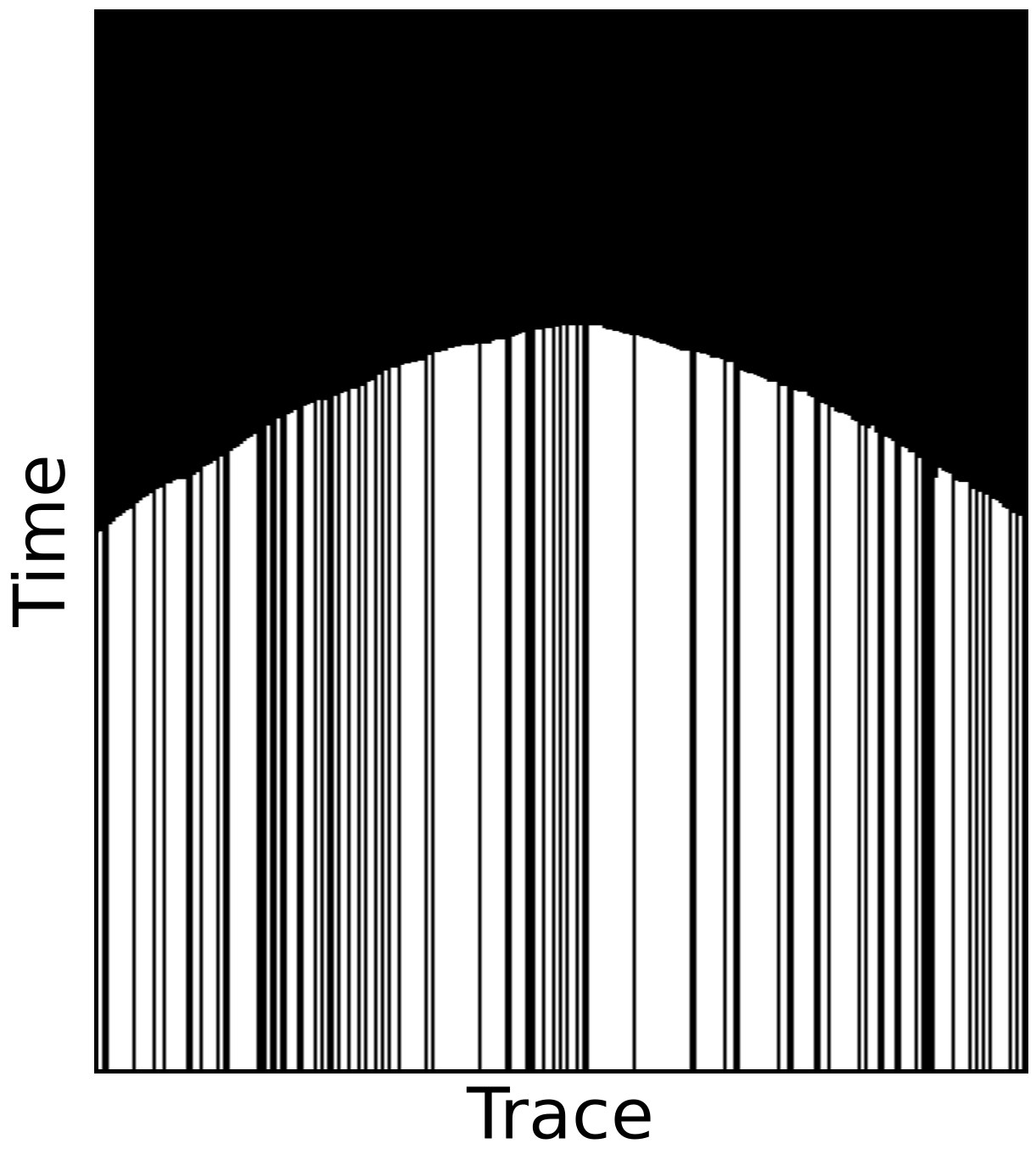} &
\includegraphics[height=0.20\textheight]{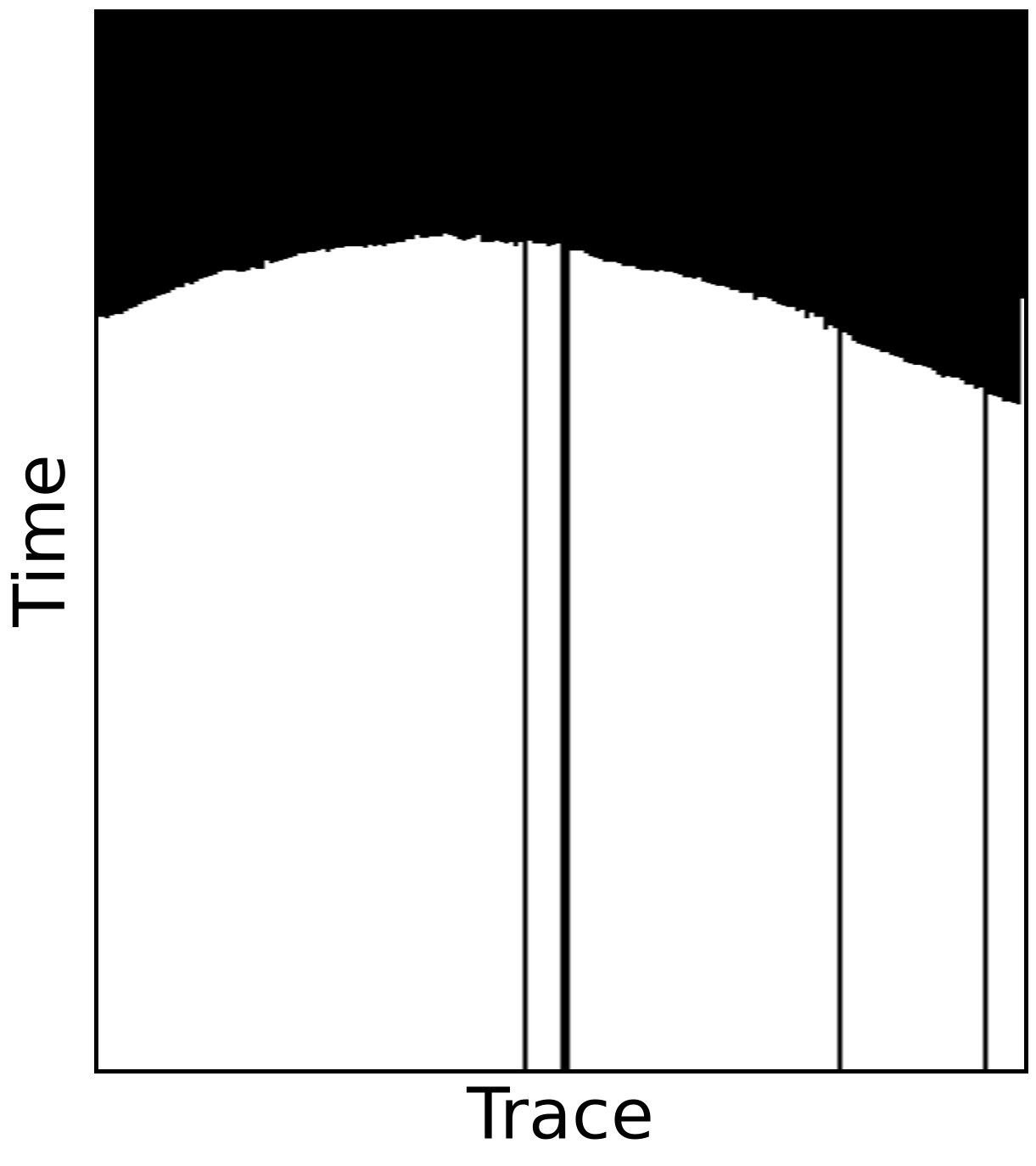} &
\includegraphics[height=0.20\textheight]{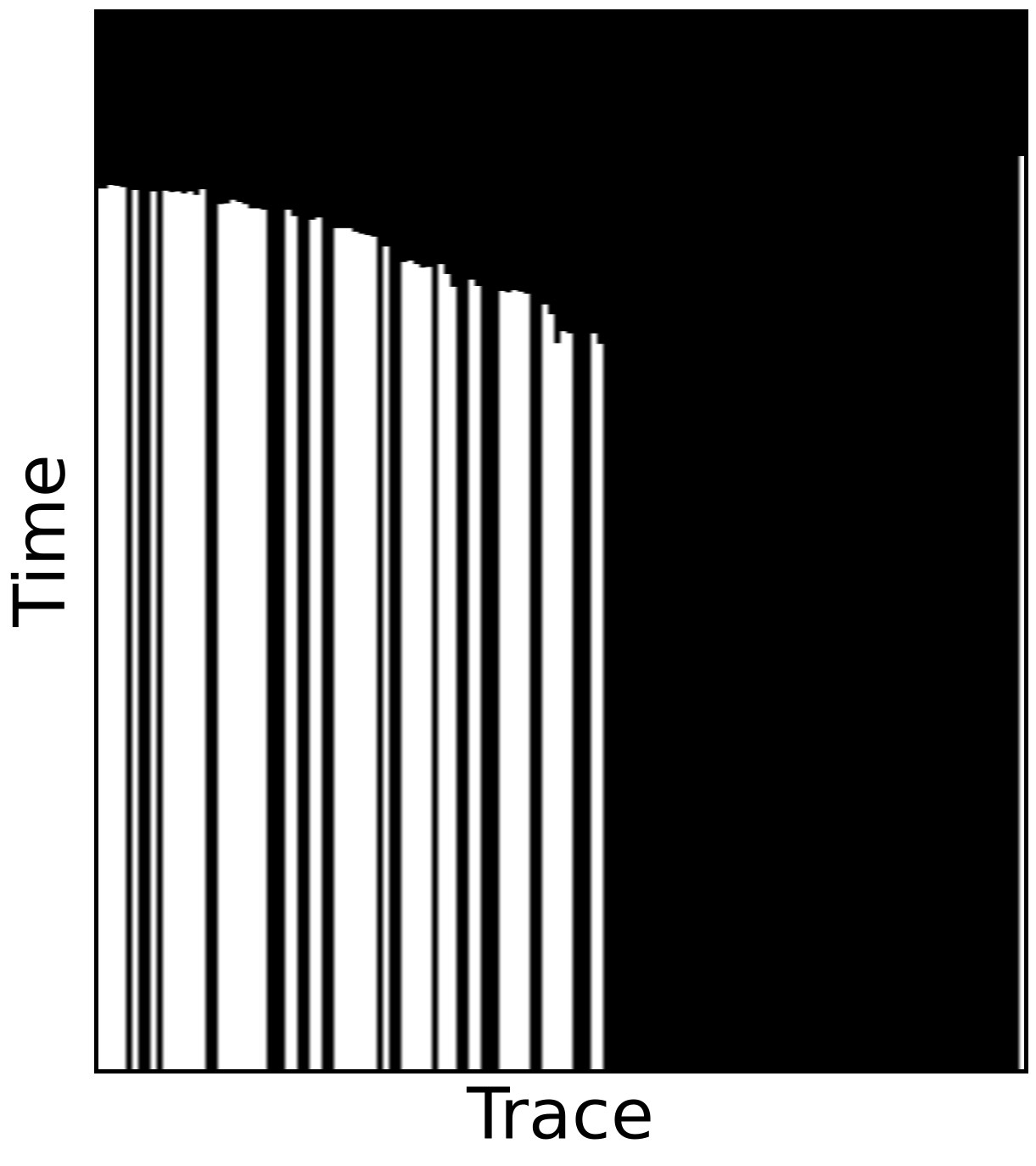} \\
{\small\textbf{(d) Brunswick label}} & {\small\textbf{(e) Halfmile label}} & {\small\textbf{(f) Lalor label}} \\
\end{tabular}
\caption{Example first-arrival picking data. Top row: input shot gathers from the Brunswick, Halfmile, and Lalor surveys. Bottom row: the corresponding binary step-mask labels, which switch from zero to one at the annotated first arrival.}
\label{fig:picking-examples}
\end{figure}

\paragraph{First-arrival picking datasets.}
The picking collection is built on HardPicks~\citep{stcharles2023hardpicks}, a public benchmark of hardrock seismic reflection data collected from 3D surveys at mining sites in Canada and Finland, containing millions of traces with labeled first-break picks. \textsc{SPBench} uses its Brunswick, Halfmile, and Lalor surveys, which were acquired with dynamite sources, and evaluates them both per survey and as a combined set. Each input SEG-Y file is paired with a binary step mask $w(t) = H(t - t^*)$, where $H$ is the Heaviside step function and $t^*$ is the annotated first arrival, so the mask is zero before $t^*$ and one from the arrival onward. Figure~\ref{fig:picking-examples} shows example inputs and labels.

\input{tables/reproduced_methods.tex}

\FloatBarrier
\section{Experiment Design}
\label{app:experiment-design}

\paragraph{Seismic interpolation.}
Models take a masked shot gather as input and reconstruct the complete gather. Missing-trace experiments are organized into three families. Random missing removes traces at random positions, uniform missing removes traces at a fixed regular interval, and consecutive missing removes contiguous blocks of traces. Each family has fixed levels: 30\%, 50\%, and 70\% for random missing, 50\% and 75\% for uniform missing, and gaps of 20, 30, and 40 traces for consecutive missing. All families run on both SEG C3 and Mobil AVO. The exact mask patterns and all parameter settings are available in the released code. Splits are made at the shot level to prevent leakage, and masks are generated only after the split is fixed. Observed traces are retained during evaluation. Most models train on patches of 256 time samples by 128 traces, or 256 by 120 on Mobil AVO, whose spatial extent is at most 120 traces. Data are normalized globally by the absolute maximum of the training set, and a few models use patch-wise normalization instead. Evaluation is performed on the full gather. Each run is executed with three random seeds. This task contributes 16 settings.

\paragraph{Random noise attenuation.}
Models take a noise-contaminated gather as input and predict the clean gather. Gaussian or Poisson noise is added to the clean data until the input SNR drops to $-5$, $0$, or $5$ dB, using experiment-specific random seeds, on SEG C3 and Mobil AVO. The exact noise realizations and all parameter settings are available in the released code. Shot-level train/validation/test partitions follow a $7{:}1{:}1$ split to prevent leakage. Patching, normalization, and evaluation follow the interpolation protocol: most models train on patches of 256 time samples by 128 traces, or 256 by 120 on Mobil AVO, data are normalized globally by the absolute maximum of the training set, and evaluation is performed on the full gather. Each run is executed with three random seeds. This task contributes 12 settings.

\paragraph{Ground-roll noise suppression.}
Models predict the additive ground-roll component. The estimate is subtracted from the contaminated input to recover the clean signal, and all metrics are computed on the recovered result. The synthetic settings scale the modeled ground roll to five amplitude multipliers, denoted as levels 1, 3, 5, 7, and 9. The \textsc{SPBench} field ground-roll set contributes a single setting with the same evaluation suite. Its labels are produced by expert processing rather than being ideal clean references, so results on this setting should be interpreted with care. Each run is executed with three random seeds. This task contributes 6 settings.

\paragraph{Multiple suppression.}
Models take a multiple-contaminated gather as input and estimate the additive multiple component, whose labels are obtained directly from physical modeling. Primary preservation can therefore be evaluated separately from noise removal by subtracting the estimate from the input. Shot gathers contain 638 traces and 2000 time samples, and models train on patches of 256 time samples by 512 traces. Shot-level train/validation/test partitions follow a $7{:}1{:}1$ split, and data are normalized globally by the absolute maximum of the training set. Each run is executed with three random seeds. This task contributes a single setting.

\paragraph{Deblending.}
Models take a blended common-receiver gather as input and recover the unblended reference. Writing the blending process as $\bm{b} = \bm{B}\bm{U}$, the task is to recover the unblended single-source responses $\bm{U}$ from $\bm{b}$ given the blending operator $\bm{B}$. The synthetic settings cover three blending levels defined by the blending factor $bf \in \{0.2, 0.4, 0.6\}$, corresponding to weak, moderate, and strong blending. For each level, the firing schedule is generated recursively as $\tau_{1} = 0$ and $\tau_{i+1} = \tau_i + \lfloor \epsilon_i \, n_t / bf \cdot d_t \rfloor$, with the firing times $\tau_i$ in milliseconds, $n_t$ the number of time samples per trace, $d_t$ the sampling interval, and $\epsilon_i$ a uniform random variate on $(0,1)$, so a larger $bf$ produces shorter average inter-shot intervals and stronger overlap. The Mobil AVO setting uses $bf = 5$ with a fixed 500-sample delay plus a random jitter of up to $n_t/5$ samples, mimicking streamer-style acquisition timing. Each configuration records the source combination, dither schedule, random seed, and split identifier so that every method is evaluated on identical interference patterns. Each run is executed with three random seeds. This task contributes 4 settings.

\paragraph{First-arrival picking.}
Models take a receiver-line patch as input and predict a binary step mask, from which the first-arrival pick is read at the mask transition. FFID-level splitting prevents gather leakage, and receiver-line segmentation prevents patches from crossing distant acquisition lines. All models use the same $128\times512$ patch geometry, and hit rates are reported at tolerances of 1, 3, 5, 7, and 9 samples. Each run is executed with three random seeds. This task contributes 4 settings.

Training schedules, optimizers, and stopping rules for every run are recorded per configuration group in the released artifacts.

\FloatBarrier
\section{Evaluation Metrics}
\label{app:metrics}

\paragraph{Rank correlation.}
To compare the rankings produced by different metrics or data domains, we use the Kendall rank correlation
\begin{equation*}
\tau = \frac{n_c - n_d}{\binom{n}{2}},
\end{equation*}
where $n_c$ and $n_d$ count concordant and discordant pairs among $n$ ranked methods. $\tau$ ranges from $-1$ (fully reversed rankings) through $0$ (no association) to $+1$ (identical rankings). This is the no-ties form. Where the compared scores contain ties, as in the per-survey model rankings of Section~\ref{sec:metrics-validation}, we use the tie-corrected form (Kendall $\tau_b$) computed by standard libraries, which reduces to the no-ties form on continuous scores. Sections~\ref{sec:transfer}, \ref{sec:robustness}, and~\ref{sec:metrics-validation} use it to quantify ranking agreement between settings and between metrics.

All metrics are computed between a prediction $p$ and a reference $r$, both treated as vectors in $\mathbb{R}^N$ with samples $p_i$ and $r_i$. Each metric is first evaluated per test record, then averaged over records, and finally over the three random seeds of each run, so every reported value is a mean over seeds. Unless noted otherwise, larger values indicate better recovery.

\paragraph{Signal-to-noise ratios.}
The output signal-to-noise ratio, the standard global fidelity measure in seismic processing,
\begin{equation*}
\mathrm{SNR} = 10\log_{10}\frac{\|r\|_2^2}{\|p - r\|_2^2}\;(\mathrm{dB}),
\end{equation*}
compares the energy of the reference with the energy of the residual error. Higher is better, and every $+6$ dB roughly halves the residual amplitude. The peak signal-to-noise ratio
\begin{equation*}
\mathrm{PSNR} = 10\log_{10}\frac{N\,\mathrm{MAX}^2}{\|p - r\|_2^2}\;(\mathrm{dB})
\end{equation*}
normalizes the residual energy by the peak absolute amplitude $\mathrm{MAX} = \max_i |r_i|$ of the reference, so it is more comparable across datasets with different amplitude scales; higher is better. Since $\mathrm{MSE} = \|p - r\|_2^2 / N$, this equals the common form $10\log_{10}(\mathrm{MAX}^2/\mathrm{MSE})$. Within one setting, $\mathrm{MAX}$ is fixed by the reference data, so PSNR and SNR differ only by a constant and produce identical rankings. The result tables therefore report SNR, and the PSNR values are available in the released results.

\paragraph{Pointwise error metrics.}
The mean squared error $\mathrm{MSE} = \frac{1}{N}\|p - r\|_2^2 = \frac{1}{N}\sum_{i} (p_i - r_i)^2$ measures the average squared deviation between prediction and reference, and the root mean squared error $\mathrm{RMSE} = \sqrt{\mathrm{MSE}}$ expresses it on the signal-amplitude scale; smaller is better for both. Both weight errors by squared amplitude, so they are dominated by strong events. The mean absolute error $\mathrm{MAE} = \frac{1}{N}\sum_i |p_i - r_i|$ measures the average absolute deviation, weights all errors linearly, and is less sensitive to large outliers than MSE; smaller is better.

\paragraph{Structural similarity.}
The structural similarity index (SSIM) compares prediction and reference through luminance, contrast, and structure terms,
\begin{equation*}
\mathrm{SSIM}(p, r) = \frac{(2\mu_p\mu_r + c_1)(2\sigma_{pr} + c_2)}{(\mu_p^2 + \mu_r^2 + c_1)(\sigma_p^2 + \sigma_r^2 + c_2)},
\end{equation*}
where $\mu_p$ and $\mu_r$ denote local means, $\sigma_p^2$ and $\sigma_r^2$ local variances, and $\sigma_{pr}$ the local covariance, and $c_1$, $c_2$ are small stabilizing constants. SSIM is upper-bounded by 1, with larger values indicating greater structural similarity. It can be negative on sign-alternating seismic amplitudes, and it captures perceptual and structural fidelity rather than pointwise error, so it can rank models differently from SNR.

\paragraph{Component-resolved fidelity construction (SCoRE).}
FB-FRE partitions the reference by frequency. Given a reference shot record $r$, prediction $p$, and sampling interval $dt$, we compute the real Fourier transform along the time axis and estimate the average reference power spectrum
\begin{equation*}
P(f_k) = \frac{1}{N_{\mathrm{avg}}}\sum_{\mathrm{non\text{-}time\ axes}} |\mathrm{RFFT}(r)(f_k)|^2,\quad
f_k = \mathrm{rfftfreq}(N_t, d=dt).
\end{equation*}
Here $N_t$ is the number of time samples per trace, and the average runs over the $N_{\mathrm{avg}}$ non-time slices of the record. The effective frequency band is defined by the frequencies whose power exceeds a relative threshold $\eta=10^{-3}$ of the peak power:
\begin{equation*}
f_{\min}=\min\{f_k: P(f_k)\geq \eta \max_f P(f)\},\quad
f_{\max}=\max\{f_k: P(f_k)\geq \eta \max_f P(f)\}.
\end{equation*}
This band is divided into four contiguous subbands, low, mid, high, and very high, using relative widths $(0.20,0.30,0.30,0.20)$. The band content and the per-band scores follow Section~\ref{sec:domain-aware-metrics}, and the leaderboard columns labeled FB-FRE report the per-band partition SNR, $\mathrm{SNR}_{\mathcal{S}}$ of Section~\ref{sec:domain-aware-metrics}.

EB-FE partitions the reference by local energy. It builds a smoothed reference-energy map
\begin{equation*}
E_i=\sqrt{\mathrm{gaussian\_filter}(r_i^2,\sigma)},
\end{equation*}
removes strictly zero-reference samples, and sorts the remaining $N_{\mathrm{valid}}$ samples by $E_i$. The filter width $\sigma$ is recorded in the released configuration. For an energy-percentile bin $\mathcal{B}(p_L,p_H)$, samples whose sorted ranks fall in
\begin{equation*}
\left[N_{\mathrm{valid}}\frac{p_L}{100},\; N_{\mathrm{valid}}\frac{p_H}{100}\right)
\end{equation*}
are evaluated separately. The default bins are very weak $(5,20)$, weak $(20,40)$, medium $(40,70)$, and strong $(70,100)$. The leaderboard columns labeled EB-FE report the per-bin partition SNR, likewise $\mathrm{SNR}_{\mathcal{S}}$. The main-text tables report the medium and strong bins, and all bins are available in the released metric summaries. The released metric summaries additionally report per-partition normalized errors, $\mathrm{NE}_b = \|p_b-r_b\|_2/(\|r_b\|_2+\varepsilon)$ and the analogously defined $\mathrm{NE}_{\mathcal{B}}$, and the reference energy ratio $\rho_b = \|r_b\|_2^2/(\|r\|_2^2+\varepsilon)$.

\paragraph{Partition-ranking agreement with SNR.}
For each task, we compute the Kendall rank correlation between the SNR ranking and each partition-SNR ranking and average it over the task's settings (Table~\ref{tab:score-tau}). The agreement is moderate to strong but never perfect, so the partition rankings carry information that the aggregate score does not. The very-weak and weak energy bins are excluded from this comparison because their low reference energy makes the partition SNR unreliable.

\begin{table}[H]
\centering
\footnotesize
\setlength{\tabcolsep}{4pt}
\caption{Mean Kendall rank correlation between the SNR ranking and each SCoRE partition ranking, averaged over the settings of each task.}
\label{tab:score-tau}
\begin{tabular}{@{}lccccc@{}}
\toprule
Partition & Noise & Interp. & Ground roll & Multiple & Deblend \\
\midrule
FB-FRE Low & 0.752 & 0.699 & 0.848 & 0.500 & 0.644 \\
FB-FRE Mid & 0.854 & 0.820 & 0.838 & 0.944 & 0.900 \\
FB-FRE High & 0.818 & 0.800 & 0.696 & 0.944 & 0.944 \\
FB-FRE VHigh & 0.800 & 0.521 & 0.740 & 0.722 & 0.489 \\
EB-FE Med & 0.738 & 0.505 & 0.791 & 0.889 & 0.822 \\
EB-FE Strong & 0.716 & 0.900 & 0.922 & 0.722 & 0.878 \\
\bottomrule
\end{tabular}
\end{table}

\paragraph{Picking metrics.}
For first-arrival picking, let $t_i^*$ be the reference pick on trace $i$ and $\hat{t}_i$ the predicted pick, both in samples, over $N_{\mathrm{tr}}$ traces, and write $\Delta_i = \hat{t}_i - t_i^*$ for the signed pick error. The mean absolute error
\begin{equation*}
\mathrm{MAE} = \frac{1}{N_{\mathrm{tr}}}\sum_{i=1}^{N_{\mathrm{tr}}} |\Delta_i|
\end{equation*}
measures the average pick displacement in samples, and multiplying by the sampling interval converts it to milliseconds; smaller is better. The root mean squared pick error follows as $\mathrm{RMSE}_{\mathrm{pick}} = \sqrt{\frac{1}{N_{\mathrm{tr}}} \sum_i(\hat t_i-t_i^*)^2}$, also in samples; smaller is better. The mean bias error $\mathrm{MBE} = \frac{1}{N_{\mathrm{tr}}}\sum_i \Delta_i$ preserves the sign and therefore reveals systematic early or late picking that absolute errors cancel out. The hit rate at tolerance $k$,
\begin{equation*}
\mathrm{H@}k = \frac{\#\{ i : |\Delta_i| < k \}}{N_{\mathrm{tr}}},\qquad k \in \{1,3,5,7,9\},
\end{equation*}
is the fraction of picks with an error below $k$ samples, where $\#\{\cdot\}$ counts the traces satisfying the condition; higher is better, and the curve $\mathrm{H@}k$ versus $k$ shows how quickly picks concentrate around the reference. For detection quality, precision, recall, and F1 are computed pixel-wise between the predicted binary step mask and the reference mask. With $\mathrm{TP}$ the number of samples where both masks equal one, and $\mathrm{FP}$ and $\mathrm{FN}$ defined accordingly, $\mathrm{P} = \mathrm{TP}/(\mathrm{TP}+\mathrm{FP})$, $\mathrm{R} = \mathrm{TP}/(\mathrm{TP}+\mathrm{FN})$, and $\mathrm{F1} = 2\mathrm{PR}/(\mathrm{P}+\mathrm{R})$; higher is better. This mask-level F1 measures overlap of the predicted step mask and is not comparable to a pick-level F1 at a fixed error tolerance, which would coincide with the corresponding hit rate.

\paragraph{Ridge-continuity construction.}
The model outputs a first-arrival probability $p_i(s)$ for each trace $i$ at time sample $s$. With the threshold $\gamma = 0.5$, the predicted pick in samples is $\hat{s}_i = \min\{s : p_i(s) \ge \gamma\}$, the same quantity denoted $\hat{t}_i$ in the pick-error metrics above, and traces without any sample above the threshold are excluded. Let $\mathcal{V}$ denote the set of valid picks and $N_v = |\mathcal{V}|$. Valid picks are sorted by absolute source-receiver offset $x_i$. A robust offset-time ridge is then estimated in three steps. First, the offset range is divided into $B = \max\!\left(3,\, \min\!\left(50,\, \lfloor N_v/5 \rfloor,\, N_v\right)\right)$ equal bins, and each bin is represented by the median pick time of its members, so that isolated gross mispicks do not bend the ridge. Second, empty bins are filled by linear interpolation from adjacent valid bins. Third, the bin sequence is smoothed by a moving average whose window is $0.07B$, adjusted to an odd number not exceeding $B$, which suppresses high-frequency fluctuations caused by the finite sample count, and the smoothed ridge is recentered so that the median residual over the valid picks is zero. Let $\{(\tilde{x}_j,\tilde{t}_j)\}_{j=1}^{B}$ denote the resulting ridge at the offset-bin centers. Its discrete slope and slope change are
\begin{equation*}
d_j=\frac{\tilde{t}_j-\tilde{t}_{j-1}}{\tilde{x}_j-\tilde{x}_{j-1}},\qquad j=2,\ldots,B,
\qquad
c_j=\frac{d_j-d_{j-1}}{\tilde{x}_j-\tilde{x}_{j-1}},\qquad j=3,\ldots,B,
\end{equation*}
and the ridge curvature is the mean absolute slope change, $\mathrm{RC} = \frac{1}{B-2} \sum_{j=3}^{B}|c_j|$. This quantity is a discrete second derivative of travel time with respect to offset and carries units of time per offset squared. The normalized ridge curvature removes first-order differences in spatial and temporal scale through $\mathrm{RC}_{\mathrm{norm}} = \mathrm{RC}\,X_{\mathrm{range}}^{2}/T_{\mathrm{range}}$, where $X_{\mathrm{range}}=\max_i x_i-\min_i x_i$ and $T_{\mathrm{range}}=\max_j\tilde{t}_j-\min_j\tilde{t}_j$, which makes $\mathrm{RC}_{\mathrm{norm}}$ dimensionless. Under rescalings $x' = \alpha x$ and $t' = \beta t$, the slopes scale as $d'_j = (\beta/\alpha)\,d_j$ and the slope changes as $c'_j = (\beta/\alpha^{2})\,c_j$, so $\mathrm{RC}$ changes by $|\beta|/\alpha^{2}$ while the factor $X_{\mathrm{range}}^{2}/T_{\mathrm{range}}$ changes by $\alpha^{2}/|\beta|$, and $\mathrm{RC}_{\mathrm{norm}}$ is invariant. A numerically flat ridge, with $T_{\mathrm{range}} \le \varepsilon_t$ and $\mathrm{RC} = 0$, is assigned $\mathrm{RC}_{\mathrm{norm}} = 0$, where $\varepsilon_t$ is a numerical zero tolerance scaled to the sampling interval.

\FloatBarrier
\section{Validation of the Reference-Free Ridge Score}
\label{app:ridge-validation}

\paragraph{Systematic-shift failure mode.}
One important counterexample is a global systematic shift. If every predicted arrival is shifted by a constant $b$, $\hat{s}_i' = \hat{s}_i + b$, the recentered ridge shifts by the same $b$, so the residual statistics and the ridge curvature remain unchanged. A smooth curve that consistently follows the wrong event can therefore still obtain good scores. This failure mode can only be identified through manual labels, waveform evidence, or other physical constraints.

\paragraph{Tolerance sweep of the ridge-score agreement.}
Table~\ref{tab:rcnorm-sweep} extends the validation of Section~\ref{sec:metrics-validation} to all five released hit-rate tolerances. The agreement with the labeled rankings rises monotonically with the tolerance, from 0.244 at H@1 to 0.862 at H@9, approaching the MAE-based agreement of 0.881. Strict tolerances penalize every sample of deviation, including a constant offset shared by all traces, while the ridge score is invariant to such shifts. The gap between the strict-tolerance rows and the MAE row therefore quantifies the systematic-shift blindness described above. The main text reports H@5 as the middle tolerance of the released hit-rate set.

\begin{table}[H]
\centering
\footnotesize
\setlength{\tabcolsep}{4pt}
\caption{Kendall rank correlation of the reference-free ridge score with labeled accuracy rankings and between its two aggregation rules, per field picking survey over the ten evaluated methods.}
\label{tab:rcnorm-sweep}
\begin{tabular}{@{}lcccc@{}}
\toprule
Ranking pair & Brunswick & Halfmile & Lalor & Mean \\
\midrule
$\mathrm{RC}_{\mathrm{norm}}$ vs pick error (MAE) & 0.778 & 0.911 & 0.956 & 0.881 \\
$\mathrm{RC}_{\mathrm{norm}}$ vs $1-$H@1 & 0.244 & 0.422 & 0.067 & 0.244 \\
$\mathrm{RC}_{\mathrm{norm}}$ vs $1-$H@3 & 0.556 & 0.644 & 0.333 & 0.511 \\
$\mathrm{RC}_{\mathrm{norm}}$ vs $1-$H@5 & 0.689 & 0.899 & 0.822 & 0.803 \\
$\mathrm{RC}_{\mathrm{norm}}$ vs $1-$H@7 & 0.719 & 0.911 & 0.911 & 0.847 \\
$\mathrm{RC}_{\mathrm{norm}}$ vs $1-$H@9 & 0.764 & 0.911 & 0.911 & 0.862 \\
\midrule
Shot macro vs trace weighted & 0.956 & 0.956 & 1.000 & 0.970 \\
\bottomrule
\end{tabular}
\end{table}

\paragraph{Aggregation choices.}
The main-text scores use shot-level macro averaging, and weighting by the valid trace count per shot instead produces nearly identical model rankings (Table~\ref{tab:rcnorm-sweep}, last row).

\FloatBarrier
\input{tables/full_results_appendix.tex}

\end{document}

%% file: math_commands.tex
\usepackage{amsmath,amsfonts,bm}

\def\eqref#1{equation~\ref{#1}}
\def\1{\bm{1}}

\DeclareMathAlphabet{\mathsfit}{\encodingdefault}{\sfdefault}{m}{sl}
\SetMathAlphabet{\mathsfit}{bold}{\encodingdefault}{\sfdefault}{bx}{n}

%% file: tables/main_results_tables.tex
\begin{table}[p]
\caption{Representative reconstruction results on five benchmark settings (top eight methods by output SNR per setting). All scores are means over three seeds with standard deviations in small type. Rows are ordered by increasing SNR; shading is computed per block and metric. Setting labels: G0 is Gaussian noise at 0 dB input SNR, R50 is random missing of 50\% of traces, L5 is ground-roll multiplier level 5 of five levels (L1--L9), and mod is the moderate of three blending levels.}
\label{tab:main-results}
\centering
\scriptsize
\setlength{\tabcolsep}{2.6pt}
\renewcommand{\arraystretch}{0.85}
\resizebox{\linewidth}{!}{%
\begin{tabular}{lcccccccc}
\toprule
Method & Params & SNR & \multicolumn{4}{c}{FB-FRE} & \multicolumn{2}{c}{EB-FE} \\
\cmidrule(lr){4-7}\cmidrule(lr){8-9}
 & & & Low & Mid & High & VHigh & Med. & Strong \\
\midrule
    \multicolumn{9}{@{}l}{\cellcolor{black!6}\textbf{\textcolor{seisRandomHeat}{Random noise (SEG C3, G0)}}} \\
Attention UNet & 7.85M & \heatbase{12.80{\scriptsize$\pm$0.10}} & \heatlow{12.00{\scriptsize$\pm$0.15}} & \heatlow{15.73{\scriptsize$\pm$0.07}} & \heatbase{11.98{\scriptsize$\pm$0.08}} & \heatbase{2.18{\scriptsize$\pm$0.10}} & \heatbase{1.26{\scriptsize$\pm$0.09}} & \heatbase{14.82{\scriptsize$\pm$0.10}} \\
FBResNet & 0.45M & \heatlow{12.97{\scriptsize$\pm$0.00}} & \heatbase{11.99{\scriptsize$\pm$0.04}} & \heatbase{15.45{\scriptsize$\pm$0.01}} & \heatlow{12.12{\scriptsize$\pm$0.02}} & \heatmid{2.91{\scriptsize$\pm$0.01}} & \heatmid{1.47{\scriptsize$\pm$0.02}} & \heatlow{15.00{\scriptsize$\pm$0.01}} \\
QUNet & 21.88M & \heatlow{13.08{\scriptsize$\pm$0.08}} & \heatlow{12.59{\scriptsize$\pm$0.16}} & \heatlow{15.81{\scriptsize$\pm$0.04}} & \heatlow{12.18{\scriptsize$\pm$0.05}} & \heatlow{2.61{\scriptsize$\pm$0.05}} & \heatlow{1.34{\scriptsize$\pm$0.06}} & \heatlow{15.19{\scriptsize$\pm$0.09}} \\
UNet-L & 31.04M & \heatmid{13.13{\scriptsize$\pm$0.04}} & \heatmid{12.78{\scriptsize$\pm$0.06}} & \heatmid{15.91{\scriptsize$\pm$0.06}} & \heatmid{12.23{\scriptsize$\pm$0.04}} & \heatlow{2.60{\scriptsize$\pm$0.06}} & \heatlow{1.44{\scriptsize$\pm$0.03}} & \heatmid{15.25{\scriptsize$\pm$0.05}} \\
UNet++ & 9.05M & \heatmid{13.47{\scriptsize$\pm$0.04}} & \heatmid{12.78{\scriptsize$\pm$0.06}} & \heatmid{15.90{\scriptsize$\pm$0.04}} & \heatgood{12.62{\scriptsize$\pm$0.04}} & \heatmid{3.26{\scriptsize$\pm$0.08}} & \heatgood{1.74{\scriptsize$\pm$0.09}} & \heatmid{15.63{\scriptsize$\pm$0.03}} \\
CBD-RDN & 1.49M & \heatgood{13.58{\scriptsize$\pm$0.45}} & \heatgood{13.23{\scriptsize$\pm$0.59}} & \heatgood{15.92{\scriptsize$\pm$0.37}} & \heatmid{12.57{\scriptsize$\pm$0.42}} & \heatgood{3.53{\scriptsize$\pm$0.53}} & \heatmid{1.67{\scriptsize$\pm$0.24}} & \heatgood{16.02{\scriptsize$\pm$0.53}} \\
DnCNN & 0.56M & \heatgood{14.10{\scriptsize$\pm$0.06}} & \heatgood{13.75{\scriptsize$\pm$0.13}} & \heatgood{16.40{\scriptsize$\pm$0.08}} & \heatgood{13.16{\scriptsize$\pm$0.05}} & \heatgood{3.93{\scriptsize$\pm$0.03}} & \heatgood{1.91{\scriptsize$\pm$0.05}} & \heatgood{16.67{\scriptsize$\pm$0.11}} \\
SCRN & 0.43M & \heatbest{14.38{\scriptsize$\pm$0.01}} & \heatbest{14.01{\scriptsize$\pm$0.00}} & \heatbest{16.60{\scriptsize$\pm$0.02}} & \heatbest{13.40{\scriptsize$\pm$0.01}} & \heatbest{4.29{\scriptsize$\pm$0.01}} & \heatbest{2.19{\scriptsize$\pm$0.00}} & \heatbest{16.92{\scriptsize$\pm$0.01}} \\
\midrule
    \multicolumn{9}{@{}l}{\cellcolor{black!6}\textbf{\textcolor{seisInterpolationHeat}{Interpolation (SEG C3, R50)}}} \\
DnCNN & 0.14M & \heatbase{16.11{\scriptsize$\pm$0.02}} & \heatbase{20.39{\scriptsize$\pm$0.09}} & \heatbase{20.61{\scriptsize$\pm$0.05}} & \heatbase{12.74{\scriptsize$\pm$0.02}} & \heatbase{10.15{\scriptsize$\pm$0.02}} & \heatmid{11.58{\scriptsize$\pm$0.00}} & \heatbase{18.88{\scriptsize$\pm$0.06}} \\
ResUNet-L & 32.44M & \heatlow{16.23{\scriptsize$\pm$0.06}} & \heatbest{24.92{\scriptsize$\pm$0.41}} & \heatlow{20.83{\scriptsize$\pm$0.18}} & \heatlow{12.79{\scriptsize$\pm$0.03}} & \heatmid{10.50{\scriptsize$\pm$0.02}} & \heatbase{11.42{\scriptsize$\pm$0.02}} & \heatlow{19.61{\scriptsize$\pm$0.16}} \\
CFunet & 7.35M & \heatlow{16.23{\scriptsize$\pm$0.09}} & \heatlow{22.69{\scriptsize$\pm$0.11}} & \heatlow{20.72{\scriptsize$\pm$0.09}} & \heatmid{12.88{\scriptsize$\pm$0.10}} & \heatgood{10.56{\scriptsize$\pm$0.13}} & \heatlow{11.48{\scriptsize$\pm$0.08}} & \heatlow{19.42{\scriptsize$\pm$0.13}} \\
UNet-L & 31.04M & \heatmid{16.26{\scriptsize$\pm$0.03}} & \heatgood{23.82{\scriptsize$\pm$0.29}} & \heatmid{20.95{\scriptsize$\pm$0.03}} & \heatlow{12.81{\scriptsize$\pm$0.05}} & \heatlow{10.38{\scriptsize$\pm$0.05}} & \heatbase{11.42{\scriptsize$\pm$0.05}} & \heatmid{19.63{\scriptsize$\pm$0.03}} \\
Attention UNet-L & 31.39M & \heatmid{16.31{\scriptsize$\pm$0.13}} & \heatmid{23.71{\scriptsize$\pm$0.09}} & \heatmid{20.94{\scriptsize$\pm$0.18}} & \heatmid{12.88{\scriptsize$\pm$0.12}} & \heatlow{10.48{\scriptsize$\pm$0.10}} & \heatlow{11.48{\scriptsize$\pm$0.13}} & \heatgood{19.64{\scriptsize$\pm$0.13}} \\
WRDL & 35.83M & \heatgood{16.54{\scriptsize$\pm$0.46}} & \heatgood{24.84{\scriptsize$\pm$1.42}} & \heatbest{21.51{\scriptsize$\pm$0.43}} & \heatgood{13.20{\scriptsize$\pm$0.30}} & \heatmid{10.49{\scriptsize$\pm$0.43}} & \heatbest{12.22{\scriptsize$\pm$0.08}} & \heatbest{20.10{\scriptsize$\pm$0.08}} \\
CA-Unet & 7.79M & \heatgood{16.55{\scriptsize$\pm$0.11}} & \heatmid{22.84{\scriptsize$\pm$0.40}} & \heatgood{20.96{\scriptsize$\pm$0.14}} & \heatbest{13.27{\scriptsize$\pm$0.10}} & \heatbest{10.73{\scriptsize$\pm$0.15}} & \heatmid{11.84{\scriptsize$\pm$0.14}} & \heatgood{19.76{\scriptsize$\pm$0.03}} \\
DnCNN-L & 0.56M & \heatbest{16.56{\scriptsize$\pm$0.01}} & \heatlow{22.39{\scriptsize$\pm$0.16}} & \heatgood{21.25{\scriptsize$\pm$0.10}} & \heatgood{13.13{\scriptsize$\pm$0.04}} & \heatgood{10.59{\scriptsize$\pm$0.07}} & \heatgood{11.88{\scriptsize$\pm$0.04}} & \heatmid{19.62{\scriptsize$\pm$0.04}} \\
\midrule
    \multicolumn{9}{@{}l}{\cellcolor{black!6}\textbf{\textcolor{seisSurfaceHeat}{Ground roll (SPBench, L5)}}} \\
SANet & 0.23M & \heatbase{14.26{\scriptsize$\pm$0.14}} & \heatlow{5.25{\scriptsize$\pm$0.34}} & \heatlow{17.62{\scriptsize$\pm$0.13}} & \heatbase{5.49{\scriptsize$\pm$0.07}} & \heatbase{3.73{\scriptsize$\pm$0.08}} & \heatbase{3.23{\scriptsize$\pm$0.14}} & \heatbase{15.61{\scriptsize$\pm$0.15}} \\
DnCNN & 0.14M & \heatlow{15.57{\scriptsize$\pm$0.22}} & \heatlow{9.13{\scriptsize$\pm$0.39}} & \heatbase{16.65{\scriptsize$\pm$0.21}} & \heatlow{10.90{\scriptsize$\pm$0.38}} & \heatlow{9.40{\scriptsize$\pm$0.50}} & \heatlow{9.83{\scriptsize$\pm$0.33}} & \heatlow{15.84{\scriptsize$\pm$0.23}} \\
ResUNet & 8.11M & \heatlow{18.11{\scriptsize$\pm$2.12}} & \heatbase{4.96{\scriptsize$\pm$3.67}} & \heatgood{23.94{\scriptsize$\pm$1.49}} & \heatbest{20.25{\scriptsize$\pm$1.45}} & \heatbest{19.23{\scriptsize$\pm$0.52}} & \heatlow{5.59{\scriptsize$\pm$2.49}} & \heatlow{20.16{\scriptsize$\pm$1.95}} \\
ResUNet-L & 32.44M & \heatmid{19.52{\scriptsize$\pm$0.29}} & \heatgood{11.98{\scriptsize$\pm$0.15}} & \heatlow{20.64{\scriptsize$\pm$0.38}} & \heatmid{16.39{\scriptsize$\pm$0.00}} & \heatmid{14.99{\scriptsize$\pm$0.13}} & \heatmid{10.57{\scriptsize$\pm$0.27}} & \heatmid{20.23{\scriptsize$\pm$0.31}} \\
UNet & 7.76M & \heatmid{20.22{\scriptsize$\pm$0.32}} & \heatmid{11.46{\scriptsize$\pm$0.17}} & \heatmid{21.76{\scriptsize$\pm$0.27}} & \heatgood{16.78{\scriptsize$\pm$1.04}} & \heatgood{15.37{\scriptsize$\pm$1.25}} & \heatgood{11.25{\scriptsize$\pm$0.30}} & \heatmid{20.92{\scriptsize$\pm$0.31}} \\
Attention UNet-L & 31.39M & \heatgood{20.23{\scriptsize$\pm$0.33}} & \heatgood{11.55{\scriptsize$\pm$0.12}} & \heatmid{21.92{\scriptsize$\pm$0.35}} & \heatlow{15.73{\scriptsize$\pm$1.29}} & \heatlow{14.21{\scriptsize$\pm$1.39}} & \heatgood{11.09{\scriptsize$\pm$0.27}} & \heatgood{20.98{\scriptsize$\pm$0.36}} \\
UNet-L & 31.04M & \heatgood{20.39{\scriptsize$\pm$0.13}} & \heatmid{11.44{\scriptsize$\pm$0.31}} & \heatgood{22.12{\scriptsize$\pm$0.24}} & \heatmid{16.33{\scriptsize$\pm$0.21}} & \heatmid{14.49{\scriptsize$\pm$0.22}} & \heatmid{10.81{\scriptsize$\pm$0.30}} & \heatgood{21.25{\scriptsize$\pm$0.17}} \\
Attention UNet & 7.85M & \heatbest{21.96{\scriptsize$\pm$1.06}} & \heatbest{12.06{\scriptsize$\pm$2.17}} & \heatbest{24.22{\scriptsize$\pm$0.92}} & \heatgood{18.43{\scriptsize$\pm$0.45}} & \heatgood{17.55{\scriptsize$\pm$0.36}} & \heatbest{11.97{\scriptsize$\pm$1.79}} & \heatbest{22.96{\scriptsize$\pm$0.92}} \\
\midrule
    \multicolumn{9}{@{}l}{\cellcolor{black!6}\textbf{\textcolor{seisMultipleHeat}{Multiple (SPBench)}}} \\
DNNDAT & 37.07M & \heatbase{18.51{\scriptsize$\pm$0.46}} & \heatbase{19.18{\scriptsize$\pm$1.12}} & \heatbase{19.21{\scriptsize$\pm$0.68}} & \heatbase{18.71{\scriptsize$\pm$0.40}} & \heatlow{15.36{\scriptsize$\pm$1.19}} & \heatbase{-13.83{\scriptsize$\pm$0.77}} & \heatbase{25.03{\scriptsize$\pm$0.24}} \\
UNet & 7.76M & \heatlow{20.52{\scriptsize$\pm$0.58}} & \heatmid{22.65{\scriptsize$\pm$0.44}} & \heatlow{21.69{\scriptsize$\pm$0.63}} & \heatlow{20.05{\scriptsize$\pm$0.64}} & \heatbase{15.17{\scriptsize$\pm$0.35}} & \heatlow{-11.49{\scriptsize$\pm$0.72}} & \heatlow{26.35{\scriptsize$\pm$0.32}} \\
ResUNet & 8.11M & \heatlow{20.77{\scriptsize$\pm$0.71}} & \heatmid{22.64{\scriptsize$\pm$0.47}} & \heatlow{21.92{\scriptsize$\pm$0.67}} & \heatlow{20.82{\scriptsize$\pm$0.86}} & \heatmid{15.99{\scriptsize$\pm$1.06}} & \heatmid{-10.95{\scriptsize$\pm$0.86}} & \heatlow{25.93{\scriptsize$\pm$0.59}} \\
Attention UNet & 7.85M & \heatmid{20.85{\scriptsize$\pm$0.88}} & \heatlow{22.08{\scriptsize$\pm$0.67}} & \heatmid{22.08{\scriptsize$\pm$0.92}} & \heatmid{20.94{\scriptsize$\pm$0.94}} & \heatlow{15.80{\scriptsize$\pm$0.90}} & \heatlow{-11.13{\scriptsize$\pm$1.06}} & \heatgood{27.03{\scriptsize$\pm$0.16}} \\
ResUNet-L & 32.44M & \heatmid{21.61{\scriptsize$\pm$0.54}} & \heatgood{23.91{\scriptsize$\pm$0.48}} & \heatmid{22.49{\scriptsize$\pm$0.50}} & \heatmid{21.73{\scriptsize$\pm$0.49}} & \heatgood{17.81{\scriptsize$\pm$0.74}} & \heatmid{-10.08{\scriptsize$\pm$0.61}} & \heatmid{26.98{\scriptsize$\pm$0.48}} \\
Attention UNet-L & 31.39M & \heatgood{22.33{\scriptsize$\pm$0.46}} & \heatgood{23.77{\scriptsize$\pm$0.54}} & \heatgood{23.23{\scriptsize$\pm$0.54}} & \heatgood{22.71{\scriptsize$\pm$0.86}} & \heatbest{18.53{\scriptsize$\pm$0.83}} & \heatgood{-8.94{\scriptsize$\pm$0.89}} & \heatgood{27.44{\scriptsize$\pm$0.37}} \\
SAGAN & 32.34M & \heatgood{22.44{\scriptsize$\pm$1.46}} & \heatlow{20.27{\scriptsize$\pm$1.53}} & \heatbest{23.75{\scriptsize$\pm$1.23}} & \heatgood{22.29{\scriptsize$\pm$2.10}} & \heatmid{16.42{\scriptsize$\pm$1.71}} & \heatgood{-9.08{\scriptsize$\pm$2.13}} & \heatmid{26.89{\scriptsize$\pm$0.61}} \\
UNet-L & 31.04M & \heatbest{22.75{\scriptsize$\pm$0.64}} & \heatbest{24.18{\scriptsize$\pm$0.47}} & \heatgood{23.68{\scriptsize$\pm$0.57}} & \heatbest{22.93{\scriptsize$\pm$0.77}} & \heatgood{18.18{\scriptsize$\pm$0.63}} & \heatbest{-8.37{\scriptsize$\pm$0.90}} & \heatbest{27.51{\scriptsize$\pm$0.25}} \\
\midrule
    \multicolumn{9}{@{}l}{\cellcolor{black!6}\textbf{\textcolor{seisDeblendingHeat}{Deblending (SPBench, mod)}}} \\
Attention UNet & 7.85M & \heatbase{14.56{\scriptsize$\pm$0.14}} & \heatlow{11.38{\scriptsize$\pm$0.17}} & \heatlow{15.10{\scriptsize$\pm$0.15}} & \heatbase{15.20{\scriptsize$\pm$0.17}} & \heatbase{9.55{\scriptsize$\pm$0.31}} & \heatlow{6.85{\scriptsize$\pm$0.07}} & \heatbase{15.30{\scriptsize$\pm$0.18}} \\
UNet & 7.76M & \heatlow{14.70{\scriptsize$\pm$0.24}} & \heatbase{11.37{\scriptsize$\pm$0.06}} & \heatbase{15.07{\scriptsize$\pm$0.54}} & \heatlow{15.48{\scriptsize$\pm$0.21}} & \heatlow{10.10{\scriptsize$\pm$0.72}} & \heatbase{6.50{\scriptsize$\pm$0.33}} & \heatlow{15.53{\scriptsize$\pm$0.28}} \\
UNet++ & 9.05M & \heatlow{16.13{\scriptsize$\pm$0.10}} & \heatmid{13.24{\scriptsize$\pm$0.17}} & \heatlow{15.95{\scriptsize$\pm$0.07}} & \heatlow{16.88{\scriptsize$\pm$0.27}} & \heatbest{13.26{\scriptsize$\pm$0.09}} & \heatlow{8.38{\scriptsize$\pm$0.19}} & \heatlow{16.71{\scriptsize$\pm$0.13}} \\
UNet-L & 31.04M & \heatmid{16.68{\scriptsize$\pm$0.64}} & \heatgood{13.61{\scriptsize$\pm$0.61}} & \heatmid{16.58{\scriptsize$\pm$0.71}} & \heatmid{17.49{\scriptsize$\pm$0.55}} & \heatgood{13.13{\scriptsize$\pm$0.16}} & \heatmid{8.45{\scriptsize$\pm$0.59}} & \heatmid{17.38{\scriptsize$\pm$0.66}} \\
FBResNet & 0.45M & \heatmid{17.01{\scriptsize$\pm$0.00}} & \heatlow{11.41{\scriptsize$\pm$0.00}} & \heatmid{18.05{\scriptsize$\pm$0.00}} & \heatmid{17.43{\scriptsize$\pm$0.00}} & \heatlow{10.44{\scriptsize$\pm$0.00}} & \heatgood{9.27{\scriptsize$\pm$0.00}} & \heatmid{17.88{\scriptsize$\pm$0.00}} \\
cDDPM & 33.3M & \heatgood{17.24{\scriptsize$\pm$0.23}} & \heatbest{15.25{\scriptsize$\pm$0.23}} & \heatgood{19.50{\scriptsize$\pm$0.13}} & \heatgood{18.77{\scriptsize$\pm$0.16}} & \heatmid{11.60{\scriptsize$\pm$0.32}} & \heatmid{8.80{\scriptsize$\pm$0.24}} & \heatgood{19.51{\scriptsize$\pm$0.15}} \\
DnCNN & 0.56M & \heatgood{17.40{\scriptsize$\pm$0.05}} & \heatmid{11.44{\scriptsize$\pm$0.07}} & \heatgood{18.48{\scriptsize$\pm$0.06}} & \heatgood{17.94{\scriptsize$\pm$0.06}} & \heatmid{10.60{\scriptsize$\pm$0.10}} & \heatgood{9.97{\scriptsize$\pm$0.07}} & \heatgood{18.19{\scriptsize$\pm$0.05}} \\
SCRN & 0.43M & \heatbest{19.36{\scriptsize$\pm$0.04}} & \heatgood{14.43{\scriptsize$\pm$0.28}} & \heatbest{20.16{\scriptsize$\pm$0.10}} & \heatbest{19.40{\scriptsize$\pm$0.08}} & \heatgood{13.12{\scriptsize$\pm$0.17}} & \heatbest{11.93{\scriptsize$\pm$0.09}} & \heatbest{19.94{\scriptsize$\pm$0.03}} \\
\bottomrule
\end{tabular}%
}

\vspace{3mm}
\caption{Representative first-arrival picking results on the Lalor survey, listing the top eight methods by MAE. All scores are means over three seeds with standard deviations in small type. Hit rates (H@$k$) measure the fraction of picks within $k$ samples, and $\mathrm{RC}_{\mathrm{norm}}$ measures the continuity of the predicted first-arrival ridge. Cell shading is computed independently for each metric.}
\label{tab:main-picking-results}
\centering
\scriptsize
\setlength{\tabcolsep}{3.0pt}
\renewcommand{\arraystretch}{0.85}
\resizebox{\linewidth}{!}{%
\begin{tabular}{llcccccc}
\toprule
Setting & Method & Params & MAE & $\mathrm{RC}_{\mathrm{norm}}$ & H@1 & H@5 & H@9 \\
\midrule
\taskheat{seisFirstArrivalHeat}{First arrival} & STUNet & 71.55M & \heatbest{2.27{\scriptsize$\pm$0.31}} & \heatbest{18.78{\scriptsize$\pm$0.94}} & \heatbase{0.236{\scriptsize$\pm$0.022}} & \heatgood{0.869{\scriptsize$\pm$0.021}} & \heatbest{0.962{\scriptsize$\pm$0.009}} \\
 & DSU-Net & 1.99M & \heatgood{5.69{\scriptsize$\pm$0.72}} & \heatgood{21.85{\scriptsize$\pm$4.43}} & \heatmid{0.545{\scriptsize$\pm$0.011}} & \heatbest{0.885{\scriptsize$\pm$0.006}} & \heatgood{0.949{\scriptsize$\pm$0.005}} \\
 & Attention UNet & 7.85M & \heatgood{13.18{\scriptsize$\pm$0.83}} & \heatgood{31.88{\scriptsize$\pm$0.44}} & \heatgood{0.616{\scriptsize$\pm$0.022}} & \heatgood{0.860{\scriptsize$\pm$0.009}} & \heatgood{0.920{\scriptsize$\pm$0.004}} \\
 & Attention UNet-L & 31.39M & \heatmid{15.63{\scriptsize$\pm$1.85}} & \heatmid{41.21{\scriptsize$\pm$3.26}} & \heatbest{0.672{\scriptsize$\pm$0.042}} & \heatmid{0.858{\scriptsize$\pm$0.011}} & \heatmid{0.912{\scriptsize$\pm$0.010}} \\
 & HUNet & 7.76M & \heatmid{17.05{\scriptsize$\pm$8.10}} & \heatmid{38.08{\scriptsize$\pm$26.88}} & \heatlow{0.507{\scriptsize$\pm$0.120}} & \heatlow{0.840{\scriptsize$\pm$0.050}} & \heatmid{0.904{\scriptsize$\pm$0.041}} \\
 & UNet & 7.76M & \heatlow{17.93{\scriptsize$\pm$4.32}} & \heatlow{42.52{\scriptsize$\pm$11.56}} & \heatlow{0.544{\scriptsize$\pm$0.031}} & \heatlow{0.836{\scriptsize$\pm$0.012}} & \heatlow{0.897{\scriptsize$\pm$0.015}} \\
 & ResUNet-L & 32.44M & \heatlow{18.35{\scriptsize$\pm$1.51}} & \heatlow{47.30{\scriptsize$\pm$3.71}} & \heatgood{0.633{\scriptsize$\pm$0.042}} & \heatmid{0.842{\scriptsize$\pm$0.011}} & \heatlow{0.898{\scriptsize$\pm$0.009}} \\
 & UNet-L & 31.04M & \heatbase{19.37{\scriptsize$\pm$12.79}} & \heatbase{50.52{\scriptsize$\pm$35.13}} & \heatmid{0.583{\scriptsize$\pm$0.103}} & \heatbase{0.836{\scriptsize$\pm$0.058}} & \heatbase{0.893{\scriptsize$\pm$0.055}} \\
\bottomrule
\end{tabular}%
}
\end{table}

%% file: tables/reproduced_methods.tex
\section{Reproduced Methods}
\label{app:methods}

This appendix lists the reproduced methods evaluated in each benchmark task. Four widely used architectures, U-Net, ResUNet, DnCNN, and attention-gated U-Net, are evaluated in every task as shared baselines. Beyond these baselines, the reproduced methods were selected from the task literature with clearly documented reproduction procedures. Table~\ref{tab:methods} summarizes them per task, with a short description derived from the source paper. Width-modified variants of the baselines (marked with an -L suffix on the leaderboard) follow the same architecture with increased base channels. Architecture notes, training schedules, and hyperparameters are recorded in the released configuration groups and the online artifact documentation.

\begin{table}[!tb]
\caption{Reproduced methods per benchmark task beyond the four shared baselines.}
\label{tab:methods}
\centering
\scriptsize
\setlength{\tabcolsep}{4pt}
\renewcommand{\arraystretch}{1.3}
\begin{tabularx}{\linewidth}{@{}L{0.17\linewidth}Y@{}}
\toprule
Method & Description \\
\midrule
\rowcolor{black!8}\multicolumn{2}{@{}l}{\textbf{Random noise attenuation and deblending}} \\
CBD-RDN & Residual dense network adapted for seismic denoising and upscaling, combining local and global residual learning~\citep{wang2022cbdrdn}. \\
FBResNet & Feedback residual network that iteratively refines strong-noise estimates~\citep{liao2023fbresnet}. \\
cDDPM & Conditional denoising diffusion probabilistic model, evaluated on random-noise attenuation and adapted to blending-noise suppression~\citep{li2024cddpm}. \\
QUNet & U-Net with quadratic neurons for richer feature representations, used for random-noise attenuation and transferred to deblending~\citep{wang2025qunet}. \\
SCRN & Swin-Transformer convolutional residual network for joint denoising and interpolation, transferred to deblending~\citep{gao2024scrn}. \\
U-Net++ & Nested U-Net++ used within a two-step prediction strategy, transferred from denoising to deblending~\citep{zhang2024unetpp}. \\
\midrule
\rowcolor{black!8}\multicolumn{2}{@{}l}{\textbf{Seismic interpolation}} \\
SPNet & Sparse-prior network that integrates the POCS algorithm into a deep architecture for trace interpolation~\citep{wu2024spnet}. \\
ANet & Attention-guided network with a hybrid loss for consecutively missing traces~\citep{yu2022anet}. \\
CA-Unet & Coordinate-attention U-Net for consecutive-missing interpolation~\citep{li2022caunet}. \\
CFunet & Coarse-refine network with upsampling and a Fourier-domain loss~\citep{park2022cfunet}. \\
WRDL & Wavelet-based residual deep learning for seismic reconstruction~\citep{liu2022wrdl}. \\
\midrule
\rowcolor{black!8}\multicolumn{2}{@{}l}{\textbf{Ground-roll noise suppression}} \\
cDDPM & Conditional denoising diffusion probabilistic model for ground-roll attenuation~\citep{li2024cddpm}. \\
Pix2Pix cGAN & Conditional GAN with a U-Net generator for ground-roll suppression~\citep{yuan2020gan}. \\
Physics CNN & Physics-constrained network combining learned and analytic separation filters~\citep{pham2022physics}. \\
SANet & Soft-attention network for small-scale ground-roll attenuation~\citep{yang2023softattentiongroundroll}. \\
Enhanced Atten-UNet & Attention U-Net with an adaptive frequency-modulation loss~\citep{sun2024enhancedunet}. \\
\midrule
\rowcolor{black!8}\multicolumn{2}{@{}l}{\textbf{Multiple suppression}} \\
DNNDAT & Deep neural network with data augmentation for marine multiple suppression~\citep{wang2022dnndat}. \\
SAGAN & GAN with a U-Net generator, self-attention, and a Markovian discriminator~\citep{tao2022sagan}. \\
\midrule
\rowcolor{black!8}\multicolumn{2}{@{}l}{\textbf{First-arrival picking}} \\
DSU-Net & U-Net with dynamic snake convolution for 2-D first-break picking~\citep{wang2024dsunet}. \\
HUNet & Uncertainty-aware picking network with prediction refinement~\citep{pu2024hunet}. \\
STUNet & U-shaped network with Swin Transformer feature extraction~\citep{jiang2023stunet}. \\
\bottomrule
\end{tabularx}
\end{table}

%% file: tables/full_results_appendix.tex
\section{Complete Leaderboard Results}
\label{app:complete-results}
The following tables report 482 result records across all benchmark settings. The remaining 13 records of architecture variants are omitted from the tables and included in the released results. Reconstruction rows are ordered by increasing output SNR, and first-arrival rows are ordered by increasing MAE. All scores are means over three random seeds with standard deviations shown in small type. PSNR differs from SNR by a per-setting constant and yields identical rankings, so it is omitted from the tables and reported in the released results. A dash indicates that the record does not report that field. Cell shading is computed independently within each setting and metric. Per-seed results and complete machine-readable metric dictionaries are available on the project website.
\begingroup
\setlength{\intextsep}{3pt plus 1pt minus 1pt}
\setlength{\abovecaptionskip}{2pt}
\setlength{\belowcaptionskip}{0pt}
\subsection{Random-noise attenuation}

\begin{table}[H]
\caption{SEGC3 Random Noise Gaussian SNR -5 dB.}
\label{tab:full-segc3-random-noise-gaussian-snrneg5}
\centering
\scriptsize
\setlength{\tabcolsep}{2.4pt}
\renewcommand{\arraystretch}{1.0}
\taskheat{seisRandomHeat}{}
\begin{tabular*}{0.92\linewidth}{@{\extracolsep{\fill}}lcccccccc@{}}
\toprule
Method & Params & SNR & \multicolumn{4}{c}{FB-FRE} & \multicolumn{2}{c}{EB-FE} \\
\cmidrule(lr){4-7}\cmidrule(lr){8-9}
 & & & Low & Mid & High & VHigh & Med. & Strong \\
\midrule
FBResNet & 0.45M & \heatbase{10.35{\scriptsize$\pm$0.06}} & \heatbase{9.09{\scriptsize$\pm$0.08}} & \heatbase{12.46{\scriptsize$\pm$0.07}} & \heatbase{9.55{\scriptsize$\pm$0.05}} & \heatmid{1.01{\scriptsize$\pm$0.04}} & \heatmid{0.17{\scriptsize$\pm$0.04}} & \heatbase{11.68{\scriptsize$\pm$0.08}} \\
cDDPM & 33.3M & \heatbase{10.41{\scriptsize$\pm$0.07}} & \heatbase{9.92{\scriptsize$\pm$0.15}} & \heatbase{12.78{\scriptsize$\pm$0.06}} & \heatmid{9.92{\scriptsize$\pm$0.06}} & \heatlow{0.89{\scriptsize$\pm$0.08}} & \heatbase{-0.72{\scriptsize$\pm$0.04}} & \heatgood{12.65{\scriptsize$\pm$0.10}} \\
ResUNet & 8.11M & \heatlow{10.50{\scriptsize$\pm$0.12}} & \heatlow{10.07{\scriptsize$\pm$0.30}} & \heatlow{12.95{\scriptsize$\pm$0.22}} & \heatbase{9.62{\scriptsize$\pm$0.08}} & \heatbase{0.66{\scriptsize$\pm$0.06}} & \heatbase{-0.05{\scriptsize$\pm$0.04}} & \heatbase{11.89{\scriptsize$\pm$0.18}} \\
UNet & 7.76M & \heatlow{10.72{\scriptsize$\pm$0.16}} & \heatlow{10.08{\scriptsize$\pm$0.33}} & \heatlow{13.19{\scriptsize$\pm$0.20}} & \heatlow{9.82{\scriptsize$\pm$0.14}} & \heatbase{0.81{\scriptsize$\pm$0.10}} & \heatlow{0.12{\scriptsize$\pm$0.04}} & \heatlow{12.12{\scriptsize$\pm$0.21}} \\
Attention UNet & 7.85M & \heatmid{10.76{\scriptsize$\pm$0.04}} & \heatmid{10.08{\scriptsize$\pm$0.07}} & \heatmid{13.22{\scriptsize$\pm$0.08}} & \heatlow{9.89{\scriptsize$\pm$0.03}} & \heatlow{0.82{\scriptsize$\pm$0.01}} & \heatmid{0.15{\scriptsize$\pm$0.04}} & \heatlow{12.16{\scriptsize$\pm$0.05}} \\
\bottomrule
\end{tabular*}
\end{table}

\begin{table}[H]
\caption{SEGC3 Random Noise Gaussian SNR 0 dB.}
\label{tab:full-segc3-random-noise-gaussian-snr0}
\centering
\scriptsize
\setlength{\tabcolsep}{2.4pt}
\renewcommand{\arraystretch}{1.0}
\taskheat{seisRandomHeat}{}
\begin{tabular*}{0.92\linewidth}{@{\extracolsep{\fill}}lcccccccc@{}}
\toprule
Method & Params & SNR & \multicolumn{4}{c}{FB-FRE} & \multicolumn{2}{c}{EB-FE} \\
\cmidrule(lr){4-7}\cmidrule(lr){8-9}
 & & & Low & Mid & High & VHigh & Med. & Strong \\
\midrule
cDDPM & 33.3M & \heatbase{12.38{\scriptsize$\pm$0.04}} & \heatbase{11.58{\scriptsize$\pm$0.09}} & \heatbase{15.27{\scriptsize$\pm$0.09}} & \heatlow{11.95{\scriptsize$\pm$0.05}} & \heatlow{2.44{\scriptsize$\pm$0.07}} & \heatbase{0.55{\scriptsize$\pm$0.03}} & \heatmid{15.57{\scriptsize$\pm$0.11}} \\
ResUNet & 8.11M & \heatbase{12.56{\scriptsize$\pm$0.11}} & \heatmid{12.05{\scriptsize$\pm$0.17}} & \heatlow{15.57{\scriptsize$\pm$0.08}} & \heatbase{11.73{\scriptsize$\pm$0.10}} & \heatbase{1.90{\scriptsize$\pm$0.14}} & \heatbase{1.04{\scriptsize$\pm$0.12}} & \heatbase{14.55{\scriptsize$\pm$0.11}} \\
UNet & 7.76M & \heatlow{12.73{\scriptsize$\pm$0.20}} & \heatbase{11.99{\scriptsize$\pm$0.32}} & \heatlow{15.69{\scriptsize$\pm$0.10}} & \heatbase{11.90{\scriptsize$\pm$0.19}} & \heatbase{2.09{\scriptsize$\pm$0.24}} & \heatlow{1.17{\scriptsize$\pm$0.21}} & \heatbase{14.73{\scriptsize$\pm$0.22}} \\
Attention UNet & 7.85M & \heatlow{12.80{\scriptsize$\pm$0.10}} & \heatlow{12.00{\scriptsize$\pm$0.15}} & \heatmid{15.73{\scriptsize$\pm$0.07}} & \heatlow{11.98{\scriptsize$\pm$0.08}} & \heatlow{2.18{\scriptsize$\pm$0.10}} & \heatlow{1.26{\scriptsize$\pm$0.09}} & \heatlow{14.82{\scriptsize$\pm$0.10}} \\
FBResNet & 0.45M & \heatmid{12.97{\scriptsize$\pm$0.00}} & \heatlow{11.99{\scriptsize$\pm$0.04}} & \heatbase{15.45{\scriptsize$\pm$0.01}} & \heatmid{12.12{\scriptsize$\pm$0.02}} & \heatmid{2.91{\scriptsize$\pm$0.01}} & \heatmid{1.47{\scriptsize$\pm$0.02}} & \heatlow{15.00{\scriptsize$\pm$0.01}} \\
\bottomrule
\end{tabular*}
\end{table}

\begin{table}[H]
\caption{SEGC3 Random Noise Gaussian SNR 5 dB.}
\label{tab:full-segc3-random-noise-gaussian-snr5}
\centering
\scriptsize
\setlength{\tabcolsep}{2.4pt}
\renewcommand{\arraystretch}{1.0}
\taskheat{seisRandomHeat}{}
\begin{tabular*}{0.92\linewidth}{@{\extracolsep{\fill}}lcccccccc@{}}
\toprule
Method & Params & SNR & \multicolumn{4}{c}{FB-FRE} & \multicolumn{2}{c}{EB-FE} \\
\cmidrule(lr){4-7}\cmidrule(lr){8-9}
 & & & Low & Mid & High & VHigh & Med. & Strong \\
\midrule
cDDPM & 33.3M & \heatbase{14.31{\scriptsize$\pm$0.23}} & \heatbase{13.01{\scriptsize$\pm$0.29}} & \heatbase{17.87{\scriptsize$\pm$0.14}} & \heatbase{14.00{\scriptsize$\pm$0.18}} & \heatlow{4.26{\scriptsize$\pm$0.17}} & \heatbase{2.35{\scriptsize$\pm$0.17}} & \heatmid{18.56{\scriptsize$\pm$0.14}} \\
UNet & 7.76M & \heatbase{14.84{\scriptsize$\pm$0.44}} & \heatlow{13.78{\scriptsize$\pm$0.59}} & \heatbase{18.19{\scriptsize$\pm$0.35}} & \heatbase{13.96{\scriptsize$\pm$0.42}} & \heatbase{3.96{\scriptsize$\pm$0.45}} & \heatbase{3.04{\scriptsize$\pm$0.43}} & \heatbase{17.30{\scriptsize$\pm$0.48}} \\
Attention UNet & 7.85M & \heatlow{14.89{\scriptsize$\pm$0.23}} & \heatbase{13.61{\scriptsize$\pm$0.29}} & \heatlow{18.21{\scriptsize$\pm$0.18}} & \heatlow{14.08{\scriptsize$\pm$0.19}} & \heatlow{4.08{\scriptsize$\pm$0.27}} & \heatlow{3.09{\scriptsize$\pm$0.20}} & \heatlow{17.38{\scriptsize$\pm$0.24}} \\
ResUNet & 8.11M & \heatlow{14.92{\scriptsize$\pm$0.05}} & \heatlow{14.15{\scriptsize$\pm$0.11}} & \heatlow{18.47{\scriptsize$\pm$0.03}} & \heatlow{14.05{\scriptsize$\pm$0.03}} & \heatbase{3.98{\scriptsize$\pm$0.04}} & \heatlow{3.18{\scriptsize$\pm$0.07}} & \heatbase{17.37{\scriptsize$\pm$0.04}} \\
FBResNet & 0.45M & \heatmid{15.84{\scriptsize$\pm$0.02}} & \heatmid{14.76{\scriptsize$\pm$0.04}} & \heatmid{18.65{\scriptsize$\pm$0.04}} & \heatmid{14.87{\scriptsize$\pm$0.00}} & \heatmid{5.43{\scriptsize$\pm$0.01}} & \heatmid{3.72{\scriptsize$\pm$0.00}} & \heatmid{18.56{\scriptsize$\pm$0.02}} \\
\bottomrule
\end{tabular*}
\end{table}

\begin{table}[H]
\caption{SEGC3 Random Noise Poisson SNR -5 dB.}
\label{tab:full-segc3-random-noise-poisson-snrneg5}
\centering
\scriptsize
\setlength{\tabcolsep}{2.4pt}
\renewcommand{\arraystretch}{1.0}
\taskheat{seisRandomHeat}{}
\begin{tabular*}{0.92\linewidth}{@{\extracolsep{\fill}}lcccccccc@{}}
\toprule
Method & Params & SNR & \multicolumn{4}{c}{FB-FRE} & \multicolumn{2}{c}{EB-FE} \\
\cmidrule(lr){4-7}\cmidrule(lr){8-9}
 & & & Low & Mid & High & VHigh & Med. & Strong \\
\midrule
FBResNet & 0.45M & \heatbase{10.37{\scriptsize$\pm$0.04}} & \heatbase{9.17{\scriptsize$\pm$0.04}} & \heatbase{12.49{\scriptsize$\pm$0.06}} & \heatbase{9.59{\scriptsize$\pm$0.04}} & \heatmid{0.99{\scriptsize$\pm$0.01}} & \heatgood{0.15{\scriptsize$\pm$0.02}} & \heatbase{11.72{\scriptsize$\pm$0.07}} \\
cDDPM & 33.3M & \heatbase{10.43{\scriptsize$\pm$0.06}} & \heatbase{9.92{\scriptsize$\pm$0.11}} & \heatbase{12.75{\scriptsize$\pm$0.04}} & \heatmid{10.00{\scriptsize$\pm$0.05}} & \heatlow{0.93{\scriptsize$\pm$0.06}} & \heatbase{-0.69{\scriptsize$\pm$0.05}} & \heatgood{12.64{\scriptsize$\pm$0.06}} \\
ResUNet & 8.11M & \heatlow{10.53{\scriptsize$\pm$0.17}} & \heatlow{10.10{\scriptsize$\pm$0.36}} & \heatlow{12.96{\scriptsize$\pm$0.27}} & \heatbase{9.66{\scriptsize$\pm$0.12}} & \heatbase{0.65{\scriptsize$\pm$0.08}} & \heatbase{-0.08{\scriptsize$\pm$0.02}} & \heatbase{11.94{\scriptsize$\pm$0.23}} \\
UNet & 7.76M & \heatlow{10.72{\scriptsize$\pm$0.17}} & \heatlow{10.06{\scriptsize$\pm$0.39}} & \heatlow{13.19{\scriptsize$\pm$0.17}} & \heatlow{9.83{\scriptsize$\pm$0.16}} & \heatbase{0.78{\scriptsize$\pm$0.15}} & \heatlow{0.06{\scriptsize$\pm$0.10}} & \heatlow{12.14{\scriptsize$\pm$0.22}} \\
Attention UNet & 7.85M & \heatmid{10.84{\scriptsize$\pm$0.07}} & \heatmid{10.19{\scriptsize$\pm$0.14}} & \heatgood{13.30{\scriptsize$\pm$0.10}} & \heatlow{9.98{\scriptsize$\pm$0.08}} & \heatlow{0.85{\scriptsize$\pm$0.03}} & \heatgood{0.15{\scriptsize$\pm$0.05}} & \heatlow{12.28{\scriptsize$\pm$0.11}} \\
\bottomrule
\end{tabular*}
\end{table}

\begin{table}[H]
\caption{SEGC3 Random Noise Poisson SNR 0 dB.}
\label{tab:full-segc3-random-noise-poisson-snr0}
\centering
\scriptsize
\setlength{\tabcolsep}{2.4pt}
\renewcommand{\arraystretch}{1.0}
\taskheat{seisRandomHeat}{}
\begin{tabular*}{0.92\linewidth}{@{\extracolsep{\fill}}lcccccccc@{}}
\toprule
Method & Params & SNR & \multicolumn{4}{c}{FB-FRE} & \multicolumn{2}{c}{EB-FE} \\
\cmidrule(lr){4-7}\cmidrule(lr){8-9}
 & & & Low & Mid & High & VHigh & Med. & Strong \\
\midrule
cDDPM & 33.3M & \heatbase{12.37{\scriptsize$\pm$0.02}} & \heatbase{11.49{\scriptsize$\pm$0.02}} & \heatbase{15.24{\scriptsize$\pm$0.00}} & \heatlow{11.97{\scriptsize$\pm$0.03}} & \heatlow{2.46{\scriptsize$\pm$0.02}} & \heatbase{0.54{\scriptsize$\pm$0.02}} & \heatgood{15.53{\scriptsize$\pm$0.01}} \\
ResUNet & 8.11M & \heatbase{12.57{\scriptsize$\pm$0.09}} & \heatmid{12.04{\scriptsize$\pm$0.16}} & \heatlow{15.57{\scriptsize$\pm$0.08}} & \heatbase{11.75{\scriptsize$\pm$0.09}} & \heatbase{1.91{\scriptsize$\pm$0.13}} & \heatbase{1.01{\scriptsize$\pm$0.09}} & \heatbase{14.58{\scriptsize$\pm$0.10}} \\
UNet & 7.76M & \heatlow{12.73{\scriptsize$\pm$0.19}} & \heatlow{11.99{\scriptsize$\pm$0.31}} & \heatlow{15.66{\scriptsize$\pm$0.09}} & \heatbase{11.91{\scriptsize$\pm$0.19}} & \heatbase{2.08{\scriptsize$\pm$0.24}} & \heatlow{1.14{\scriptsize$\pm$0.23}} & \heatbase{14.74{\scriptsize$\pm$0.21}} \\
Attention UNet & 7.85M & \heatlow{12.80{\scriptsize$\pm$0.12}} & \heatlow{12.01{\scriptsize$\pm$0.16}} & \heatmid{15.74{\scriptsize$\pm$0.08}} & \heatlow{11.98{\scriptsize$\pm$0.11}} & \heatlow{2.17{\scriptsize$\pm$0.13}} & \heatlow{1.24{\scriptsize$\pm$0.09}} & \heatlow{14.83{\scriptsize$\pm$0.14}} \\
FBResNet & 0.45M & \heatmid{12.98{\scriptsize$\pm$0.02}} & \heatbase{11.97{\scriptsize$\pm$0.06}} & \heatbase{15.45{\scriptsize$\pm$0.03}} & \heatmid{12.14{\scriptsize$\pm$0.01}} & \heatgood{2.91{\scriptsize$\pm$0.01}} & \heatgood{1.46{\scriptsize$\pm$0.01}} & \heatlow{15.02{\scriptsize$\pm$0.03}} \\
\bottomrule
\end{tabular*}
\end{table}

\begin{table}[H]
\caption{SEGC3 Random Noise Poisson SNR 5 dB.}
\label{tab:full-segc3-random-noise-poisson-snr5}
\centering
\scriptsize
\setlength{\tabcolsep}{2.4pt}
\renewcommand{\arraystretch}{1.0}
\taskheat{seisRandomHeat}{}
\begin{tabular*}{0.92\linewidth}{@{\extracolsep{\fill}}lcccccccc@{}}
\toprule
Method & Params & SNR & \multicolumn{4}{c}{FB-FRE} & \multicolumn{2}{c}{EB-FE} \\
\cmidrule(lr){4-7}\cmidrule(lr){8-9}
 & & & Low & Mid & High & VHigh & Med. & Strong \\
\midrule
cDDPM & 33.3M & \heatbase{14.36{\scriptsize$\pm$0.09}} & \heatbase{13.03{\scriptsize$\pm$0.14}} & \heatbase{17.89{\scriptsize$\pm$0.07}} & \heatbase{14.03{\scriptsize$\pm$0.07}} & \heatlow{4.30{\scriptsize$\pm$0.06}} & \heatbase{2.38{\scriptsize$\pm$0.03}} & \heatgood{18.60{\scriptsize$\pm$0.06}} \\
UNet & 7.76M & \heatbase{14.89{\scriptsize$\pm$0.37}} & \heatlow{13.81{\scriptsize$\pm$0.49}} & \heatlow{18.26{\scriptsize$\pm$0.26}} & \heatbase{14.00{\scriptsize$\pm$0.37}} & \heatbase{4.01{\scriptsize$\pm$0.39}} & \heatlow{3.11{\scriptsize$\pm$0.34}} & \heatbase{17.35{\scriptsize$\pm$0.41}} \\
Attention UNet & 7.85M & \heatlow{14.92{\scriptsize$\pm$0.22}} & \heatbase{13.60{\scriptsize$\pm$0.29}} & \heatbase{18.24{\scriptsize$\pm$0.17}} & \heatlow{14.11{\scriptsize$\pm$0.18}} & \heatlow{4.13{\scriptsize$\pm$0.25}} & \heatbase{3.10{\scriptsize$\pm$0.19}} & \heatlow{17.41{\scriptsize$\pm$0.23}} \\
ResUNet & 8.11M & \heatlow{14.92{\scriptsize$\pm$0.05}} & \heatlow{14.14{\scriptsize$\pm$0.10}} & \heatlow{18.47{\scriptsize$\pm$0.03}} & \heatlow{14.05{\scriptsize$\pm$0.02}} & \heatbase{3.96{\scriptsize$\pm$0.05}} & \heatlow{3.17{\scriptsize$\pm$0.07}} & \heatbase{17.38{\scriptsize$\pm$0.04}} \\
QUNet & 21.88M & \heatmid{15.77{\scriptsize$\pm$0.04}} & \heatmid{14.97{\scriptsize$\pm$0.05}} & \heatgood{18.88{\scriptsize$\pm$0.02}} & \heatmid{14.72{\scriptsize$\pm$0.05}} & \heatmid{5.01{\scriptsize$\pm$0.03}} & \heatmid{3.74{\scriptsize$\pm$0.02}} & \heatlow{18.45{\scriptsize$\pm$0.01}} \\
\bottomrule
\end{tabular*}
\end{table}

\begin{table}[H]
\caption{Mobile AVO Random Noise Gaussian SNR -5 dB.}
\label{tab:full-mobile-avo-random-noise-gaussian-snrneg5}
\centering
\scriptsize
\setlength{\tabcolsep}{2.4pt}
\renewcommand{\arraystretch}{1.0}
\taskheat{seisRandomHeat}{}
\begin{tabular*}{0.92\linewidth}{@{\extracolsep{\fill}}lcccccccc@{}}
\toprule
Method & Params & SNR & \multicolumn{4}{c}{FB-FRE} & \multicolumn{2}{c}{EB-FE} \\
\cmidrule(lr){4-7}\cmidrule(lr){8-9}
 & & & Low & Mid & High & VHigh & Med. & Strong \\
\midrule
cDDPM & 33.3M & \heatbase{7.16{\scriptsize$\pm$0.16}} & \heatmid{9.75{\scriptsize$\pm$0.12}} & \heatbase{9.55{\scriptsize$\pm$0.13}} & \heatbase{6.83{\scriptsize$\pm$0.18}} & \heatbase{-11.45{\scriptsize$\pm$0.25}} & \heatbase{-15.45{\scriptsize$\pm$0.16}} & \heatgood{10.26{\scriptsize$\pm$0.10}} \\
ResUNet & 8.11M & \heatbase{8.01{\scriptsize$\pm$0.00}} & \heatlow{9.51{\scriptsize$\pm$0.01}} & \heatbase{9.78{\scriptsize$\pm$0.01}} & \heatbase{7.10{\scriptsize$\pm$0.01}} & \heatbase{-9.12{\scriptsize$\pm$0.01}} & \heatbase{-10.31{\scriptsize$\pm$0.00}} & \heatbase{8.76{\scriptsize$\pm$0.00}} \\
UNet & 7.76M & \heatlow{8.94{\scriptsize$\pm$0.00}} & \heatbase{9.12{\scriptsize$\pm$0.01}} & \heatlow{10.22{\scriptsize$\pm$0.00}} & \heatmid{8.62{\scriptsize$\pm$0.01}} & \heatmid{-5.21{\scriptsize$\pm$0.01}} & \heatlow{-7.65{\scriptsize$\pm$0.00}} & \heatlow{9.40{\scriptsize$\pm$0.00}} \\
DnCNN & 0.56M & \heatlow{8.97{\scriptsize$\pm$0.00}} & \heatbase{9.47{\scriptsize$\pm$0.01}} & \heatmid{10.55{\scriptsize$\pm$0.00}} & \heatmid{8.73{\scriptsize$\pm$0.00}} & \heatlow{-7.44{\scriptsize$\pm$0.01}} & \heatlow{-9.81{\scriptsize$\pm$0.00}} & \heatmid{9.79{\scriptsize$\pm$0.00}} \\
UNet++ & 9.05M & \heatmid{9.23{\scriptsize$\pm$0.11}} & \heatgood{9.77{\scriptsize$\pm$0.15}} & \heatlow{10.12{\scriptsize$\pm$0.13}} & \heatlow{7.89{\scriptsize$\pm$0.24}} & \heatbest{-2.73{\scriptsize$\pm$0.22}} & \heatgood{-2.34{\scriptsize$\pm$0.83}} & \heatbase{9.32{\scriptsize$\pm$0.14}} \\
\bottomrule
\end{tabular*}
\end{table}

\begin{table}[H]
\caption{Mobile AVO Random Noise Gaussian SNR 0 dB.}
\label{tab:full-mobile-avo-random-noise-gaussian-snr0}
\centering
\scriptsize
\setlength{\tabcolsep}{2.4pt}
\renewcommand{\arraystretch}{1.0}
\taskheat{seisRandomHeat}{}
\begin{tabular*}{0.92\linewidth}{@{\extracolsep{\fill}}lcccccccc@{}}
\toprule
Method & Params & SNR & \multicolumn{4}{c}{FB-FRE} & \multicolumn{2}{c}{EB-FE} \\
\cmidrule(lr){4-7}\cmidrule(lr){8-9}
 & & & Low & Mid & High & VHigh & Med. & Strong \\
\midrule
cDDPM & 33.3M & \heatbase{7.35{\scriptsize$\pm$0.13}} & \heatbase{10.54{\scriptsize$\pm$0.11}} & \heatbase{10.18{\scriptsize$\pm$0.10}} & \heatbase{6.90{\scriptsize$\pm$0.13}} & \heatbase{-12.21{\scriptsize$\pm$0.15}} & \heatbase{-15.96{\scriptsize$\pm$0.11}} & \heatlow{11.52{\scriptsize$\pm$0.12}} \\
ResUNet & 8.11M & \heatbase{9.71{\scriptsize$\pm$0.00}} & \heatlow{11.77{\scriptsize$\pm$0.00}} & \heatbase{11.75{\scriptsize$\pm$0.01}} & \heatbase{8.88{\scriptsize$\pm$0.01}} & \heatbase{-8.47{\scriptsize$\pm$0.01}} & \heatbase{-8.65{\scriptsize$\pm$0.00}} & \heatbase{10.44{\scriptsize$\pm$0.00}} \\
UNet & 7.76M & \heatlow{10.51{\scriptsize$\pm$0.00}} & \heatbase{10.52{\scriptsize$\pm$0.01}} & \heatlow{12.05{\scriptsize$\pm$0.01}} & \heatlow{10.61{\scriptsize$\pm$0.01}} & \heatmid{-4.23{\scriptsize$\pm$0.01}} & \heatlow{-5.39{\scriptsize$\pm$0.01}} & \heatbase{10.89{\scriptsize$\pm$0.00}} \\
Attention UNet & 7.85M & \heatlow{11.14{\scriptsize$\pm$0.00}} & \heatlow{12.44{\scriptsize$\pm$0.00}} & \heatlow{12.71{\scriptsize$\pm$0.00}} & \heatlow{10.10{\scriptsize$\pm$0.01}} & \heatlow{-5.89{\scriptsize$\pm$0.01}} & \heatmid{-2.85{\scriptsize$\pm$0.01}} & \heatlow{11.32{\scriptsize$\pm$0.00}} \\
DnCNN & 0.56M & \heatmid{11.95{\scriptsize$\pm$0.00}} & \heatmid{12.76{\scriptsize$\pm$0.01}} & \heatgood{13.48{\scriptsize$\pm$0.00}} & \heatgood{11.42{\scriptsize$\pm$0.01}} & \heatlow{-4.65{\scriptsize$\pm$0.01}} & \heatlow{-6.59{\scriptsize$\pm$0.00}} & \heatgood{12.68{\scriptsize$\pm$0.00}} \\
\bottomrule
\end{tabular*}
\end{table}

\begin{table}[H]
\caption{Mobile AVO Random Noise Gaussian SNR +5 dB.}
\label{tab:full-mobile-avo-random-noise-gaussian-snr5}
\centering
\scriptsize
\setlength{\tabcolsep}{2.4pt}
\renewcommand{\arraystretch}{1.0}
\taskheat{seisRandomHeat}{}
\begin{tabular*}{0.92\linewidth}{@{\extracolsep{\fill}}lcccccccc@{}}
\toprule
Method & Params & SNR & \multicolumn{4}{c}{FB-FRE} & \multicolumn{2}{c}{EB-FE} \\
\cmidrule(lr){4-7}\cmidrule(lr){8-9}
 & & & Low & Mid & High & VHigh & Med. & Strong \\
\midrule
cDDPM & 33.3M & \heatbase{7.68{\scriptsize$\pm$0.18}} & \heatbase{11.19{\scriptsize$\pm$0.17}} & \heatbase{10.92{\scriptsize$\pm$0.16}} & \heatbase{7.29{\scriptsize$\pm$0.18}} & \heatbase{-12.22{\scriptsize$\pm$0.19}} & \heatbase{-15.99{\scriptsize$\pm$0.14}} & \heatlow{12.58{\scriptsize$\pm$0.17}} \\
ResUNet & 8.11M & \heatbase{10.83{\scriptsize$\pm$0.00}} & \heatlow{13.38{\scriptsize$\pm$0.00}} & \heatbase{13.23{\scriptsize$\pm$0.00}} & \heatbase{10.21{\scriptsize$\pm$0.01}} & \heatbase{-8.46{\scriptsize$\pm$0.01}} & \heatbase{-6.49{\scriptsize$\pm$0.00}} & \heatbase{11.38{\scriptsize$\pm$0.00}} \\
Attention UNet & 7.85M & \heatlow{12.35{\scriptsize$\pm$0.01}} & \heatlow{14.06{\scriptsize$\pm$0.00}} & \heatlow{14.24{\scriptsize$\pm$0.00}} & \heatlow{11.32{\scriptsize$\pm$0.01}} & \heatlow{-5.41{\scriptsize$\pm$0.00}} & \heatgood{-2.29{\scriptsize$\pm$0.00}} & \heatbase{12.56{\scriptsize$\pm$0.01}} \\
UNet & 7.76M & \heatlow{12.72{\scriptsize$\pm$0.00}} & \heatbase{13.35{\scriptsize$\pm$0.01}} & \heatlow{14.36{\scriptsize$\pm$0.00}} & \heatlow{12.46{\scriptsize$\pm$0.00}} & \heatlow{-3.54{\scriptsize$\pm$0.01}} & \heatlow{-3.41{\scriptsize$\pm$0.00}} & \heatlow{13.10{\scriptsize$\pm$0.00}} \\
QUNet & 21.88M & \heatmid{14.15{\scriptsize$\pm$0.76}} & \heatmid{15.65{\scriptsize$\pm$0.62}} & \heatmid{15.68{\scriptsize$\pm$0.53}} & \heatmid{12.99{\scriptsize$\pm$0.64}} & \heatmid{-2.98{\scriptsize$\pm$1.39}} & \heatmid{-2.42{\scriptsize$\pm$2.30}} & \heatmid{14.55{\scriptsize$\pm$0.49}} \\
\bottomrule
\end{tabular*}
\end{table}

\begin{table}[H]
\caption{Mobile AVO Random Noise Poisson SNR -5 dB.}
\label{tab:full-mobile-avo-random-noise-poisson-snrneg5}
\centering
\scriptsize
\setlength{\tabcolsep}{2.4pt}
\renewcommand{\arraystretch}{1.0}
\taskheat{seisRandomHeat}{}
\begin{tabular*}{0.92\linewidth}{@{\extracolsep{\fill}}lcccccccc@{}}
\toprule
Method & Params & SNR & \multicolumn{4}{c}{FB-FRE} & \multicolumn{2}{c}{EB-FE} \\
\cmidrule(lr){4-7}\cmidrule(lr){8-9}
 & & & Low & Mid & High & VHigh & Med. & Strong \\
\midrule
cDDPM & 33.3M & \heatbase{6.87{\scriptsize$\pm$0.24}} & \heatbase{9.37{\scriptsize$\pm$0.26}} & \heatbase{9.20{\scriptsize$\pm$0.23}} & \heatbase{6.46{\scriptsize$\pm$0.29}} & \heatbase{-11.92{\scriptsize$\pm$0.37}} & \heatbase{-15.76{\scriptsize$\pm$0.23}} & \heatgood{10.04{\scriptsize$\pm$0.15}} \\
ResUNet & 8.11M & \heatbase{8.06{\scriptsize$\pm$0.00}} & \heatlow{9.54{\scriptsize$\pm$0.01}} & \heatbase{9.84{\scriptsize$\pm$0.01}} & \heatbase{7.17{\scriptsize$\pm$0.00}} & \heatbase{-9.08{\scriptsize$\pm$0.01}} & \heatbase{-10.31{\scriptsize$\pm$0.00}} & \heatbase{8.82{\scriptsize$\pm$0.00}} \\
UNet & 7.76M & \heatlow{8.99{\scriptsize$\pm$0.00}} & \heatbase{9.20{\scriptsize$\pm$0.01}} & \heatlow{10.25{\scriptsize$\pm$0.01}} & \heatmid{8.66{\scriptsize$\pm$0.00}} & \heatmid{-5.20{\scriptsize$\pm$0.00}} & \heatlow{-7.64{\scriptsize$\pm$0.01}} & \heatlow{9.45{\scriptsize$\pm$0.00}} \\
DnCNN & 0.56M & \heatlow{9.15{\scriptsize$\pm$0.00}} & \heatgood{9.69{\scriptsize$\pm$0.01}} & \heatmid{10.62{\scriptsize$\pm$0.00}} & \heatmid{8.83{\scriptsize$\pm$0.00}} & \heatlow{-7.03{\scriptsize$\pm$0.00}} & \heatlow{-9.42{\scriptsize$\pm$0.01}} & \heatmid{9.94{\scriptsize$\pm$0.01}} \\
Attention UNet & 7.85M & \heatmid{9.27{\scriptsize$\pm$0.01}} & \heatgood{9.99{\scriptsize$\pm$0.02}} & \heatmid{10.48{\scriptsize$\pm$0.01}} & \heatlow{8.41{\scriptsize$\pm$0.01}} & \heatlow{-5.87{\scriptsize$\pm$0.01}} & \heatmid{-3.27{\scriptsize$\pm$0.02}} & \heatlow{9.42{\scriptsize$\pm$0.01}} \\
\bottomrule
\end{tabular*}
\end{table}

\begin{table}[H]
\caption{Mobile AVO Random Noise Poisson SNR 0 dB.}
\label{tab:full-mobile-avo-random-noise-poisson-snr0}
\centering
\scriptsize
\setlength{\tabcolsep}{2.4pt}
\renewcommand{\arraystretch}{1.0}
\taskheat{seisRandomHeat}{}
\begin{tabular*}{0.92\linewidth}{@{\extracolsep{\fill}}lcccccccc@{}}
\toprule
Method & Params & SNR & \multicolumn{4}{c}{FB-FRE} & \multicolumn{2}{c}{EB-FE} \\
\cmidrule(lr){4-7}\cmidrule(lr){8-9}
 & & & Low & Mid & High & VHigh & Med. & Strong \\
\midrule
cDDPM & 33.3M & \heatbase{7.05{\scriptsize$\pm$0.62}} & \heatbase{10.22{\scriptsize$\pm$0.61}} & \heatbase{9.93{\scriptsize$\pm$0.56}} & \heatbase{6.63{\scriptsize$\pm$0.60}} & \heatbase{-12.51{\scriptsize$\pm$0.66}} & \heatbase{-16.24{\scriptsize$\pm$0.55}} & \heatlow{11.28{\scriptsize$\pm$0.51}} \\
ResUNet & 8.11M & \heatbase{9.65{\scriptsize$\pm$0.00}} & \heatlow{11.67{\scriptsize$\pm$0.01}} & \heatbase{11.71{\scriptsize$\pm$0.01}} & \heatbase{8.86{\scriptsize$\pm$0.00}} & \heatbase{-8.63{\scriptsize$\pm$0.00}} & \heatbase{-8.75{\scriptsize$\pm$0.00}} & \heatbase{10.39{\scriptsize$\pm$0.00}} \\
UNet & 7.76M & \heatlow{10.27{\scriptsize$\pm$0.00}} & \heatbase{10.28{\scriptsize$\pm$0.01}} & \heatlow{11.99{\scriptsize$\pm$0.00}} & \heatlow{10.42{\scriptsize$\pm$0.01}} & \heatmid{-4.44{\scriptsize$\pm$0.01}} & \heatlow{-5.52{\scriptsize$\pm$0.00}} & \heatbase{10.63{\scriptsize$\pm$0.00}} \\
Attention UNet & 7.85M & \heatlow{11.16{\scriptsize$\pm$0.01}} & \heatlow{12.37{\scriptsize$\pm$0.01}} & \heatlow{12.69{\scriptsize$\pm$0.01}} & \heatlow{10.17{\scriptsize$\pm$0.00}} & \heatlow{-5.61{\scriptsize$\pm$0.01}} & \heatmid{-2.86{\scriptsize$\pm$0.00}} & \heatlow{11.34{\scriptsize$\pm$0.01}} \\
DnCNN & 0.56M & \heatmid{11.98{\scriptsize$\pm$0.00}} & \heatgood{12.74{\scriptsize$\pm$0.01}} & \heatgood{13.50{\scriptsize$\pm$0.00}} & \heatgood{11.52{\scriptsize$\pm$0.00}} & \heatlow{-4.66{\scriptsize$\pm$0.00}} & \heatlow{-6.35{\scriptsize$\pm$0.00}} & \heatgood{12.67{\scriptsize$\pm$0.00}} \\
\bottomrule
\end{tabular*}
\end{table}

\begin{table}[H]
\caption{Mobile AVO Random Noise Poisson SNR +5 dB.}
\label{tab:full-mobile-avo-random-noise-poisson-snr5}
\centering
\scriptsize
\setlength{\tabcolsep}{2.4pt}
\renewcommand{\arraystretch}{1.0}
\taskheat{seisRandomHeat}{}
\begin{tabular*}{0.92\linewidth}{@{\extracolsep{\fill}}lcccccccc@{}}
\toprule
Method & Params & SNR & \multicolumn{4}{c}{FB-FRE} & \multicolumn{2}{c}{EB-FE} \\
\cmidrule(lr){4-7}\cmidrule(lr){8-9}
 & & & Low & Mid & High & VHigh & Med. & Strong \\
\midrule
cDDPM & 33.3M & \heatbase{7.78{\scriptsize$\pm$0.18}} & \heatbase{11.28{\scriptsize$\pm$0.18}} & \heatbase{11.01{\scriptsize$\pm$0.17}} & \heatbase{7.39{\scriptsize$\pm$0.18}} & \heatbase{-12.12{\scriptsize$\pm$0.19}} & \heatbase{-15.89{\scriptsize$\pm$0.13}} & \heatbase{12.67{\scriptsize$\pm$0.20}} \\
ResUNet & 8.11M & \heatbase{10.82{\scriptsize$\pm$0.00}} & \heatlow{13.44{\scriptsize$\pm$0.00}} & \heatbase{13.25{\scriptsize$\pm$0.01}} & \heatbase{10.14{\scriptsize$\pm$0.01}} & \heatbase{-8.52{\scriptsize$\pm$0.00}} & \heatbase{-6.51{\scriptsize$\pm$0.00}} & \heatbase{11.37{\scriptsize$\pm$0.00}} \\
UNet & 7.76M & \heatlow{12.67{\scriptsize$\pm$0.00}} & \heatbase{13.31{\scriptsize$\pm$0.01}} & \heatlow{14.35{\scriptsize$\pm$0.00}} & \heatlow{12.44{\scriptsize$\pm$0.01}} & \heatlow{-3.62{\scriptsize$\pm$0.02}} & \heatlow{-3.41{\scriptsize$\pm$0.00}} & \heatlow{13.05{\scriptsize$\pm$0.00}} \\
Attention UNet & 7.85M & \heatlow{12.68{\scriptsize$\pm$0.00}} & \heatlow{14.28{\scriptsize$\pm$0.00}} & \heatlow{14.68{\scriptsize$\pm$0.01}} & \heatlow{11.56{\scriptsize$\pm$0.00}} & \heatlow{-4.93{\scriptsize$\pm$0.01}} & \heatgood{-2.07{\scriptsize$\pm$0.00}} & \heatlow{12.90{\scriptsize$\pm$0.00}} \\
QUNet & 21.88M & \heatmid{14.10{\scriptsize$\pm$0.76}} & \heatmid{15.67{\scriptsize$\pm$0.72}} & \heatmid{15.66{\scriptsize$\pm$0.52}} & \heatmid{12.92{\scriptsize$\pm$0.60}} & \heatmid{-3.23{\scriptsize$\pm$1.24}} & \heatmid{-2.42{\scriptsize$\pm$2.18}} & \heatmid{14.49{\scriptsize$\pm$0.51}} \\
\bottomrule
\end{tabular*}
\end{table}

\subsection{Seismic interpolation}

\begin{table}[H]
\caption{Mobile AVO Continuous Missing 20 Traces.}
\label{tab:full-mobile-avo-interp-continuous20tr}
\centering
\scriptsize
\setlength{\tabcolsep}{2.4pt}
\renewcommand{\arraystretch}{1.0}
\taskheat{seisInterpolationHeat}{}
\begin{tabular*}{0.92\linewidth}{@{\extracolsep{\fill}}lcccccccc@{}}
\toprule
Method & Params & SNR & \multicolumn{4}{c}{FB-FRE} & \multicolumn{2}{c}{EB-FE} \\
\cmidrule(lr){4-7}\cmidrule(lr){8-9}
 & & & Low & Mid & High & VHigh & Med. & Strong \\
\midrule
ANet & 7.05M & \heatbase{9.76{\scriptsize$\pm$0.70}} & \heatbase{11.36{\scriptsize$\pm$1.03}} & \heatbase{9.41{\scriptsize$\pm$0.45}} & \heatbase{8.75{\scriptsize$\pm$0.64}} & \heatbase{1.52{\scriptsize$\pm$1.17}} & \heatmid{7.04{\scriptsize$\pm$0.47}} & \heatbase{9.88{\scriptsize$\pm$0.71}} \\
CFunet & 7.35M & \heatbase{10.92{\scriptsize$\pm$0.06}} & \heatlow{12.63{\scriptsize$\pm$0.06}} & \heatbase{10.12{\scriptsize$\pm$0.04}} & \heatbase{10.55{\scriptsize$\pm$0.32}} & \heatlow{5.74{\scriptsize$\pm$0.45}} & \heatbest{8.19{\scriptsize$\pm$0.08}} & \heatbase{11.04{\scriptsize$\pm$0.06}} \\
CA-Unet & 7.79M & \heatlow{11.63{\scriptsize$\pm$0.33}} & \heatmid{13.55{\scriptsize$\pm$0.72}} & \heatlow{10.72{\scriptsize$\pm$0.22}} & \heatlow{11.54{\scriptsize$\pm$0.12}} & \heatgood{7.64{\scriptsize$\pm$0.10}} & \heatbest{8.72{\scriptsize$\pm$0.25}} & \heatlow{11.76{\scriptsize$\pm$0.36}} \\
WRDL & 35.83M & \heatlow{11.72{\scriptsize$\pm$0.22}} & \heatmid{14.33{\scriptsize$\pm$0.39}} & \heatlow{10.34{\scriptsize$\pm$0.16}} & \heatlow{11.25{\scriptsize$\pm$0.21}} & \heatmid{6.97{\scriptsize$\pm$0.22}} & \heatgood{7.87{\scriptsize$\pm$0.21}} & \heatlow{11.91{\scriptsize$\pm$0.23}} \\
DnCNN & 0.14M & \heatlow{14.05{\scriptsize$\pm$0.14}} & \heatbase{12.20{\scriptsize$\pm$0.17}} & \heatmid{15.26{\scriptsize$\pm$0.08}} & \heatgood{15.34{\scriptsize$\pm$0.05}} & \heatbest{10.28{\scriptsize$\pm$0.08}} & \heatbase{6.28{\scriptsize$\pm$0.26}} & \heatlow{14.08{\scriptsize$\pm$0.13}} \\
\bottomrule
\end{tabular*}
\end{table}

\begin{table}[H]
\caption{Mobile AVO Continuous Missing 30 Traces.}
\label{tab:full-mobile-avo-interp-continuous30tr}
\centering
\scriptsize
\setlength{\tabcolsep}{2.4pt}
\renewcommand{\arraystretch}{1.0}
\taskheat{seisInterpolationHeat}{}
\begin{tabular*}{0.92\linewidth}{@{\extracolsep{\fill}}lcccccccc@{}}
\toprule
Method & Params & SNR & \multicolumn{4}{c}{FB-FRE} & \multicolumn{2}{c}{EB-FE} \\
\cmidrule(lr){4-7}\cmidrule(lr){8-9}
 & & & Low & Mid & High & VHigh & Med. & Strong \\
\midrule
ANet & 7.05M & \heatbase{7.49{\scriptsize$\pm$0.90}} & \heatbase{8.32{\scriptsize$\pm$1.04}} & \heatbase{7.46{\scriptsize$\pm$0.32}} & \heatbase{7.61{\scriptsize$\pm$0.87}} & \heatbase{1.15{\scriptsize$\pm$1.33}} & \heatmid{4.99{\scriptsize$\pm$0.73}} & \heatbase{7.69{\scriptsize$\pm$0.74}} \\
CFunet & 7.35M & \heatbase{8.18{\scriptsize$\pm$0.19}} & \heatbase{8.76{\scriptsize$\pm$0.14}} & \heatbase{7.92{\scriptsize$\pm$0.21}} & \heatbase{9.12{\scriptsize$\pm$0.32}} & \heatlow{4.93{\scriptsize$\pm$0.28}} & \heatbest{6.23{\scriptsize$\pm$0.06}} & \heatbase{8.25{\scriptsize$\pm$0.20}} \\
CA-Unet & 7.79M & \heatlow{9.40{\scriptsize$\pm$0.41}} & \heatmid{10.90{\scriptsize$\pm$0.90}} & \heatlow{8.47{\scriptsize$\pm$0.18}} & \heatlow{9.75{\scriptsize$\pm$0.31}} & \heatmid{6.32{\scriptsize$\pm$0.24}} & \heatbest{6.29{\scriptsize$\pm$0.37}} & \heatlow{9.54{\scriptsize$\pm$0.45}} \\
WRDL & 35.83M & \heatlow{9.71{\scriptsize$\pm$0.28}} & \heatmid{11.58{\scriptsize$\pm$0.31}} & \heatlow{8.49{\scriptsize$\pm$0.22}} & \heatlow{9.79{\scriptsize$\pm$0.36}} & \heatmid{5.79{\scriptsize$\pm$0.13}} & \heatgood{5.38{\scriptsize$\pm$0.15}} & \heatlow{9.93{\scriptsize$\pm$0.30}} \\
DnCNN & 0.14M & \heatlow{11.63{\scriptsize$\pm$0.12}} & \heatlow{9.22{\scriptsize$\pm$0.05}} & \heatlow{13.18{\scriptsize$\pm$0.17}} & \heatmid{13.55{\scriptsize$\pm$0.20}} & \heatbest{8.68{\scriptsize$\pm$0.06}} & \heatlow{4.63{\scriptsize$\pm$0.23}} & \heatlow{11.66{\scriptsize$\pm$0.12}} \\
\bottomrule
\end{tabular*}
\end{table}

\begin{table}[H]
\caption{Mobile AVO Continuous Missing 40 Traces.}
\label{tab:full-mobile-avo-interp-continuous40tr}
\centering
\scriptsize
\setlength{\tabcolsep}{2.4pt}
\renewcommand{\arraystretch}{1.0}
\taskheat{seisInterpolationHeat}{}
\begin{tabular*}{0.92\linewidth}{@{\extracolsep{\fill}}lcccccccc@{}}
\toprule
Method & Params & SNR & \multicolumn{4}{c}{FB-FRE} & \multicolumn{2}{c}{EB-FE} \\
\cmidrule(lr){4-7}\cmidrule(lr){8-9}
 & & & Low & Mid & High & VHigh & Med. & Strong \\
\midrule
ANet & 7.05M & \heatbase{6.40{\scriptsize$\pm$0.57}} & \heatbase{6.70{\scriptsize$\pm$0.71}} & \heatbase{6.48{\scriptsize$\pm$0.34}} & \heatbase{6.94{\scriptsize$\pm$0.67}} & \heatbase{0.90{\scriptsize$\pm$1.36}} & \heatgood{4.25{\scriptsize$\pm$0.30}} & \heatbase{6.48{\scriptsize$\pm$0.58}} \\
CFunet & 7.35M & \heatbase{6.64{\scriptsize$\pm$0.16}} & \heatbase{6.75{\scriptsize$\pm$0.14}} & \heatbase{6.64{\scriptsize$\pm$0.17}} & \heatbase{8.11{\scriptsize$\pm$0.33}} & \heatlow{4.29{\scriptsize$\pm$0.31}} & \heatbest{4.88{\scriptsize$\pm$0.08}} & \heatbase{6.70{\scriptsize$\pm$0.16}} \\
WRDL & 35.83M & \heatlow{7.96{\scriptsize$\pm$0.12}} & \heatmid{9.41{\scriptsize$\pm$0.11}} & \heatlow{6.80{\scriptsize$\pm$0.13}} & \heatlow{8.53{\scriptsize$\pm$0.14}} & \heatmid{4.98{\scriptsize$\pm$0.16}} & \heatlow{3.46{\scriptsize$\pm$0.10}} & \heatlow{8.20{\scriptsize$\pm$0.12}} \\
CA-Unet & 7.79M & \heatlow{8.00{\scriptsize$\pm$0.06}} & \heatmid{9.30{\scriptsize$\pm$0.11}} & \heatlow{7.01{\scriptsize$\pm$0.01}} & \heatlow{8.56{\scriptsize$\pm$0.08}} & \heatmid{5.20{\scriptsize$\pm$0.16}} & \heatgood{4.29{\scriptsize$\pm$0.10}} & \heatlow{8.16{\scriptsize$\pm$0.06}} \\
DnCNN & 0.14M & \heatlow{10.18{\scriptsize$\pm$0.09}} & \heatlow{7.46{\scriptsize$\pm$0.02}} & \heatlow{11.88{\scriptsize$\pm$0.13}} & \heatgood{12.38{\scriptsize$\pm$0.15}} & \heatbest{7.69{\scriptsize$\pm$0.18}} & \heatlow{3.49{\scriptsize$\pm$0.41}} & \heatlow{10.21{\scriptsize$\pm$0.10}} \\
\bottomrule
\end{tabular*}
\end{table}

\begin{table}[H]
\caption{Mobile AVO Random Missing 30\%.}
\label{tab:full-mobile-avo-interp-random30}
\centering
\scriptsize
\setlength{\tabcolsep}{2.4pt}
\renewcommand{\arraystretch}{1.0}
\taskheat{seisInterpolationHeat}{}
\begin{tabular*}{0.92\linewidth}{@{\extracolsep{\fill}}lcccccccc@{}}
\toprule
Method & Params & SNR & \multicolumn{4}{c}{FB-FRE} & \multicolumn{2}{c}{EB-FE} \\
\cmidrule(lr){4-7}\cmidrule(lr){8-9}
 & & & Low & Mid & High & VHigh & Med. & Strong \\
\midrule
ANet & 7.05M & \heatbase{12.79{\scriptsize$\pm$0.64}} & \heatbase{18.01{\scriptsize$\pm$0.52}} & \heatbase{12.27{\scriptsize$\pm$0.59}} & \heatbase{8.55{\scriptsize$\pm$0.81}} & \heatbase{0.57{\scriptsize$\pm$0.49}} & \heatlow{8.69{\scriptsize$\pm$0.30}} & \heatbase{12.99{\scriptsize$\pm$0.67}} \\
CFunet & 7.35M & \heatbase{14.45{\scriptsize$\pm$0.39}} & \heatmid{20.57{\scriptsize$\pm$0.44}} & \heatbase{13.34{\scriptsize$\pm$0.34}} & \heatbase{10.28{\scriptsize$\pm$0.62}} & \heatmid{4.50{\scriptsize$\pm$0.37}} & \heatmid{10.13{\scriptsize$\pm$0.21}} & \heatbase{14.67{\scriptsize$\pm$0.41}} \\
DnCNN & 0.14M & \heatlow{15.36{\scriptsize$\pm$0.13}} & \heatbase{17.99{\scriptsize$\pm$0.09}} & \heatlow{16.01{\scriptsize$\pm$0.18}} & \heatmid{11.56{\scriptsize$\pm$0.15}} & \heatmid{4.02{\scriptsize$\pm$0.16}} & \heatlow{8.63{\scriptsize$\pm$0.00}} & \heatlow{15.15{\scriptsize$\pm$0.15}} \\
ResUNet & 8.11M & \heatlow{15.59{\scriptsize$\pm$0.86}} & \heatlow{19.05{\scriptsize$\pm$1.14}} & \heatlow{16.92{\scriptsize$\pm$0.94}} & \heatlow{10.76{\scriptsize$\pm$1.03}} & \heatbase{0.71{\scriptsize$\pm$1.34}} & \heatbase{7.26{\scriptsize$\pm$1.56}} & \heatlow{15.08{\scriptsize$\pm$1.09}} \\
Attention UNet & 7.85M & \heatlow{15.95{\scriptsize$\pm$0.31}} & \heatlow{19.03{\scriptsize$\pm$0.45}} & \heatlow{16.90{\scriptsize$\pm$0.49}} & \heatlow{11.45{\scriptsize$\pm$0.24}} & \heatlow{2.15{\scriptsize$\pm$0.32}} & \heatbase{8.38{\scriptsize$\pm$0.67}} & \heatlow{15.54{\scriptsize$\pm$0.37}} \\
\bottomrule
\end{tabular*}
\end{table}

\begin{table}[H]
\caption{Mobile AVO Random Missing 50\%.}
\label{tab:full-mobile-avo-interp-random50}
\centering
\scriptsize
\setlength{\tabcolsep}{2.4pt}
\renewcommand{\arraystretch}{1.0}
\taskheat{seisInterpolationHeat}{}
\begin{tabular*}{0.92\linewidth}{@{\extracolsep{\fill}}lcccccccc@{}}
\toprule
Method & Params & SNR & \multicolumn{4}{c}{FB-FRE} & \multicolumn{2}{c}{EB-FE} \\
\cmidrule(lr){4-7}\cmidrule(lr){8-9}
 & & & Low & Mid & High & VHigh & Med. & Strong \\
\midrule
ANet & 7.05M & \heatbase{10.44{\scriptsize$\pm$0.18}} & \heatbase{15.44{\scriptsize$\pm$0.16}} & \heatbase{9.62{\scriptsize$\pm$0.12}} & \heatbase{6.38{\scriptsize$\pm$0.24}} & \heatbase{-0.36{\scriptsize$\pm$0.19}} & \heatlow{6.12{\scriptsize$\pm$0.13}} & \heatbase{10.66{\scriptsize$\pm$0.18}} \\
ResUNet & 8.11M & \heatbase{11.64{\scriptsize$\pm$0.54}} & \heatlow{15.73{\scriptsize$\pm$0.22}} & \heatmid{13.36{\scriptsize$\pm$0.07}} & \heatbase{7.05{\scriptsize$\pm$0.14}} & \heatbase{-1.49{\scriptsize$\pm$0.36}} & \heatbase{4.77{\scriptsize$\pm$0.68}} & \heatbase{11.62{\scriptsize$\pm$0.11}} \\
CFunet & 7.35M & \heatlow{11.81{\scriptsize$\pm$0.19}} & \heatmid{17.66{\scriptsize$\pm$0.16}} & \heatbase{10.58{\scriptsize$\pm$0.13}} & \heatlow{7.66{\scriptsize$\pm$0.46}} & \heatgood{2.91{\scriptsize$\pm$0.24}} & \heatgood{7.32{\scriptsize$\pm$0.09}} & \heatlow{12.04{\scriptsize$\pm$0.21}} \\
DnCNN & 0.14M & \heatlow{11.87{\scriptsize$\pm$0.16}} & \heatbase{14.49{\scriptsize$\pm$0.22}} & \heatlow{12.48{\scriptsize$\pm$0.14}} & \heatlow{8.36{\scriptsize$\pm$0.16}} & \heatmid{2.12{\scriptsize$\pm$0.04}} & \heatlow{5.84{\scriptsize$\pm$0.46}} & \heatlow{11.79{\scriptsize$\pm$0.14}} \\
SPNet & 0.67M & \heatlow{12.90{\scriptsize$\pm$0.14}} & \heatlow{15.75{\scriptsize$\pm$0.13}} & \heatmid{13.42{\scriptsize$\pm$0.17}} & \heatmid{9.77{\scriptsize$\pm$0.18}} & \heatgood{3.05{\scriptsize$\pm$0.17}} & \heatmid{6.67{\scriptsize$\pm$0.08}} & \heatmid{12.95{\scriptsize$\pm$0.14}} \\
\bottomrule
\end{tabular*}
\end{table}

\begin{table}[H]
\caption{Mobile AVO Random Missing 70\%.}
\label{tab:full-mobile-avo-interp-random70}
\centering
\scriptsize
\setlength{\tabcolsep}{2.4pt}
\renewcommand{\arraystretch}{1.0}
\taskheat{seisInterpolationHeat}{}
\begin{tabular*}{0.92\linewidth}{@{\extracolsep{\fill}}lcccccccc@{}}
\toprule
Method & Params & SNR & \multicolumn{4}{c}{FB-FRE} & \multicolumn{2}{c}{EB-FE} \\
\cmidrule(lr){4-7}\cmidrule(lr){8-9}
 & & & Low & Mid & High & VHigh & Med. & Strong \\
\midrule
DnCNN & 0.14M & \heatbase{8.22{\scriptsize$\pm$0.22}} & \heatbase{9.93{\scriptsize$\pm$0.34}} & \heatbase{8.68{\scriptsize$\pm$0.27}} & \heatlow{5.30{\scriptsize$\pm$0.30}} & \heatgood{0.45{\scriptsize$\pm$0.16}} & \heatbase{2.63{\scriptsize$\pm$0.64}} & \heatbase{8.14{\scriptsize$\pm$0.29}} \\
SPNet & 0.67M & \heatbase{8.42{\scriptsize$\pm$0.10}} & \heatbase{10.21{\scriptsize$\pm$0.05}} & \heatbase{8.86{\scriptsize$\pm$0.09}} & \heatmid{5.86{\scriptsize$\pm$0.14}} & \heatbest{0.85{\scriptsize$\pm$0.13}} & \heatbest{3.58{\scriptsize$\pm$0.05}} & \heatbase{8.45{\scriptsize$\pm$0.10}} \\
ResUNet & 8.11M & \heatlow{9.08{\scriptsize$\pm$0.65}} & \heatmid{12.74{\scriptsize$\pm$0.76}} & \heatlow{10.14{\scriptsize$\pm$0.60}} & \heatbase{4.80{\scriptsize$\pm$0.73}} & \heatbase{-2.32{\scriptsize$\pm$0.77}} & \heatmid{2.81{\scriptsize$\pm$0.73}} & \heatlow{8.97{\scriptsize$\pm$0.70}} \\
Attention UNet & 7.85M & \heatmid{9.60{\scriptsize$\pm$0.24}} & \heatmid{12.99{\scriptsize$\pm$0.37}} & \heatmid{10.63{\scriptsize$\pm$0.16}} & \heatlow{5.84{\scriptsize$\pm$0.16}} & \heatbase{-1.37{\scriptsize$\pm$0.03}} & \heatlow{2.68{\scriptsize$\pm$0.59}} & \heatmid{9.70{\scriptsize$\pm$0.17}} \\
DnCNN-L & 0.56M & \heatmid{10.09{\scriptsize$\pm$0.21}} & \heatlow{12.12{\scriptsize$\pm$0.32}} & \heatmid{10.34{\scriptsize$\pm$0.17}} & \heatbest{7.61{\scriptsize$\pm$0.18}} & \heatbest{1.47{\scriptsize$\pm$0.10}} & \heatmid{3.22{\scriptsize$\pm$0.38}} & \heatmid{10.07{\scriptsize$\pm$0.22}} \\
\bottomrule
\end{tabular*}
\end{table}

\begin{table}[H]
\caption{Mobile AVO Uniform Missing 50\%.}
\label{tab:full-mobile-avo-interp-uniform50}
\centering
\scriptsize
\setlength{\tabcolsep}{2.4pt}
\renewcommand{\arraystretch}{1.0}
\taskheat{seisInterpolationHeat}{}
\begin{tabular*}{0.92\linewidth}{@{\extracolsep{\fill}}lcccccccc@{}}
\toprule
Method & Params & SNR & \multicolumn{4}{c}{FB-FRE} & \multicolumn{2}{c}{EB-FE} \\
\cmidrule(lr){4-7}\cmidrule(lr){8-9}
 & & & Low & Mid & High & VHigh & Med. & Strong \\
\midrule
CFunet & 7.35M & \heatbase{13.37{\scriptsize$\pm$0.28}} & \heatbase{20.35{\scriptsize$\pm$0.38}} & \heatbase{12.27{\scriptsize$\pm$0.24}} & \heatbase{8.75{\scriptsize$\pm$0.60}} & \heatmid{3.24{\scriptsize$\pm$0.26}} & \heatmid{9.14{\scriptsize$\pm$0.18}} & \heatbase{13.58{\scriptsize$\pm$0.30}} \\
ANet & 7.05M & \heatbase{14.17{\scriptsize$\pm$0.58}} & \heatlow{20.99{\scriptsize$\pm$0.98}} & \heatbase{13.39{\scriptsize$\pm$0.49}} & \heatbase{9.70{\scriptsize$\pm$0.57}} & \heatbase{1.49{\scriptsize$\pm$0.56}} & \heatlow{8.94{\scriptsize$\pm$0.26}} & \heatbase{14.46{\scriptsize$\pm$0.61}} \\
DnCNN & 0.14M & \heatlow{16.35{\scriptsize$\pm$0.27}} & \heatbase{20.29{\scriptsize$\pm$0.62}} & \heatlow{17.42{\scriptsize$\pm$0.27}} & \heatlow{12.05{\scriptsize$\pm$0.27}} & \heatlow{2.72{\scriptsize$\pm$0.25}} & \heatbase{8.37{\scriptsize$\pm$0.34}} & \heatlow{16.34{\scriptsize$\pm$0.30}} \\
ResUNet & 8.11M & \heatlow{16.36{\scriptsize$\pm$0.18}} & \heatlow{21.67{\scriptsize$\pm$0.37}} & \heatmid{18.70{\scriptsize$\pm$0.39}} & \heatlow{11.58{\scriptsize$\pm$0.50}} & \heatbase{0.58{\scriptsize$\pm$0.03}} & \heatbase{7.08{\scriptsize$\pm$0.78}} & \heatlow{16.28{\scriptsize$\pm$0.43}} \\
SPNet & 0.67M & \heatlow{17.43{\scriptsize$\pm$0.31}} & \heatlow{21.45{\scriptsize$\pm$0.18}} & \heatmid{18.36{\scriptsize$\pm$0.25}} & \heatlow{13.40{\scriptsize$\pm$0.46}} & \heatmid{3.98{\scriptsize$\pm$0.37}} & \heatmid{9.21{\scriptsize$\pm$0.08}} & \heatlow{17.50{\scriptsize$\pm$0.33}} \\
\bottomrule
\end{tabular*}
\end{table}

\begin{table}[H]
\caption{Mobile AVO Uniform Missing 75\%.}
\label{tab:full-mobile-avo-interp-uniform75}
\centering
\scriptsize
\setlength{\tabcolsep}{2.4pt}
\renewcommand{\arraystretch}{1.0}
\taskheat{seisInterpolationHeat}{}
\begin{tabular*}{0.92\linewidth}{@{\extracolsep{\fill}}lcccccccc@{}}
\toprule
Method & Params & SNR & \multicolumn{4}{c}{FB-FRE} & \multicolumn{2}{c}{EB-FE} \\
\cmidrule(lr){4-7}\cmidrule(lr){8-9}
 & & & Low & Mid & High & VHigh & Med. & Strong \\
\midrule
ANet & 7.05M & \heatbase{11.04{\scriptsize$\pm$1.23}} & \heatlow{17.93{\scriptsize$\pm$1.35}} & \heatbase{10.41{\scriptsize$\pm$0.89}} & \heatbase{6.14{\scriptsize$\pm$1.61}} & \heatbase{-0.72{\scriptsize$\pm$1.22}} & \heatgood{6.05{\scriptsize$\pm$0.25}} & \heatbase{11.32{\scriptsize$\pm$1.32}} \\
CFunet & 7.35M & \heatbase{11.60{\scriptsize$\pm$0.26}} & \heatlow{17.74{\scriptsize$\pm$0.28}} & \heatbase{10.46{\scriptsize$\pm$0.13}} & \heatbase{7.21{\scriptsize$\pm$0.46}} & \heatgood{2.13{\scriptsize$\pm$0.23}} & \heatgood{6.56{\scriptsize$\pm$0.07}} & \heatbase{11.88{\scriptsize$\pm$0.28}} \\
DnCNN & 0.14M & \heatlow{11.91{\scriptsize$\pm$0.21}} & \heatbase{15.65{\scriptsize$\pm$0.41}} & \heatlow{12.69{\scriptsize$\pm$0.21}} & \heatlow{7.84{\scriptsize$\pm$0.22}} & \heatlow{0.47{\scriptsize$\pm$0.12}} & \heatbase{4.82{\scriptsize$\pm$0.17}} & \heatlow{11.89{\scriptsize$\pm$0.22}} \\
SPNet & 0.67M & \heatlow{12.37{\scriptsize$\pm$0.01}} & \heatbase{15.78{\scriptsize$\pm$0.23}} & \heatlow{13.08{\scriptsize$\pm$0.01}} & \heatlow{8.64{\scriptsize$\pm$0.07}} & \heatmid{1.00{\scriptsize$\pm$0.08}} & \heatlow{4.99{\scriptsize$\pm$0.11}} & \heatlow{12.43{\scriptsize$\pm$0.01}} \\
ResUNet & 8.11M & \heatlow{12.58{\scriptsize$\pm$0.12}} & \heatmid{18.24{\scriptsize$\pm$0.21}} & \heatmid{14.24{\scriptsize$\pm$0.12}} & \heatlow{7.95{\scriptsize$\pm$0.18}} & \heatbase{-1.35{\scriptsize$\pm$0.25}} & \heatbase{4.72{\scriptsize$\pm$0.32}} & \heatlow{12.64{\scriptsize$\pm$0.05}} \\
\bottomrule
\end{tabular*}
\end{table}

\begin{table}[H]
\caption{SEGC3 Random Missing 30\%.}
\label{tab:full-segc3-interp-random30}
\centering
\scriptsize
\setlength{\tabcolsep}{2.4pt}
\renewcommand{\arraystretch}{1.0}
\taskheat{seisInterpolationHeat}{}
\begin{tabular*}{0.92\linewidth}{@{\extracolsep{\fill}}lcccccccc@{}}
\toprule
Method & Params & SNR & \multicolumn{4}{c}{FB-FRE} & \multicolumn{2}{c}{EB-FE} \\
\cmidrule(lr){4-7}\cmidrule(lr){8-9}
 & & & Low & Mid & High & VHigh & Med. & Strong \\
\midrule
ANet & 7.05M & \heatbase{19.06{\scriptsize$\pm$0.09}} & \heatbase{21.28{\scriptsize$\pm$0.41}} & \heatbase{24.27{\scriptsize$\pm$0.13}} & \heatbase{15.74{\scriptsize$\pm$0.04}} & \heatbase{12.16{\scriptsize$\pm$0.18}} & \heatbase{15.12{\scriptsize$\pm$0.15}} & \heatbase{21.29{\scriptsize$\pm$0.07}} \\
CFunet & 7.35M & \heatbase{20.08{\scriptsize$\pm$0.10}} & \heatlow{24.44{\scriptsize$\pm$0.23}} & \heatbase{26.44{\scriptsize$\pm$0.12}} & \heatbase{16.31{\scriptsize$\pm$0.11}} & \heatlow{13.65{\scriptsize$\pm$0.06}} & \heatbase{15.38{\scriptsize$\pm$0.11}} & \heatlow{23.38{\scriptsize$\pm$0.11}} \\
ResUNet & 8.11M & \heatlow{20.39{\scriptsize$\pm$0.36}} & \heatmid{26.15{\scriptsize$\pm$0.16}} & \heatlow{26.99{\scriptsize$\pm$0.48}} & \heatlow{16.60{\scriptsize$\pm$0.35}} & \heatlow{13.52{\scriptsize$\pm$0.35}} & \heatlow{15.67{\scriptsize$\pm$0.31}} & \heatlow{23.68{\scriptsize$\pm$0.46}} \\
Attention UNet & 7.85M & \heatlow{20.44{\scriptsize$\pm$0.13}} & \heatbase{23.71{\scriptsize$\pm$0.09}} & \heatlow{26.71{\scriptsize$\pm$0.13}} & \heatlow{16.61{\scriptsize$\pm$0.13}} & \heatlow{13.25{\scriptsize$\pm$0.23}} & \heatlow{15.67{\scriptsize$\pm$0.15}} & \heatbase{23.25{\scriptsize$\pm$0.13}} \\
UNet & 7.76M & \heatlow{20.45{\scriptsize$\pm$0.13}} & \heatlow{24.11{\scriptsize$\pm$0.37}} & \heatlow{26.89{\scriptsize$\pm$0.15}} & \heatlow{16.61{\scriptsize$\pm$0.15}} & \heatbase{13.23{\scriptsize$\pm$0.17}} & \heatlow{15.67{\scriptsize$\pm$0.15}} & \heatlow{23.40{\scriptsize$\pm$0.09}} \\
\bottomrule
\end{tabular*}
\end{table}

\begin{table}[H]
\caption{SEGC3 Random Missing 50\%.}
\label{tab:full-segc3-interp-random50}
\centering
\scriptsize
\setlength{\tabcolsep}{2.4pt}
\renewcommand{\arraystretch}{1.0}
\taskheat{seisInterpolationHeat}{}
\begin{tabular*}{0.92\linewidth}{@{\extracolsep{\fill}}lcccccccc@{}}
\toprule
Method & Params & SNR & \multicolumn{4}{c}{FB-FRE} & \multicolumn{2}{c}{EB-FE} \\
\cmidrule(lr){4-7}\cmidrule(lr){8-9}
 & & & Low & Mid & High & VHigh & Med. & Strong \\
\midrule
ANet & 7.05M & \heatbase{14.68{\scriptsize$\pm$0.67}} & \heatbase{18.14{\scriptsize$\pm$1.10}} & \heatbase{18.72{\scriptsize$\pm$0.79}} & \heatbase{11.65{\scriptsize$\pm$0.59}} & \heatbase{8.25{\scriptsize$\pm$0.85}} & \heatbase{10.43{\scriptsize$\pm$0.54}} & \heatbase{17.14{\scriptsize$\pm$0.88}} \\
SPNet & 0.67M & \heatbase{15.62{\scriptsize$\pm$0.07}} & \heatlow{21.39{\scriptsize$\pm$0.19}} & \heatlow{20.11{\scriptsize$\pm$0.10}} & \heatbase{12.31{\scriptsize$\pm$0.05}} & \heatbase{9.36{\scriptsize$\pm$0.06}} & \heatmid{11.25{\scriptsize$\pm$0.07}} & \heatlow{18.54{\scriptsize$\pm$0.08}} \\
ResUNet & 8.11M & \heatlow{15.66{\scriptsize$\pm$0.04}} & \heatmid{22.23{\scriptsize$\pm$0.16}} & \heatbase{19.87{\scriptsize$\pm$0.07}} & \heatlow{12.45{\scriptsize$\pm$0.06}} & \heatlow{9.70{\scriptsize$\pm$0.07}} & \heatbase{11.02{\scriptsize$\pm$0.04}} & \heatlow{18.66{\scriptsize$\pm$0.04}} \\
UNet & 7.76M & \heatlow{15.68{\scriptsize$\pm$0.07}} & \heatlow{20.67{\scriptsize$\pm$0.24}} & \heatlow{19.92{\scriptsize$\pm$0.09}} & \heatlow{12.46{\scriptsize$\pm$0.06}} & \heatlow{9.45{\scriptsize$\pm$0.06}} & \heatlow{11.04{\scriptsize$\pm$0.08}} & \heatlow{18.54{\scriptsize$\pm$0.05}} \\
Attention UNet & 7.85M & \heatlow{15.71{\scriptsize$\pm$0.17}} & \heatlow{20.48{\scriptsize$\pm$0.26}} & \heatlow{19.90{\scriptsize$\pm$0.29}} & \heatlow{12.50{\scriptsize$\pm$0.14}} & \heatlow{9.55{\scriptsize$\pm$0.14}} & \heatlow{11.09{\scriptsize$\pm$0.15}} & \heatbase{18.50{\scriptsize$\pm$0.27}} \\
\bottomrule
\end{tabular*}
\end{table}

\begin{table}[H]
\caption{SEGC3 Random Missing 70\%.}
\label{tab:full-segc3-interp-random70}
\centering
\scriptsize
\setlength{\tabcolsep}{2.4pt}
\renewcommand{\arraystretch}{1.0}
\taskheat{seisInterpolationHeat}{}
\begin{tabular*}{0.92\linewidth}{@{\extracolsep{\fill}}lcccccccc@{}}
\toprule
Method & Params & SNR & \multicolumn{4}{c}{FB-FRE} & \multicolumn{2}{c}{EB-FE} \\
\cmidrule(lr){4-7}\cmidrule(lr){8-9}
 & & & Low & Mid & High & VHigh & Med. & Strong \\
\midrule
SPNet & 0.67M & \heatbase{10.27{\scriptsize$\pm$0.05}} & \heatbase{14.23{\scriptsize$\pm$0.22}} & \heatbase{13.05{\scriptsize$\pm$0.03}} & \heatbase{7.72{\scriptsize$\pm$0.05}} & \heatbase{4.72{\scriptsize$\pm$0.10}} & \heatbase{6.32{\scriptsize$\pm$0.05}} & \heatbase{12.34{\scriptsize$\pm$0.05}} \\
DnCNN & 0.14M & \heatbase{10.60{\scriptsize$\pm$0.02}} & \heatbase{14.11{\scriptsize$\pm$0.10}} & \heatbase{13.27{\scriptsize$\pm$0.04}} & \heatbase{8.07{\scriptsize$\pm$0.02}} & \heatlow{5.33{\scriptsize$\pm$0.03}} & \heatbase{6.56{\scriptsize$\pm$0.01}} & \heatbase{12.56{\scriptsize$\pm$0.04}} \\
UNet & 7.76M & \heatlow{10.93{\scriptsize$\pm$0.02}} & \heatmid{16.00{\scriptsize$\pm$0.25}} & \heatlow{13.38{\scriptsize$\pm$0.06}} & \heatlow{8.55{\scriptsize$\pm$0.02}} & \heatmid{5.68{\scriptsize$\pm$0.04}} & \heatlow{6.68{\scriptsize$\pm$0.04}} & \heatlow{13.18{\scriptsize$\pm$0.05}} \\
Attention UNet & 7.85M & \heatmid{10.96{\scriptsize$\pm$0.10}} & \heatmid{15.84{\scriptsize$\pm$0.26}} & \heatmid{13.42{\scriptsize$\pm$0.15}} & \heatmid{8.58{\scriptsize$\pm$0.08}} & \heatlow{5.67{\scriptsize$\pm$0.09}} & \heatmid{6.69{\scriptsize$\pm$0.07}} & \heatmid{13.23{\scriptsize$\pm$0.14}} \\
ResUNet & 8.11M & \heatmid{11.00{\scriptsize$\pm$0.02}} & \heatmid{16.86{\scriptsize$\pm$0.29}} & \heatmid{13.51{\scriptsize$\pm$0.05}} & \heatmid{8.57{\scriptsize$\pm$0.03}} & \heatgood{5.83{\scriptsize$\pm$0.05}} & \heatmid{6.71{\scriptsize$\pm$0.03}} & \heatmid{13.31{\scriptsize$\pm$0.08}} \\
\bottomrule
\end{tabular*}
\end{table}

\begin{table}[H]
\caption{SEGC3 Uniform Missing 50\%.}
\label{tab:full-segc3-interp-uniform50}
\centering
\scriptsize
\setlength{\tabcolsep}{2.4pt}
\renewcommand{\arraystretch}{1.0}
\taskheat{seisInterpolationHeat}{}
\begin{tabular*}{0.92\linewidth}{@{\extracolsep{\fill}}lcccccccc@{}}
\toprule
Method & Params & SNR & \multicolumn{4}{c}{FB-FRE} & \multicolumn{2}{c}{EB-FE} \\
\cmidrule(lr){4-7}\cmidrule(lr){8-9}
 & & & Low & Mid & High & VHigh & Med. & Strong \\
\midrule
CFunet & 7.35M & \heatbase{16.52{\scriptsize$\pm$0.23}} & \heatlow{24.29{\scriptsize$\pm$0.45}} & \heatbase{27.95{\scriptsize$\pm$0.15}} & \heatbase{12.30{\scriptsize$\pm$0.28}} & \heatlow{9.58{\scriptsize$\pm$0.16}} & \heatbase{11.44{\scriptsize$\pm$0.26}} & \heatbase{20.17{\scriptsize$\pm$0.25}} \\
DnCNN & 0.14M & \heatbase{16.66{\scriptsize$\pm$0.06}} & \heatbase{22.77{\scriptsize$\pm$0.11}} & \heatlow{28.90{\scriptsize$\pm$0.22}} & \heatbase{12.47{\scriptsize$\pm$0.06}} & \heatbase{9.10{\scriptsize$\pm$0.02}} & \heatbase{11.49{\scriptsize$\pm$0.05}} & \heatlow{20.36{\scriptsize$\pm$0.10}} \\
ANet & 7.05M & \heatlow{16.78{\scriptsize$\pm$0.20}} & \heatbase{22.12{\scriptsize$\pm$0.92}} & \heatbase{26.19{\scriptsize$\pm$0.49}} & \heatlow{12.91{\scriptsize$\pm$0.22}} & \heatlow{9.38{\scriptsize$\pm$0.43}} & \heatlow{11.92{\scriptsize$\pm$0.26}} & \heatbase{20.14{\scriptsize$\pm$0.12}} \\
DnCNN-L & 0.56M & \heatlow{16.89{\scriptsize$\pm$0.06}} & \heatlow{24.90{\scriptsize$\pm$0.55}} & \heatlow{29.69{\scriptsize$\pm$0.93}} & \heatlow{12.74{\scriptsize$\pm$0.05}} & \heatbase{8.86{\scriptsize$\pm$0.04}} & \heatlow{11.69{\scriptsize$\pm$0.04}} & \heatlow{20.88{\scriptsize$\pm$0.13}} \\
SPNet & 0.67M & \heatlow{17.12{\scriptsize$\pm$0.04}} & \heatmid{27.93{\scriptsize$\pm$0.19}} & \heatlow{31.35{\scriptsize$\pm$0.17}} & \heatlow{12.79{\scriptsize$\pm$0.06}} & \heatlow{9.84{\scriptsize$\pm$0.03}} & \heatlow{12.14{\scriptsize$\pm$0.04}} & \heatlow{21.37{\scriptsize$\pm$0.04}} \\
\bottomrule
\end{tabular*}
\end{table}

\begin{table}[H]
\caption{SEGC3 Uniform Missing 75\%.}
\label{tab:full-segc3-interp-uniform75}
\centering
\scriptsize
\setlength{\tabcolsep}{2.4pt}
\renewcommand{\arraystretch}{1.0}
\taskheat{seisInterpolationHeat}{}
\begin{tabular*}{0.92\linewidth}{@{\extracolsep{\fill}}lcccccccc@{}}
\toprule
Method & Params & SNR & \multicolumn{4}{c}{FB-FRE} & \multicolumn{2}{c}{EB-FE} \\
\cmidrule(lr){4-7}\cmidrule(lr){8-9}
 & & & Low & Mid & High & VHigh & Med. & Strong \\
\midrule
DnCNN & 0.14M & \heatbase{12.29{\scriptsize$\pm$0.05}} & \heatbase{19.03{\scriptsize$\pm$0.12}} & \heatbase{17.41{\scriptsize$\pm$0.06}} & \heatbase{8.96{\scriptsize$\pm$0.05}} & \heatbase{5.55{\scriptsize$\pm$0.06}} & \heatbase{7.39{\scriptsize$\pm$0.05}} & \heatbase{15.37{\scriptsize$\pm$0.09}} \\
SPNet & 0.67M & \heatbase{12.47{\scriptsize$\pm$0.08}} & \heatlow{21.55{\scriptsize$\pm$0.10}} & \heatbase{17.66{\scriptsize$\pm$0.08}} & \heatbase{9.15{\scriptsize$\pm$0.07}} & \heatbase{5.71{\scriptsize$\pm$0.15}} & \heatbase{7.58{\scriptsize$\pm$0.05}} & \heatbase{15.66{\scriptsize$\pm$0.17}} \\
DnCNN-L & 0.56M & \heatlow{12.84{\scriptsize$\pm$0.07}} & \heatbase{21.38{\scriptsize$\pm$0.11}} & \heatlow{17.87{\scriptsize$\pm$0.06}} & \heatlow{9.52{\scriptsize$\pm$0.07}} & \heatlow{6.04{\scriptsize$\pm$0.11}} & \heatlow{7.81{\scriptsize$\pm$0.05}} & \heatlow{16.27{\scriptsize$\pm$0.12}} \\
ResUNet-L & 32.44M & \heatlow{13.26{\scriptsize$\pm$0.05}} & \heatbest{26.48{\scriptsize$\pm$0.05}} & \heatmid{17.94{\scriptsize$\pm$0.05}} & \heatlow{9.95{\scriptsize$\pm$0.07}} & \heatmid{7.20{\scriptsize$\pm$0.08}} & \heatlow{8.12{\scriptsize$\pm$0.06}} & \heatgood{17.14{\scriptsize$\pm$0.04}} \\
UNet-L & 31.04M & \heatlow{13.26{\scriptsize$\pm$0.05}} & \heatgood{25.44{\scriptsize$\pm$0.08}} & \heatlow{17.80{\scriptsize$\pm$0.02}} & \heatmid{10.00{\scriptsize$\pm$0.06}} & \heatgood{7.27{\scriptsize$\pm$0.07}} & \heatlow{8.12{\scriptsize$\pm$0.05}} & \heatmid{17.09{\scriptsize$\pm$0.05}} \\
\bottomrule
\end{tabular*}
\end{table}

\begin{table}[H]
\caption{SEGC3 Continuous Missing 20 Traces.}
\label{tab:full-segc3-interp-continuous20tr}
\centering
\scriptsize
\setlength{\tabcolsep}{2.4pt}
\renewcommand{\arraystretch}{1.0}
\taskheat{seisInterpolationHeat}{}
\begin{tabular*}{0.92\linewidth}{@{\extracolsep{\fill}}lcccccccc@{}}
\toprule
Method & Params & SNR & \multicolumn{4}{c}{FB-FRE} & \multicolumn{2}{c}{EB-FE} \\
\cmidrule(lr){4-7}\cmidrule(lr){8-9}
 & & & Low & Mid & High & VHigh & Med. & Strong \\
\midrule
ANet & 7.05M & \heatbase{14.61{\scriptsize$\pm$0.42}} & \heatbase{16.39{\scriptsize$\pm$0.75}} & \heatbase{16.34{\scriptsize$\pm$0.28}} & \heatbase{13.02{\scriptsize$\pm$0.45}} & \heatbase{10.49{\scriptsize$\pm$0.63}} & \heatbase{11.55{\scriptsize$\pm$0.40}} & \heatbase{16.06{\scriptsize$\pm$0.50}} \\
DnCNN & 0.14M & \heatbase{14.88{\scriptsize$\pm$0.10}} & \heatbase{16.49{\scriptsize$\pm$0.16}} & \heatbase{16.28{\scriptsize$\pm$0.14}} & \heatbase{13.16{\scriptsize$\pm$0.07}} & \heatbase{11.72{\scriptsize$\pm$0.07}} & \heatbase{11.83{\scriptsize$\pm$0.02}} & \heatbase{16.06{\scriptsize$\pm$0.15}} \\
CFunet & 7.35M & \heatlow{15.12{\scriptsize$\pm$0.10}} & \heatlow{18.22{\scriptsize$\pm$0.07}} & \heatlow{16.49{\scriptsize$\pm$0.08}} & \heatlow{13.53{\scriptsize$\pm$0.17}} & \heatlow{11.99{\scriptsize$\pm$0.06}} & \heatlow{11.92{\scriptsize$\pm$0.03}} & \heatbase{16.51{\scriptsize$\pm$0.18}} \\
SPNet & 0.67M & \heatlow{15.45{\scriptsize$\pm$0.05}} & \heatlow{18.27{\scriptsize$\pm$0.09}} & \heatmid{17.08{\scriptsize$\pm$0.04}} & \heatlow{13.59{\scriptsize$\pm$0.08}} & \heatlow{11.96{\scriptsize$\pm$0.00}} & \heatlow{12.02{\scriptsize$\pm$0.05}} & \heatlow{17.06{\scriptsize$\pm$0.07}} \\
Attention UNet & 7.85M & \heatlow{15.50{\scriptsize$\pm$0.09}} & \heatlow{17.62{\scriptsize$\pm$0.28}} & \heatlow{16.81{\scriptsize$\pm$0.03}} & \heatlow{13.76{\scriptsize$\pm$0.17}} & \heatlow{12.04{\scriptsize$\pm$0.13}} & \heatmid{12.11{\scriptsize$\pm$0.03}} & \heatlow{16.78{\scriptsize$\pm$0.15}} \\
\bottomrule
\end{tabular*}
\end{table}

\begin{table}[H]
\caption{SEGC3 Continuous Missing 30 Traces.}
\label{tab:full-segc3-interp-continuous30tr}
\centering
\scriptsize
\setlength{\tabcolsep}{2.4pt}
\renewcommand{\arraystretch}{1.0}
\taskheat{seisInterpolationHeat}{}
\begin{tabular*}{0.92\linewidth}{@{\extracolsep{\fill}}lcccccccc@{}}
\toprule
Method & Params & SNR & \multicolumn{4}{c}{FB-FRE} & \multicolumn{2}{c}{EB-FE} \\
\cmidrule(lr){4-7}\cmidrule(lr){8-9}
 & & & Low & Mid & High & VHigh & Med. & Strong \\
\midrule
ANet & 7.05M & \heatbase{11.08{\scriptsize$\pm$2.05}} & \heatbase{12.55{\scriptsize$\pm$2.96}} & \heatbase{12.23{\scriptsize$\pm$2.25}} & \heatbase{9.87{\scriptsize$\pm$1.77}} & \heatbase{7.69{\scriptsize$\pm$2.35}} & \heatbase{8.68{\scriptsize$\pm$1.29}} & \heatbase{11.94{\scriptsize$\pm$2.39}} \\
CFunet & 7.35M & \heatbase{11.41{\scriptsize$\pm$0.13}} & \heatlow{14.09{\scriptsize$\pm$0.07}} & \heatbase{12.26{\scriptsize$\pm$0.12}} & \heatbase{10.22{\scriptsize$\pm$0.16}} & \heatlow{9.61{\scriptsize$\pm$0.05}} & \heatbase{9.26{\scriptsize$\pm$0.05}} & \heatlow{12.15{\scriptsize$\pm$0.17}} \\
DnCNN & 0.14M & \heatlow{11.42{\scriptsize$\pm$0.06}} & \heatbase{12.70{\scriptsize$\pm$0.18}} & \heatbase{12.23{\scriptsize$\pm$0.08}} & \heatlow{10.27{\scriptsize$\pm$0.04}} & \heatlow{9.53{\scriptsize$\pm$0.02}} & \heatlow{9.32{\scriptsize$\pm$0.02}} & \heatbase{12.10{\scriptsize$\pm$0.09}} \\
SPNet & 0.67M & \heatlow{12.06{\scriptsize$\pm$0.05}} & \heatlow{14.16{\scriptsize$\pm$0.07}} & \heatlow{13.31{\scriptsize$\pm$0.06}} & \heatlow{10.48{\scriptsize$\pm$0.05}} & \heatbase{9.52{\scriptsize$\pm$0.07}} & \heatlow{9.35{\scriptsize$\pm$0.03}} & \heatlow{13.14{\scriptsize$\pm$0.07}} \\
DnCNN-L & 0.56M & \heatlow{12.29{\scriptsize$\pm$0.04}} & \heatlow{14.44{\scriptsize$\pm$0.08}} & \heatlow{13.28{\scriptsize$\pm$0.09}} & \heatmid{10.97{\scriptsize$\pm$0.01}} & \heatgood{10.05{\scriptsize$\pm$0.09}} & \heatmid{9.64{\scriptsize$\pm$0.03}} & \heatlow{13.25{\scriptsize$\pm$0.08}} \\
\bottomrule
\end{tabular*}
\end{table}

\begin{table}[H]
\caption{SEGC3 Continuous Missing 40 Traces.}
\label{tab:full-segc3-interp-continuous40tr}
\centering
\scriptsize
\setlength{\tabcolsep}{2.4pt}
\renewcommand{\arraystretch}{1.0}
\taskheat{seisInterpolationHeat}{}
\begin{tabular*}{0.92\linewidth}{@{\extracolsep{\fill}}lcccccccc@{}}
\toprule
Method & Params & SNR & \multicolumn{4}{c}{FB-FRE} & \multicolumn{2}{c}{EB-FE} \\
\cmidrule(lr){4-7}\cmidrule(lr){8-9}
 & & & Low & Mid & High & VHigh & Med. & Strong \\
\midrule
CFunet & 7.35M & \heatbase{9.01{\scriptsize$\pm$0.05}} & \heatbase{10.82{\scriptsize$\pm$0.08}} & \heatbase{9.46{\scriptsize$\pm$0.01}} & \heatbase{8.32{\scriptsize$\pm$0.09}} & \heatlow{8.12{\scriptsize$\pm$0.07}} & \heatbase{7.61{\scriptsize$\pm$0.03}} & \heatbase{9.46{\scriptsize$\pm$0.06}} \\
DnCNN & 0.14M & \heatbase{9.28{\scriptsize$\pm$0.04}} & \heatbase{10.16{\scriptsize$\pm$0.09}} & \heatbase{9.80{\scriptsize$\pm$0.05}} & \heatlow{8.49{\scriptsize$\pm$0.03}} & \heatlow{8.11{\scriptsize$\pm$0.10}} & \heatlow{7.76{\scriptsize$\pm$0.02}} & \heatbase{9.72{\scriptsize$\pm$0.05}} \\
ANet & 7.05M & \heatlow{9.51{\scriptsize$\pm$1.17}} & \heatlow{11.01{\scriptsize$\pm$1.65}} & \heatlow{10.59{\scriptsize$\pm$1.23}} & \heatbase{8.21{\scriptsize$\pm$1.08}} & \heatbase{6.46{\scriptsize$\pm$1.41}} & \heatbase{7.20{\scriptsize$\pm$0.77}} & \heatlow{10.30{\scriptsize$\pm$1.35}} \\
DnCNN-L & 0.56M & \heatlow{9.79{\scriptsize$\pm$0.04}} & \heatlow{11.06{\scriptsize$\pm$0.04}} & \heatlow{10.41{\scriptsize$\pm$0.06}} & \heatlow{8.93{\scriptsize$\pm$0.01}} & \heatgood{8.40{\scriptsize$\pm$0.02}} & \heatmid{7.95{\scriptsize$\pm$0.03}} & \heatlow{10.38{\scriptsize$\pm$0.06}} \\
SPNet & 0.67M & \heatlow{9.88{\scriptsize$\pm$0.05}} & \heatlow{11.53{\scriptsize$\pm$0.04}} & \heatlow{10.76{\scriptsize$\pm$0.06}} & \heatlow{8.68{\scriptsize$\pm$0.04}} & \heatbase{7.99{\scriptsize$\pm$0.06}} & \heatlow{7.75{\scriptsize$\pm$0.01}} & \heatlow{10.63{\scriptsize$\pm$0.06}} \\
\bottomrule
\end{tabular*}
\end{table}

\subsection{Ground-roll suppression}

\begin{table}[H]
\caption{SEGC3 Ground-Roll Noise 1.}
\label{tab:full-segc3-groundroll-noise1}
\centering
\scriptsize
\setlength{\tabcolsep}{2.4pt}
\renewcommand{\arraystretch}{1.0}
\taskheat{seisSurfaceHeat}{}
%
\end{table}

\begin{table}[H]
\caption{SEGC3 Ground-Roll Noise 3.}
\label{tab:full-segc3-groundroll-noise3}
\centering
\scriptsize
\setlength{\tabcolsep}{2.4pt}
\renewcommand{\arraystretch}{1.0}
\taskheat{seisSurfaceHeat}{}
%
\end{table}

\begin{table}[H]
\caption{SEGC3 Ground-Roll Noise 5.}
\label{tab:full-segc3-groundroll-noise5}
\centering
\scriptsize
\setlength{\tabcolsep}{2.4pt}
\renewcommand{\arraystretch}{1.0}
\taskheat{seisSurfaceHeat}{}
%
\end{table}

\begin{table}[H]
\caption{SEGC3 Ground-Roll Noise 7.}
\label{tab:full-segc3-groundroll-noise7}
\centering
\scriptsize
\setlength{\tabcolsep}{2.4pt}
\renewcommand{\arraystretch}{1.0}
\taskheat{seisSurfaceHeat}{}
%
\end{table}

\begin{table}[H]
\caption{Field Ground-Roll Noise 1.0.}
\label{tab:full-field-groundroll-noise1}
\centering
\scriptsize
\setlength{\tabcolsep}{2.4pt}
\renewcommand{\arraystretch}{1.0}
\taskheat{seisSurfaceHeat}{}
%
\end{table}

\subsection{Multiple attenuation}

\begin{table}[H]
\caption{Marine Multiples Attenuation Dataset.}
\label{tab:full-multiples-attenuation}
\centering
\scriptsize
\setlength{\tabcolsep}{2.4pt}
\renewcommand{\arraystretch}{1.0}
\taskheat{seisMultipleHeat}{}
%
\end{table}

\subsection{Deblending}

\begin{table}[H]
\caption{Common-Receiver Deblending T02\_mod.}
\label{tab:full-blending-noise-t02-mod}
\centering
\scriptsize
\setlength{\tabcolsep}{2.4pt}
\renewcommand{\arraystretch}{1.0}
\taskheat{seisDeblendingHeat}{}
%
\end{table}

\begin{table}[H]
\caption{Common-Receiver Deblending T02\_comp.}
\label{tab:full-blending-noise-t02-comp}
\centering
\scriptsize
\setlength{\tabcolsep}{2.4pt}
\renewcommand{\arraystretch}{1.0}
\taskheat{seisDeblendingHeat}{}
%
\end{table}

\begin{table}[H]
\caption{Common-Receiver Deblending T02\_simp.}
\label{tab:full-blending-noise-t02-simp}
\centering
\scriptsize
\setlength{\tabcolsep}{2.4pt}
\renewcommand{\arraystretch}{1.0}
\taskheat{seisDeblendingHeat}{}
%
\end{table}

\begin{table}[H]
\caption{AVO Common-Receiver Deblending T03\_avo\_mod.}
\label{tab:full-blending-noise-avo-t03-avo-mod}
\centering
\scriptsize
\setlength{\tabcolsep}{2.4pt}
\renewcommand{\arraystretch}{1.0}
\taskheat{seisDeblendingHeat}{}
%
\end{table}

\subsection{First-arrival picking}

\begin{table}[H]
\caption{First-Break Picking Multi-Dataset.}
\label{tab:full-fbp-geomseg-all}
\centering
\scriptsize
\setlength{\tabcolsep}{3.0pt}
\renewcommand{\arraystretch}{1.0}
\taskheat{seisFirstArrivalHeat}{}
%
\end{table}

\begin{table}[H]
\caption{Brunswick First-Break Picking.}
\label{tab:full-fbp-brunswick-valid}
\centering
\scriptsize
\setlength{\tabcolsep}{3.0pt}
\renewcommand{\arraystretch}{1.0}
\taskheat{seisFirstArrivalHeat}{}
%
\end{table}

\begin{table}[H]
\caption{Halfmile First-Break Picking.}
\label{tab:full-fbp-halfmile-valid}
\centering
\scriptsize
\setlength{\tabcolsep}{3.0pt}
\renewcommand{\arraystretch}{1.0}
\taskheat{seisFirstArrivalHeat}{}
%
\end{table}

\begin{table}[H]
\caption{Lalor First-Break Picking.}
\label{tab:full-fbp-lalor-valid}
\centering
\scriptsize
\setlength{\tabcolsep}{3.0pt}
\renewcommand{\arraystretch}{1.0}
\taskheat{seisFirstArrivalHeat}{}
%
\end{table}

\endgroup